\documentclass[11pt]{article}
\usepackage{graphicx}
\usepackage{natbib} 
\usepackage{url} 
\usepackage{amsmath}
\usepackage{amsfonts}
\usepackage{amssymb} 
\usepackage{lineno}
\usepackage{setspace}
\usepackage{lscape} 
\usepackage{float} 
\usepackage{breakcites}
\usepackage{pdfpages}
\usepackage{bm}
\usepackage{xcolor}
\usepackage{hyperref}
\usepackage{amsmath}
\usepackage{amssymb}
\usepackage{graphicx}
\usepackage{mathtools}
\usepackage{float}
\usepackage{tabularx}
\usepackage{natbib} 
\usepackage{wrapfig}
\usepackage{multirow}
\usepackage{tabu}
\usepackage{booktabs}
\usepackage{pgfplots}
\usepackage{pgfplotstable}
\usepackage{qtree}
\usepackage{xcolor}
\usepackage{listings}
\usepackage{caption}
\usepackage{siunitx}
\usepackage{makecell}
\usepackage{bm}
\usepackage{subfig}
\usepackage{titlesec}
\usepackage{url}
\usepackage{colortbl}
\usepackage{multicol}
\usepackage{dsfont}
\usepackage{tikz}
\usepackage{amsthm}
\usepackage{verbatim}
\usepackage[font=small,labelfont=bf]{caption}
\usepackage{algorithm}
\usepackage{titling}
\usepackage[algo2e]{algorithm2e} 

\definecolor{olive}{rgb}{0.3, 0.4, .1}
\definecolor{fore}{RGB}{249,242,215}
\definecolor{back}{RGB}{51,51,51}
\definecolor{title}{RGB}{255,0,90}
\definecolor{dgreen}{rgb}{0.,0.6,0.}
\definecolor{gold}{rgb}{1.,0.84,0.}
\definecolor{JungleGreen}{cmyk}{0.99,0,0.52,0}
\definecolor{BlueGreen}{cmyk}{0.85,0,0.33,0}
\definecolor{RawSienna}{cmyk}{0,0.72,1,0.45}
\definecolor{Magenta}{cmyk}{0,1,0,0}

\newcounter{savedenum}

\newcommand\bzero{\mbox{\boldmath${0}$}}
\newcommand\bbe{\mbox{\boldmath${ \beta}$}}

\newcommand\bfeta{\mbox{\boldmath${\eta}$}}
\newcommand\bga{\mbox{\boldmath${\gamma}$}}

\newcommand\bgPhi{\mbox{\boldmath${\Phi}$}}

\newcommand\btheta{\mbox{\boldmath${\theta}$}}

\newcommand\bdel{\mbox{\boldmath${\delta}$}}

\newcommand\bomega{\mbox{\boldmath${\omega}$}}

\newcommand\bSig{\mbox{\boldmath${\Sigma}$}}

\newcommand\bD{{\bf D}}

\newcommand\bH{{\bf H}}

\newcommand\bI{{\bf I}}

\newcommand\mcN{{\mathcal N}}

\newcommand\bOne{{\bf 1}}

\newcommand\bR{{\bf R}}

\newcommand\bS{{\bf S}}

\newcommand\bs{{\bf s}}

\newcommand\bv{{\bf v}}

\newcommand\bW{{\bf W}}
\newcommand\bw{{\bf w}}
\newcommand\bX{{\bf X}}

\newcommand\bZ{{\bf Z}}

\newcommand\bprob{{\bf p}}

\newcommand\tbX{\tilde {\bf X}}

\newcommand\mbR{{\mathbb R}}

\newcommand{\tb}[1]{\textbf{#1}}

\newcommand{\tblue}[1]{\textcolor{blue}{#1}}
\newcommand{\tred}[1]{\textcolor{red}{#1}}

\newcommand{\ssq}{\sigma^{2}}

\newcommand\bmu{\mbox{\boldmath${\mu}$}}
\newcommand\blamb{\mbox{\boldmath${\lambda}$}}

\newcounter{saveenumi}

\DeclareMathOperator*{\argmax}{\arg\max}

\newcommand\asi{\alpha_\sigma}
\newcommand\bsi{\beta_\sigma}

\newcommand\tbmu{{\tilde{\bmu}}}
\newcommand\tbC{\tilde {\bf C}}
\newcommand\talpha{\tilde {\alpha}}
\newcommand\tibeta{\tilde {\beta}}

\newcommand{\pr}[1]{{#1}^{\prime}}

\usetikzlibrary{shapes.geometric, arrows}

\tikzstyle{startstop} = [rectangle, rounded corners, minimum width=3cm, minimum height=1cm,text centered, draw=black, fill=red!30]

\tikzstyle{io} = [trapezium, trapezium left angle=70, trapezium right angle=110, minimum width=3cm, minimum height=1cm, text centered, draw=black, fill=blue!30]

\tikzstyle{process} = [rectangle, minimum width=3cm, minimum height=1cm, text centered, draw=black, fill=orange!30]
\tikzstyle{decision} = [diamond, minimum width=3cm, minimum height=1cm, text centered, draw=black, fill=green!30]

\tikzstyle{arrow} = [thick,->,>=stealth]

\graphicspath{{Figures/}}

\hypersetup{
    colorlinks=true,
    linkcolor=black,
    filecolor=magenta,      
    urlcolor=cyan,
    pdftitle={Overleaf Example},
    pdfpagemode=FullScreen,
    }

\usepackage{mathtools}

\newcommand{\interior}[1]{%
 {\kern0pt#1}^{\mathrm{o}}%
}
\usepackage [english]{babel}
\usepackage [autostyle, english = american]{csquotes}
\MakeOuterQuote{"}

\usepackage{caption}
\usepackage[makeroom]{cancel}

\newcommand{\blind}{1}

\pgfplotsset{compat=1.17} % To avoid errors

\let\oldcases\cases
\let\oldendcases\endcases
\renewenvironment{cases}{\setstretch{1}\oldcases}{\oldendcases}

\begin{document} 

\def\spacingset#1{\renewcommand{\baselinestretch}%
{#1}\small\normalsize} \spacingset{1}

%%%%%%%%%%%%%%%%%%%%%%%%%%%%%%%%%%%%%%%%%%%%%%%%%%%%%%%%%%%%%%%%%%%%%%%%%%%%%%
\date{}
\if1\blind
{
  \title{\bf A Scalable Variational Bayes Approach to Fit High-dimensional Spatial Generalized Linear Mixed Models}
  \author{Jin Hyung Lee \\
    Department of Statistics, George Mason University\\
    and \\
    Ben Seiyon Lee\thanks{
    Corresponding Author: slee287@gmu.edu}\hspace{.2cm}\\
    Department of Statistics, George Mason University}
  \maketitle
} \fi

\if0\blind
{
  \title{\bf  A Scalable Variational Bayes Approach to Fit High-dimensional Spatial Generalized Linear Mixed Models}
  \maketitle
} \fi

\bigskip

\thispagestyle{empty}

\begin{abstract}
Gaussian and discrete non-Gaussian spatial datasets are common across fields like public health, ecology, geosciences, and social sciences. Bayesian spatial generalized linear mixed models (SGLMMs) are a flexible class of models for analyzing such data, but they struggle to scale to large datasets. Many scalable Bayesian methods, built upon basis representations or sparse covariance matrices, still rely on posterior sampling via Markov chain Monte Carlo (MCMC).  Variational Bayes (VB) methods have been applied to SGLMMs, but only for small areal datasets. We propose two computationally efficient VB approaches for analyzing moderately sized and massive (millions of locations) Gaussian and discrete non-Gaussian spatial data in the continuous spatial domain. Our methods leverage semi-parametric approximations of latent spatial processes and parallel computing to ensure computational efficiency. The proposed methods deliver inferential and predictive performance comparable to gold-standard MCMC methods while achieving computational speedups of up to 3600 times. In most cases, our VB approaches outperform state-of-the-art alternatives such as INLA and Hamiltonian Monte Carlo. We validate our methods through a comparative numerical study and applications to real-world datasets. These VB approaches can enable practitioners to model millions of discrete non-Gaussian spatial observations on standard laptops, significantly expanding access to advanced spatial modeling tools.
\end{abstract}

\noindent%
\vfill

\newpage
\spacingset{2} % DON'T change the spacing!

\setcounter{page}{1}
\section{Introduction}
\label{sec1}
Spatially-dependent non-Gaussian datasets are prevalent in many scientific applications with examples including remotely sensed aerosol optical depth \citep{wei2019modis} or cloud cover \citep{sengupta2013hierarchical}, counts of bird species in ecological surveys \citep{guan2018computationally}, and ice thickness measurements in glaciological studies \citep{fretwell2013bedmap2}. Modern data collection mechanisms have led to exponential growth of these datasets with over millions of observation locations. In addition, these datasets exhibit complex spatial dependencies, which can be non-stationary over heterogeneous spatial domains. 

Spatial generalized linear mixed models (SGLMMs) \citep{diggle1998model,zhang2002estimation} are a flexible class of models amenable to both Gaussian and many non-Gaussian spatial observations. SGLMMs model the random effects as realizations from a latent Gaussian process (GP) with a spatial covariance function. Within the Bayesian hierarchical modeling framework, statistical inference proceeds by sampling from the posterior distributions using Markov Chain Monte Carlo (MCMC), which can be prohibitive when: (i) the latent variables are high-dimensional; (ii) spatial process models require costly multivariate Gaussian density evaluations; and (iii) the heavily cross-correlated spatially dependent random effects lead to slow mixing Markov chains. For a review of existing approaches for Gaussian spatial data, please see \citet{bradley2016comparison} and \citet{heaton2018case}. 

Scalable methods have been developed, but even these have limitations. Conjugate methods \citep{bradley2020bayesian,polson2013bayesian,albert1993bayesian,de2000bayesian} employ efficient Gibbs samplers but still require MCMC-based posterior sampling. Integrated nested Laplace Approximations (INLA) \citep{rue2009approximate}, in conjunction with stochastic partial differential equations \citep{lindgren2010explicit}, provides approximations of the marginal posterior distributions, but it can underestimate the uncertainty in estimation and predictions \citep{ferkingstad2015improving} and assumes normality of the approximate distributions \citep{han2013integrated}. The Vecchia-Laplace approximation \citep{zilber2021vecchia} bypasses sampling by using a Laplace approximation of the latent spatial random effects, but it is restricted to parametric covariance functions and moderately large datasets ($N=250k$).

Variational Bayes (VB) \citep{blei2006variational,blei2017variational,jordan1999introduction} approaches approximate the posterior distribution using a variational function that minimizes the Kullback-Leibler (KL) divergence between a itself and the posterior. Mean Field Variational Bayes (MFVB) \citep{blei2006variational}, hybrid MFVB methods~\citep{wang2013variational,tran2021practical} and Integrated non-factorized variational Bayes (INFVB) \citep{han2013integrated} are popular variants of VB that can scale to larger datasets. 

To the best of our knowledge, only a few studies have extended VB methods to non-Gaussian SGLMMs for areal spatial data, and none have tackled massive non-Gaussian point-referenced spatial datasets in the continuous spatial domain. \citet{ren2011variational} propose a VB approach to fit a Gaussian spatial random effects model in the continuous spatial domain, but these do not scale to larger datasets. \citet{wu2018fast} employs INFVB to accurately quantify the posterior variances; however, the proposed approaches apply to a narrow class of Gaussian spatial models on the discrete spatial domain (areal). Similarly, \citet{song2022bayesian} examines Gaussian spatial data on a graph structure. \citet{bansal2021fast} and \citet{parker2022computationally} model count and binary areal spatial datasets using Poly\'a Gamma mixtures to exploit conjugacy and bypass expensive expectations. \citet{parker2022computationally} utilizes a product-form MFVB, which can lead to severe underestimation of the posterior variances. \citet{bansal2021fast} employs INFVB, yet their approaches apply to a narrow class of spatial data --- areal count data observed at a small number of locations. 

We propose a computationally-efficient VB approach for modeling massive Gaussian and discrete non-Gaussian datasets within the continuous spatial domain. Our method is specifically tailored for SGLMMs, where the spatial random effects are either modeled as Gaussian processes with parametric covariance functions (full-SGLMMs) or spatial basis expansions (basis-SGLMMs). As far as we are aware, this study represents the first attempt to develop a scalable variational Bayes methods or modeling massive non-Gaussian spatial datasets ($N>1\times10^{4}$) within the continuous spatial domain. Furthermore, we include an extensive comparative analysis of our proposed VB approaches, MCMC algorithms, and INLA across a wide array of spatial datasets. A pivotal aspect of our proposed methodology is its accessibility to non-experts, facilitating easy fine-tuning of SGLMMs to suit their preferences (e.g., incorporating/excluding covariates, modifying covariance functions, or adjusting spatial basis functions).

We provide an overview of SGLMMs in Section~\ref{Sec:SGLMM}. We discuss variational Bayes methods in Section~\ref{Sec:Variational} and introduce our proposed VB approaches in Section~\ref{Sec:VB_SGLMM}. A comparative analysis is then performed across multiple competing methods using numerical studies (Section~\ref{Sec:Simulation}) and real data examples (Section~\ref{Sec:RealData}). Finally, Section~\ref{Sec:Discussion} provides a summary of the study, discusses limitations, and proposes directions for future research.

\section{Spatial Generalized Linear Mixed Models (SGLMMs)} \label{Sec:SGLMM}
Spatial generalized linear mixed models (SGLMMs) \citep{diggle1998model} are a highly flexible class of spatial models that can accommodate non-Gaussian observations, such as binary \citep{hanks2015restricted}, count \citep{guan2018computationally} and positive-valued continuous data \citep{zilber2021vecchia}. Let $\bZ=\{\bZ(s_i)\}_{i=1}^{N}$ denote the observations collected at locations $\bs_i \in \bS \subseteq \mathbb R^2$ and $\bX \in \mathbb R^{N \times p}$ be the matrix of the corresponding covariates. Spatial dependence is often induced through the spatial random effects $\bomega=\{ \bomega(\bs_i)\}_{i=1}^N\in \mathbb R^N$, which can be modeled as a zero-mean Gaussian Process with covariance function $C(\Psi)$ and the associated covariance parameters $\Psi$. For a finite set of locations, the spatial random effects $\bomega$ follow a multivariate normal distribution $\bomega\sim \mcN(\bzero,\bSig(\Psi) )$ with covariance matrix $\bSig(\Psi) \subset \mathbb R^{N \times N}$ such that $\bSig(\Psi)_{ij}=C(\Psi)_{ij}$ for locations $\bs_i$ and $\bs_j$. A popular class of stationary and isotropic covariance function is the Mat\'ern class \citep{stein1999interpolation} with covariance parameters $\Psi=\{\ssq, \phi, \nu\}$ (see supplement for details).  For SGLMMs, the resulting Bayesian hierarchical model is:
\begin{align}\label{EQ:SGLMM_Full}
&\mbox{Data Model: } & \bZ\mid \bfeta  \sim F( \cdot \mid \bfeta)  \nonumber \\
& &g(\mathbb{E}[\bZ\mid\bbe, \bomega]):=\bfeta =\bX \bbe+ \bomega \nonumber\\
&\mbox{Process Model: } & {\bomega \mid\Psi\sim\mathcal N\left(\bzero,  \bSig(\Psi)\right)}\\
&\mbox{Parameter Model: } & \bbe \sim p\left(\bbe\right), \Psi \sim p\left(\Psi\right) \nonumber .
\end{align} 

\noindent where $F(\cdot)$ is a valid probability distribution (e.g., Benoulli for binary data, Poisson for counts), 
$g(\cdot) $ is the link function, $\bfeta$ is the linear predictor and $p\left(\bbe\right)$ and $p\left(\Psi\right)$ denote the prior distributions for $\bbe$ and $\Psi$, respectively. We infer the unknown parameter sets $\bbe$ and $\Psi$ as well as the $N$-dimensional spatial random effects $\bomega$ by obtaining the posterior distribution $\pi(\bbe,\Psi, \bomega|\bZ)$, which is not typically available in analytical form. Hence, $\pi(\bbe,\Psi, \bomega|\bZ)$ is commonly approximated using a sampling-based approach such as MCMC. 

Fitting SGLMMs to large datasets (i.e., $N>1\times10^{4}$) can be computationally prohibitive due to repeated operations on large matrices and inferring the highly-correlated and $N$-dimensional spatial random effects $\bomega$. To illustrate, the process model in (\ref{EQ:SGLMM_Full}) requires evaluating $|\bSig(\Psi)|$ and $\bSig(\Psi)^{-1}$ with costs scaling at $\mathcal{O}(\frac{1}{3}N^3)$. Next, the high-dimensional and highly correlated $\bomega$ can result in slow-mixing Markov chains in MCMC \citep{haran2003accelerating,raftery1996implementing} and difficulties inferring all $N$ random variables.

\subsection{Basis Representations} \label{sec:BasisRepresentation}
To fit SGLMMs on large spatial datasets, basis expansions have been used to approximate the latent spatial random processes \citep{cressie2015statistics, cressie2022basis}, particularly for non-Gaussian spatial data \citep{sengupta2013hierarchical,bradley2016comparison,lee2022picar,lee2023scalable}. The latent spatial random process  $\{ \bomega(\bs_i)\}_{i=1}^N $ is approximated by a linear combination of $m$ spatial basis functions $\{ \bgPhi_j (\bs) \}_{j=1}^{m}$. Specifically, ${\bomega}\approx \bgPhi{\bdel}$ with $N\times m$ basis functions matrix $\bgPhi=\begin{bmatrix}
    \bgPhi_1 & \bgPhi_2 & \cdots & \bgPhi_m
\end{bmatrix}$ which consists of basis functions $\bgPhi_j\in \mbR^{N}$ with components corresponding to each location $\bs_i$ and basis coefficients $\bdel\in \mbR^m$.

The Bayesian hierarchical model for basis representation SGLMMs (basis-SGLMMs) is:
\begin{align}\label{EQ:basisSGLMM}
&\mbox{Data Model: } & \bZ\mid \bfeta  \sim F( \cdot \mid \bfeta) \nonumber \\
& &g(\mathbb{E}[\bZ \mid\bbe, \bdel]):=\bfeta =\bX\bbe+ \bgPhi{\bdel} \nonumber\\
&\mbox{Process Model: } &{\bdel \mid\zeta\sim\mathcal N\left(\bzero, \bSig(\zeta)\right)}\\
&\mbox{Parameter Model: } & \bbe \sim p\left(\bbe\right), \zeta \sim p\left(\zeta\right) \nonumber
\end{align}

\noindent where $\bSig(\zeta)$ is the prior covariance matrix for the basis coefficients $\bdel$ with covariance parameters $\zeta$. One example is the independent and identically distributed case where $\bSig(\zeta)= \mbox{Diag}_m(\tau^2)$ and $\zeta=\tau^2$. To complete the hierarchical model, we specify a prior distribution for $\zeta \sim p\left(\zeta\right)$. Since $\bdel \in \mathbb R^m$ where $m<<N$, basis representations offer substantial dimension reduction and considerably decrease the computational overhead. Additionally, the design of the basis functions $\bgPhi_j$ can help reduce correlation in the estimable basis coefficients $\bdel$, resulting in faster-mixing Markov chains \citep{haran2003accelerating}. A wide array of spatial basis functions have been explored in the literature, including bi-square (radial) basis functions \citep{cressie2008fixed, nychka2015multiresolution, katzfuss2017multi}, empirical orthogonal functions \citep{cressie2015statistics}, wavelets \citep{nychka2002multiresolution}, and multiresolution basis functions \citep{nychka2015multiresolution, katzfuss2017multi}.

Although the dimension reduction achieved through the basis representation significantly improves computational efficiency, MCMC methods are still required to sample from the posterior distributions $\pi(\bbe,\zeta, \bdel|\bZ)$. The dominating cost of basis-SGLMMs is the matrix-vector multiplication  $\bgPhi{\bdel}$ which incurs $\mathcal{O}(Nm)$ in costs. Therefore, MCMC can be computationally prohibitive for cases with large $N$ (observations in the millions) and large $m$ (basis functions).

\section{Variational Inference}\label{Sec:Variational}
Variational Bayes (VB) methods approximate a posterior distributions by optimizing a simpler tractable distribution (variational function) to be close to the target. For a family of distributions $Q$, VB methods select a distribution  $q^{*}(\btheta) \in Q$ that minimizes the Kullback-Leibler (KL) divergence \citep{jordan1999introduction,blei2006variational} to a target function like the posterior $ p(\btheta|Z)$ such that $q^{*}(\btheta) = \text{argmin}_{q \in Q} \text{KL}(q||p(\btheta|\bZ))$. The KL divergence is:

\begin{equation}\label{EQ:KL}
\begin{aligned}
    \text{KL}(q||p(\btheta|\bZ)) &=\mathbb{E}_q\Big[\log \frac{q(\btheta)}{ p(\btheta|\bZ)}\Big]= - \int q(\btheta) \log \frac{p(\bZ|\btheta) \cdot p(\btheta)}{q(\btheta)}d\btheta + \log p(\bZ).
\end{aligned}
\end{equation}
\noindent Since the KL divergence is always non-negative, minimizing $\text{KL}(q||p(\btheta|\bZ))$ is equivalent to maximizing the lower bound on $\log p(\bZ) $, or the Evidence Lower Bound (ELBO):
\begin{equation}\label{EQ:ELBO}
    ELBO(q):= \int q(\btheta) \text{log} \frac{p(\bZ|\btheta) \cdot p(\btheta)}{q(\btheta)}d\btheta = \mathbb{E}_{q} \left( \text{log} \frac{p(\bZ|\btheta) \cdot p(\btheta)}{q(\btheta)} \right).
\end{equation}

\noindent Two challenges include: (1) specifying the family for the variational function $q^{*}(\btheta)$  and (2) implementing distributional constraints  \citep{tran2021practical}. Without any constraints, the variational function that minimizes the KL divergence is merely the posterior $p(\btheta|\bZ)$, which is in itself intractable. In this study, we focus on three VB approaches - mean field variational Bayes (MFVB), Hybrid MFVB, and integrated non-factorized variational Bayes (INFVB).  

\paragraph{Mean Field Variational Bayes (MFVB)}
Mean Field Variational Bayes (MFVB) \citep{wainwright2008graphical}  constrains the variational function by imposing a product form for $q$. Here, $q^{*}(\btheta)=\prod_{l=1}^{L}q(\btheta_l)$ with parameter partitions $\btheta=(\btheta_1,...,\btheta_L)'$. The function that minimizes $\text{KL}(q(\btheta_l)||p(\btheta_l))$ is \citep{ormerod2010explaining}:
\begin{equation}\label{EQ:MFVB}
    q_l(\btheta_l)  \propto \exp \Big\{ \mathbb{E}_{\btheta_{-l}}\Big[\log   p(\btheta_l | \bZ,\btheta)\Big] \Big\},
\end{equation}

\noindent  where $\mathbb{E}_{\btheta_{-l}} $ denotes the expectation with respect to all other variables except $\btheta_{l}$. The product-form variational function $q^{*}(\pmb{\theta})$ is obtained by cycling through all $\btheta_{l}$ using a coordinate ascent-type algorithm \citep{bishop2006pattern, tran2021practical}. 

However, the MFVB approach is subject to four key limitations. First, the product form (\ref{EQ:MFVB}) assumes independence across the parameter partitions $\pmb{\theta}$, which can lead to poor approximations of the posterior. Second, conjugacy in $\btheta_l$ is often needed to obtain an analytical form for $q(\btheta_l)$. Variational inference can be difficult to derive for non-conjugate models because the form in (\ref{EQ:MFVB}) may not correspond to a known parametric family of distributions. Third, the MFVB approach may result in a variational function that drastically underestimates both the posterior and posterior predictive variances \citep{blei2006variational,han2013integrated,blei2017variational}, which can result in overconfident predictions. Finally, the first moments in (\ref{EQ:MFVB}) may not be available in closed form, which necessitates expensive Monte Carlo-based approximations. Within the context of SGLMMs, these include $\mathbb{E}[\Sigma(\Psi)^{-1|}]$ or $\mathbb{E}[\log |\Sigma(\Psi)|]$ for large covariance matrices. 

Hybrid MFVB, or fixed form variational Bayes (FFVB)~\citep{Salimans2013}, extends MFVB to nonconjugate cases \citep{wang2013variational}. Since the parametric family for $q_l(\theta_l)$ is unknown, $q(\theta_l)$ can be set as multivariate normal distribution using either Laplace approximations or the delta method. Despite its flexibility, Hybrid MFVB may be computationally prohibitive for models with many unknown parameters. 

\paragraph{Integrated Nonfactorized Variational Bayes (INFVB)} \label{Subsec:INFVB}
Integrated Nonfactorized Variational Bayes (INFVB) is an alternative VB approach that provides accurate representations of posterior variance \citep{han2013integrated} and reduces computational walltimes via parallelized computing. Note that MFVB imposes posterior independence constraints (\ref{EQ:MFVB}), which can lead to underestimating posterior variances in the presence of strong inter-block relationships \citep{blei2006variational, han2013integrated}. INFVB relaxes the constraints of this product form \citep{han2013integrated, wu2018fast,bansal2021fast} by constructing a  variational function based on a disjoint parameter space $\btheta=(\pmb{\theta}_c,{\pmb{\theta}_d})'$ resulting in the variational function $q_{\text{INFVB}}(\btheta)=q(\btheta_c|{\pmb{\theta}_d}){q(\pmb{\theta}_d)}$.
Replacing $q(\btheta)$ with $q_{\text{INFVB}}(\btheta)$ in (\ref{EQ:KL}) results in the following (see supplement~\ref*{SubSec:Supple_INFVBderivation} for derivations):  
\begin{equation}\label{EQ:INFVB_continuous}
    q^{*}_{\text{INFVB}}(\btheta)= \text{argmin}_{q_{\text{INFVB}}} \int q(\btheta_d) \left[ \int q(\btheta_c|\btheta_d) \text{log} \frac{q(\btheta_c|\btheta_d)}{p(\btheta_c,\btheta_d|\bZ)} d\btheta_c+\text{log} q(\btheta_d)\right]d\btheta_d.
\end{equation}

Due to the double integrals, the objective function in (\ref{EQ:INFVB_continuous}) can be difficult to optimize directly \citep{han2013integrated,wu2018fast}. To address this, the parameter sets can be discretized as $\btheta_d=\{\btheta_d^{(1)},\btheta_d^{(2)} \cdots \btheta_d^{(J)}\}$ in (\ref{EQ:INFVB_continuous}) where $j\in (1,2,...,J)$ denotes the indices for the discretized values with $J$ being the total number of discretized points. Partitioning the parameter space yields the $j$-conditional Evidence Lower Bounds (ELBOs) given $\btheta_d^{(j)}$ \citep{han2013integrated}:
\begin{equation}\label{EQ:ConditionalELBO}
\begin{aligned}
     ELBO^{(j)}: & \approx \int q(\btheta_{c}|\btheta_{d}^{(j)}) \text{log} \frac{p(\bZ|\btheta_{c},\btheta_{d}^{(j)}) \cdot p(\btheta_{c},\btheta_{d}^{(j)})}{q(\btheta_{c}|\btheta_{d}^{(j)})}d\btheta_{c} \\
     & = \mathbb{E}_{q(\btheta_{c}|\btheta_{d}^{(j)})} \left[ \text{log} \frac{p(\bZ|\btheta_{c},\btheta_{d}^{(j)}) \cdot p(\btheta_{c},\btheta_{d}^{(j)})}{q(\btheta_{c}|\btheta_{d}^{(j)})} \right]
\end{aligned}
\end{equation}

The variational functions for $q(\btheta_d)$ and $q(\btheta_c)$ are approximated using weighted averages of the conditional variational functions $q(\btheta_{c}|\btheta_{d}^{(j)})$ and the discretized function $q(\btheta_d^{(j)})$. The corresponding normalized weights are $A_j=\frac{ELBO^{(j)}}{\sum_{j=1}^{J}ELBO^{(j)}}$ with the $j$-th conditional $ELBO^{(j)}$ from (\ref{EQ:ConditionalELBO}). Specifically, $q(\btheta_d)$ is obtained by multiplying $A_j $ with the empirical distribution $1(\btheta_d=\btheta_d^{(j)})$. Similarly, $q(\btheta_c)$ is acquired by multiplying $A_j$ with $q(\btheta_c|\btheta_d^{(j)})$ as follows: 
\begin{equation}
     q(\btheta_c)= \sum_{j=1}^{J}A_j q(\btheta_c|\btheta_d^{(j)}), \hspace{0.3cm} q(\btheta_d)= \sum_{j=1}^{J}A_j 1(\btheta_d=\btheta_d^{(j)}) \hspace{0.3cm} \text{where} \quad A_j=\frac{ELBO^{(j)}}{\sum_{j=1}^{J}ELBO^{(j)}}
\end{equation}

\begin{figure}[ht]
 \begin{center}
\includegraphics[width=0.72\linewidth]{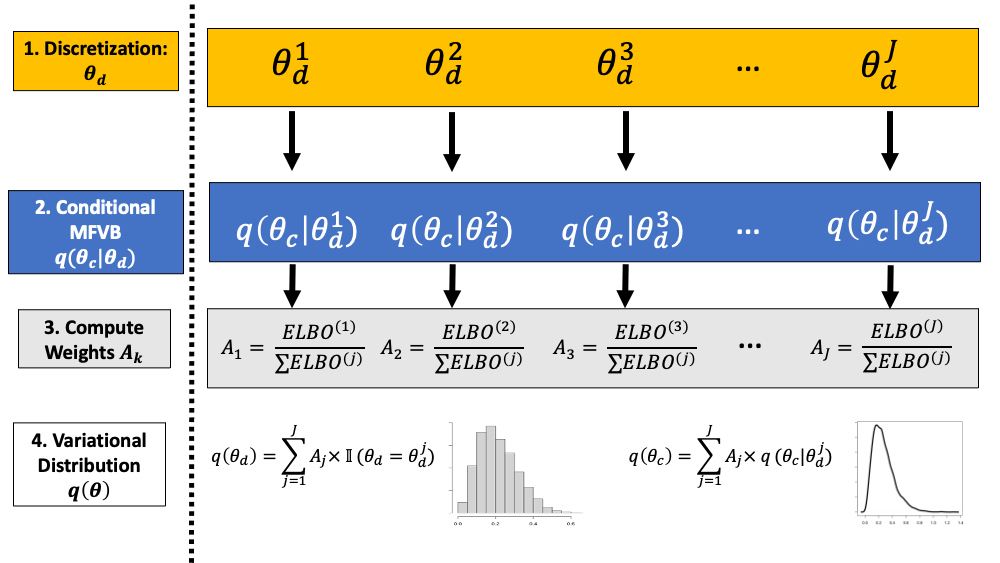}
\caption{Workflow diagram of Integrated Non-Factorized Variational Bayes (INFVB).}\label{Fig:INFVB}
\end{center}
\end{figure}

Figure~\ref{Fig:INFVB} provides an overview of the INFVB workflow and additional details for the INFVB procedure and algorithms are provided in the supplement. 

\section{Our Approach: Variational Inference for SGLMMs}\label{Sec:VB_SGLMM}
In this section, we propose two VB approaches for fitting SGLMMs. The first approach models the latent spatial process $\bomega$ as a stationary isotropic Gaussian process (full-SGLMM), consistent with traditional spatial modeling techniques, while the second approach utilizes basis expansions (basis-SGLMM) to enable scalability to massive datasets.

\subsection{Full Spatial Generalized Linear Mixed Models (full-SGLMM)}\label{Sec:VB_FullSGLMM}
 Consider the full-SGLMM (\ref{EQ:SGLMM_Full}) with the option of using one of three data models: 
\begin{align*} \label{subsec:FullSGLMM}
\tb{Data Model:} \quad 
    \textbf{\mbox{(Gaussian)}} &\qquad  \bZ| \boldsymbol{\beta},\bomega, \tau^2 \sim \mathcal{N}(\bX\boldsymbol{\beta}+\bomega,\tau^2 \bI)\\
    \textbf{\mbox{(Poisson)}} &\qquad  \bZ|\blamb \sim \mbox{Pois}(\blamb)\text{ where} \hspace{0.1cm} \blamb=\mbox{exp}(\bX\bbe+\bomega) \\
    \textbf{\mbox{(Binary)}} &\qquad  \bZ|\bprob \sim \mbox{Bern}(\bprob) \text{ where} \hspace{0.1cm} \bprob=(1+\exp\{-\bX^\prime\bbe -\bomega\})^{-1}
\end{align*}
\vspace{-1in}
\begin{align*}
\tb{Process Model:} \quad
    &\qquad  \bomega| \ssq,\phi \sim \mathcal{N}(\bzero,\ssq \bR_\phi) \\
\tb{Parameter Model:} \quad
     & \qquad \boldsymbol{\beta} \sim \mathcal{N}(\mu_{\beta},\Sigma_{\beta})
      \qquad \tau^2 \sim \mbox{IG}(\alpha_{\tau},\beta_{\tau})
      \qquad \ssq \sim \mbox{IG}(\alpha_{\sigma},\beta_{\sigma})
      \qquad \phi \sim p(\phi)
\end{align*}
\noindent Here, $\bSig(\Psi)$ in (\ref{EQ:SGLMM_Full}) becomes $\ssq\bR_\phi$ where $\ssq > 0$ is the partial sill, $\bR_\phi$ is the spatial correlation matrix parameterized by $\phi$, the range parameter in Mat\'ern class that governs the decay in spatial dependence with respect to distance. We propose two INFVB approaches differing in the choice of discretized parameter sets, specifically $\pmb{\theta}_d=\phi$ and $\pmb{\theta}_d=(\phi,\ssq)'$. In the first case ($\pmb{\theta}_d=\phi$), we begin by grouping $\bga=(\bbe,\bomega)'$ to avoid underestimating the posterior variances, which often occur in the product-form variational function \citep{blei2017variational}. Next, the parameter space is partitioned into two components: $\pmb{\theta}_c= \{\bga,\ssq\}$ (or $\pmb{\theta}_c= \{\bga,\ssq, \tau^2\}$ for the Gaussian data model) and $\pmb{\theta}_d = \phi$. We discretize $\btheta_d = \{\btheta_d^{(1)}, \btheta_d^{(2)} \cdots \btheta_d^{(J)}\}$ on a user-specified grid with $J$ points. For each $\btheta_d^{(j)}$, we separately perform MFVB on the parameter blocks within $\btheta_c$. For the $j$-th partition, initial values for $ELBO^{(j)}_{0}$, $q_{0}(\bga|\btheta_d^{(j)})$, $q_{0}(\sigma^2|\btheta_d^{(j)})$, and $q_{0}(\tau^2|\btheta_d^{(j)})$ are set. Then the INFVB algorithm proceeds until it meets the stopping criterion $|ELBO^{(j)}_{k}-ELBO^{(j)}_{k-1}|<\epsilon_{*}$ for a fixed threshold $\epsilon_{*}$. Finally, the weighted variational functions  $q(\bga)$,  $q(\ssq)$,  $q(\tau^2)$, and  $q(\phi)$ are computed using ELBO weights $A_{j}$. Additional details for constructing and updating the variational functions $q(\cdot)$ are provided in the following subsection. 

A multivariate normal prior distribution is selected for $\bga=(\bbe,\bomega)'$ where $\bga\sim \mcN(\bmu_\gamma,\bSig_\gamma)$ and $\bmu_\gamma=(\bmu_\beta, \bzero)'$ and $\Sigma_\gamma=\begin{bmatrix}
\Sigma_\beta & \bzero\\
\bzero & \ssq\bR_\phi
\end{bmatrix}$, $\ssq \sim \text{IG}(\alpha_\sigma,\beta_\sigma)$, and $\tau^2 \sim \text{IG}(\alpha_\tau,\beta_\tau)$. Discretization for the parameters in $\btheta_d$ should be reflective of the respective prior distributions (details provided in Section \ref{Subsec:Implementation}). The general INFVB model-fitting procedure is outlined in Algorithm~\ref{alg:one}. An extension using two discretized parameters $\pmb{\theta}_d=(\phi,\ssq)'$ (Algorithm~\ref*{alg:oneB}) is provided in the supplement. 

\RestyleAlgo{ruled}
\begin{algorithm}[ht]
\caption{INFVB($\phi$) Algorithm for full-SGLMMs}\label{alg:one}
\SetKw{Initialize}{Initialize}
\Initialize{Set $\epsilon_{*}$ and discretize $\{\phi_d^{(1)},\phi_d^{(2)},...,\phi_d^{(J)}\}$ for user-specified $J$.}\\
\For{j=1,...,J}{
\Initialize{Set $ELBO^{(j)}_{0}$,  $q_{0}(\bga|\phi_d^{(j)})$ , $q_{0}(\ssq|\phi_d^{(j)})$ and $k=1$}\; \\
\While{$|ELBO^{(j)}_{k}-ELBO^{(j)}_{k-1}|>\epsilon_{*}$}{
    Update $q_k(\bga|\phi_d^{(j)})$\;\\
    Update $q_k(\ssq|\phi_d^{(j)})$\;\\
    Update $ELBO^{(j)}_{k}$\;
     $k \gets k + 1$\;
}
Set $q(\bga|\phi_d^{(j)})=q_k(\bga|\phi_d^{(j)}), \quad q(\ssq|\phi_d^{(j)})=q_k(\ssq|\phi_d^{(j)}) $
}
\SetKw{Result}{Result}
\Result{Weighted Non-factorized Variational Functions}
\begin{multicols}{2}
\begin{enumerate}
    \item $A_{j}=\frac{ELBO^{(j)}}{\sum_{i=1}^{J} ELBO^{(i)}}$
    \item $q(\boldsymbol{\bga})=\Sigma_{j=1}^{J}A_{j} q(\bga|\phi_d^{(j)})$
        \item $q(\boldsymbol{\ssq})=\Sigma_{j=1}^{J}A_{j} q(\ssq|\phi_d^{(j)})$
            \item $q(\phi)=\Sigma_{j=1}^{J}A_{j} 1(\phi_d=\phi_d^{(j)})$

\end{enumerate}
\end{multicols}
\end{algorithm}

\vspace{-0.7cm}
\paragraph{Variational Functions for Gaussian, Poisson, and Bernoulli Data Models}

For Gaussian data models, conditional conjugacy exists for $\bbe$, $\bomega$ and $\ssq$; in addition, the $\bomega$ can be integrated out to reduce the number of estimable parameters. Hence, the variational functions $q(\bga|\btheta_d^{(j)})$ and $q(\ssq|\btheta_d^{(j)})$ are available in closed form (see supplement). 

On the other hand, SGLMMs with Poisson and Bernoulli data models do not benefit from conditional conjugacy for $\bga$, which can be problematic for obtaining the expectations in $q(\sigma^2)\propto \mathbb{E}_{\gamma}[\log(p(\bga,\ssq, \bZ)]$ and the $ELBO$. Monte Carlo methods can provide numerical approximations for these expectations; however, these methods can be costly as their accuracy relies on the number of Monte Carlo samples. In this study, we approximate $q(\bga)$ using a Laplace approximation \citep{wang2013variational} for Poisson data models and auxiliary variable methods \citep{jaakkola1997variational, parker2022computationally} for Bernoulli data models. Due to space limitations, details for these approximations are provided in the supplement.

\subsection{Variational Method for Basis-Representation Models} \label{SubSec:BasisSGLMM}
The Bayesian hierarchical model for basis-SGLMMs (\ref{EQ:basisSGLMM}) is provided below and the accompanying VB algorithms are provided in the supplement (Algorithms~\ref*{alg:two} and ~\ref*{alg:twoB}).
\vspace{-0.7cm}
\begin{align*}
\tb{Data Model:} \quad
     \textbf{\mbox{(Gaussian)}} &\qquad  \bZ| \boldsymbol{\beta},\bdel, \tau^2 \sim \mathcal{N}(\bX\boldsymbol{\beta}+\bgPhi\bdel,\tau^2 \bI)\\
    \textbf{\mbox{(Poisson)}} &\qquad  \bZ|\blamb \sim \mbox{Pois}(\blamb)\text{ where} \hspace{0.1cm} \blamb=\mbox{exp}(\bX\bbe+\bgPhi\bdel) \\
    \textbf{(Binary)} &\qquad  \bZ|\bprob \sim \mbox{Bern}(\bprob) \text{ where} \hspace{0.1cm} \bprob=(1+\exp\{-\bX^\prime\bbe -\bgPhi\bdel\})^{-1}
\end{align*} \vspace{-2.5cm}
\begin{align*}
\tb{Process Model:} \quad
    &\qquad  \bdel| \ssq \sim \mathcal{N}(0,\ssq \bSig_\delta) \\
\tb{Parameter Model:} \quad
     & \qquad \boldsymbol{\beta} \sim \mathcal{N}(\mu_{\beta},\Sigma_{\beta}),
    & \qquad \tau^2 \sim \mbox{IG}(\alpha_{\tau},\beta_{\tau}),
     & \qquad \ssq \sim \mbox{IG}(\alpha_{\sigma},\beta_{\sigma})
\end{align*}
\noindent We assume that the basis functions contained in $\bgPhi$ are fixed prior to model-fitting. Basis-SGLMMs can be fitted using both Hybrid MFVB and INFVB methods, as the only estimable parameters are $\bga = (\bbe, \bdel)'$ and $\ssq$ for count and binary data, and additionally $\tau^2$ for Gaussian data, all of which have amenable variational functions. For the Hybrid MFVB approach, Algorithm~\ref*{alg:two} outlines the procedural steps. For initialization, we set $q_{\gamma}^{(0)}$,$q_{\sigma^2}^{(0)}$, $\epsilon_{*}$, and $ELBO_0$. The Hybrid MFVB algorithm updates $q_{\gamma}^{(k)}$ and $q_{\sigma^2}^{(k)}$ iteratively until reaching a stopping criterion. Prior distributions are chosen similarly to the full-SGLMM case. Modifications for $q(\bga)$ are needed for count and binary data, akin to the approach outlined in Section~\ref{Sec:VB_FullSGLMM}. As in the previous subsection, we employ normal approximations for $q_k(\gamma)$; for instance, the Laplace Approximation for count data models and the quadratic approximation~\citep{jaakkola1997variational} for binary data models. The INFVB method discretizes $\pmb{\theta}_d=\ssq$ across $J$ partitions $\btheta_d = \{\btheta_d^{(1)}, \btheta_d^{(2)} \cdots \btheta_d^{(J)}\}$. Algorithm \ref*{alg:twoB} outlines the procedure, which results in weighted variational functions  $q(\bga)$ and $q(\ssq)$ from $A_{j}$. Please see the supplement for necessary derivations regarding each case (Hybrid MFVB and INFVB for the Poisson and Bernoulli data models).

\subsection{Implementation Details}\label{Subsec:Implementation}

The proposed variational methods include important tuning parameters. This includes the number of points $J$ for the discretized parameters $\btheta_d$ in INFVB, the spacing of the discretized parameter sets, and the threshold $\epsilon_{*}$ for stopping the INFVB and Hybrid MFVB algorithm. 

In both INFVB implementations for full-SGLMMs, we discretize $\phi$ into $J=1,000$ values and the parameter pairs $(\phi,\ssq)$ into $J=10,000$ values. Discretizing $\ssq$ in INFVB is a non-trivial challenge because the inverse gamma prior distribution does not have a finite upper bound. In practice, we suggest implementing both the Hybrid MFVB and INFVB approach in that order. Hybrid MFVB uses fewer computational resources and may yield similar results to the INFVB approach in some cases (e.g., basis SGLMMs). However, Hybrid MFVB may underestimate the posterior variances of the model parameters and predictions, making INFVB necessary to correctly approximate the posterior variances. For INFVB, we recommend selecting the bounds for $\ssq$ based on the quantiles from the resulting variational function $q(\ssq)$ obtained via Hybrid MFVB. For $q(\ssq)$, we set the lower and upper bounds to be 0 and 2000, respectively. 

INFVB contains embarrassingly parallel operations, which can be distributed across multiple processors. We employed 30 cores for each implementation of INFVB. If possible, we recommend using $J$ cores (i.e., one core per discretized value of $\btheta_d^{(j)}$). For basis-SGLMMs, we explored scenarios with 20, 50, and 100 basis functions. Our proposed VB methods are extremely fast; hence, the practitioner can easily test a wide range of bases and select the appropriate model. Finally, we set our stopping criterion  $\epsilon_*=1 \times 10^{-4}$. Increasing the stopping criterion would end the algorithm earlier but may compromise accuracy.

We employ a VB-subsampling approach to estimate the linear predictor for location $i$, $\bfeta_i=\bX_i'\bbe +\bomega_i$  and $\bfeta_i=\bX_i'\bbe +\bgPhi_i'\bdel$ in the full- and basis-SGLMM cases, respectively. The subsampling approach is embarrassingly parallel and yields posterior distributions comparable to those from MCMC. Further details are provided in the supplementary section~\ref*{Subsec:EstLinearPredic}.

\section{Simulation Study}\label{Sec:Simulation}
We demonstrate our proposed VB approaches through a comparative simulation study that evaluates multiple simulated datasets varying in size, strength of spatial dependence, and data type, including Gaussian, binary, and count data. We perform three types of comparative analyses: (1) an extensive comparative analysis fitting full-SGLMMs on moderately-large datasets ($N=500$); (2) basis-SGLMMs on large datasets ($N=25,000$); and (3) a separate comparison between VB-based methods and Hamiltonian Monte Carlo (HMC) using the No U-Turn Sampler (NUTS) implemented in \texttt{RStan}. The VB- and MCMC-based are implemented on a high-performance computing infrastructure with walltimes based on a single 2.4 GHz Intel Xeon Gold 6240R processor.INLA was implemented on an Apple MacBook Pro laptop equipped with an M1 Pro chip.

\subsection{Case 1: Moderately-Large Datasets with Full-SGLMMs} \label{SubSec:SimFullSGLMM}
\paragraph{Simulation Study Design}

We randomly select locations  $\bs_i \in \mathcal D=[0,1]^2 \subset \mathbb R^2$ for $i= 1, \dots, N$, where $\mathcal D$ represents the spatial domain. Each dataset consists of $N=500$ locations divided into $N_{\text{train}}=400$ for training and $N_{\text{test}}=100$ for testing. The vector of observations $\bZ=(Z(\bs_1),..., Z(\bs_N))^T $ is generated using the SGLMM framework in Section~\ref{Sec:VB_FullSGLMM} with covariates $\bX=[\bX_1, \bX_2]$ where $\bX_1, \bX_2 \sim \text{Unif}(-1,1)$ and $\bbe=(1,1)$. Four different sets of spatial random effects $\bomega={\{\bomega(s_i):s_i \in \mathcal D\}}$ are generated from a zero-mean Gaussian Process with Mat\'ern covariance function with smoothness $\nu=0.5$, partial sill $\ssq=1$, and range parameters $\phi=\{0.1, 0.3, 0.5, 0.7\}$. We generate observations $\bZ$ for the Gaussian, binary, and count datasets using the identity, logit, and log link functions, respectively.

We generated 50 replicate datasets for each of the 12 scenarios. For this simulation study, we compare the two proposed INFVB methods, the Metropolis---Hastings algorithm, and INLA. The INFVB($\phi$) approach discretizes only $\phi$, whereas INFVB($\phi, \ssq$) discretizes two parameters ($\phi, \ssq$). See Algorithm \ref{alg:one} and \ref*{alg:oneB} for details. For MCMC, we obtained 100,000 posterior samples with convergence assessed using batch means standard errors (BMSE) \citep{flegal2008markov} and visual heuristics of trace plots. For the INFVB methods, we set the stopping as threshold as $\epsilon_{*}=1\times10^{-4}$ (see Algorithm \ref{alg:one}). Details on discretizations for $\phi$ and $\ssq$ are provided in Section~\ref{Subsec:Implementation}. INLA \citep{rue2009approximate} employs stochastic partial differential equations \citep{lindgren2015bayesian} to deliver numerical Gaussian approximations of the marginal posterior distributions. Implementations are done through the R-INLA package available at \href{www.r-inla.org}{www.r-inla.org}.

To complete the hierarchical model (Section~\ref{Sec:VB_FullSGLMM}), we set parameter models (priors) $ \bbe_j \sim \mathcal{N}(0,100)$, $\tau^2 \sim \text{IG}(0.1, 0.1)$, $\ssq \sim \text{IG}(0.1, 0.1)$ and $\phi \sim \text{Unif}(0,\sqrt{2})$. We evaluate predictive performance using root mean squared prediction error $\text{RMSPE}=\left(\sqrt{\frac{1}{N_{\text{test}}}\sum_{i=1}^{N_{\text{test}}}\left(Z_{i}-\hat{Z}_{i}\right)^2}\right)$ for the Gaussian and count data and the area under the receiver operating characteristic curve (AUC) for the binary case. To evaluate interval inference, we consider the continuous ranked probability score (CRPS) \citep{hersbach2000decomposition} and compute the coverage of the the 95$\%$ credible intervals. The latter represents the frequency with which the 95\% credible interval for the linear predictor encompasses the true linear predictor. Please see supplment for the CRPS and coverage results. 

\paragraph{Results} 
The results for the binary and count data are summarized in Table~\ref{Tab1:FSGLMMBinaryCount500} and those for the Gaussian case are in the supplement~\ref*{Tab:GaussianFullsupplement}. Our proposed VI methods and MCMC have near-identical predictive performance (RMSPE and AUC). However, the VI-based methods exhibit dramatic computational speedups ranging from factors of 340-to-512 for the binary case and 18-to-20 for count data. Note that INFVB for count data requires an additional Laplace approximation step; hence, the computational speedup is less pronounced than the binary cases. 
Though INLA exhibits a substantial computational speedup for the full-SGLMM, achieving up to a 3956-fold improvement compared to MCMC-based methods (the gold standard), it yields lower AUC for binary data and higher RMSPE for count data. MCMC, INFVB($\phi$), and INFVB($\phi,\ssq$) yield comparable results for CRPS (Table~\ref*{Tab:BinaryCountCRPSCImethod}). Despite its computational speedup, INLA underperforms the other approaches in both CRPS and coverage.

\begin{table}[ht]
\begin{center} 
\scalebox{0.9}{
\begin{tabular}{ccccccccc} \toprule 
Binary &\multicolumn{4}{c}{AUC (Walltime in seconds)} && \multicolumn{3}{c}{Speedup}\\
\cmidrule{2-5}\cmidrule{7-9}
&MCMC & INFVB &INFVB &INLA &&INFVB &INFVB &INLA\\
& & ($\phi$) &($\phi,\ssq$) & &&($\phi$) &($\phi,\ssq$)\\
\arrayrulecolor{black!30}\midrule
$\phi=0.1$&   0.71 (8097.17) & 0.71 (23.81) & 0.71 (16.81) & 0.68 (2.05) & &340.01 & 481.66 & 3955.86 \\
$\phi=0.3$&  0.73 (8097.16) & 0.73 (23.79) & 0.73 (16.24) & 0.68 (2.07) & &340.32 & 498.47 & 3907.85 \\
$\phi=0.5$&  0.73 (8116.32) & 0.73 (22.86) & 0.73 (15.85) & 0.68 (2.14) & &354.97 & 511.99 & 3799.83 \\
$\phi=0.7$&  0.73 (8105.36) & 0.74 (22.22) & 0.74 (15.87) & 0.70 (2.08) & &364.80 & 510.80 & 3893.77 \\
\toprule
\toprule
Count &\multicolumn{4}{c}{RMSPE (Walltime in seconds)} && \multicolumn{3}{c}{Speedup}\\
\cmidrule{2-5}\cmidrule{7-9}
&MCMC & INFVB &INFVB &INLA &&INFVB &INFVB &INLA\\
& & ($\phi$) &($\phi,\ssq$) & &&($\phi$) &($\phi,\ssq$)\\
\arrayrulecolor{black!30}\midrule
$\phi=0.1$&   2.83 (7627.31) & 2.82 (418.20) & 2.82 (379.11) & 3.77 (2.44) & &18.24 & 20.12 & 3130.47 \\
$\phi=0.3$&   2.09 (7700.58) & 2.09 (414.15) & 2.09 (376.14) & 3.42 (2.46) & &18.59 & 20.47 & 3126.23 \\
$\phi=0.5$&   1.73 (7738.14) & 1.72 (408.98) & 1.72 (377.15) & 2.79 (2.24) & &18.92 & 20.52 & 3452.81 \\
$\phi=0.7$&   1.78 (7735.17) & 1.78 (391.20) & 1.78 (376.14) & 2.75 (2.20) & &19.77 & 20.56 & 3512.51 \\
\bottomrule 
\end{tabular}
}
    \caption{Comparison of RMSPE (Walltime in seconds) and speedup for MCMC, INFVB($\phi$), INFVB($\phi,\ssq$), and INLA for the full-SGLMMs case when $N=500$. Results for the binary (top) and count datasets (bottom) are provided. The reported values represent the averages across all replicate datasets.}\label{Tab1:FSGLMMBinaryCount500}
\end{center}
\end{table}

\begin{figure}[!ht]
 \begin{center}
\includegraphics[width=0.8\linewidth]{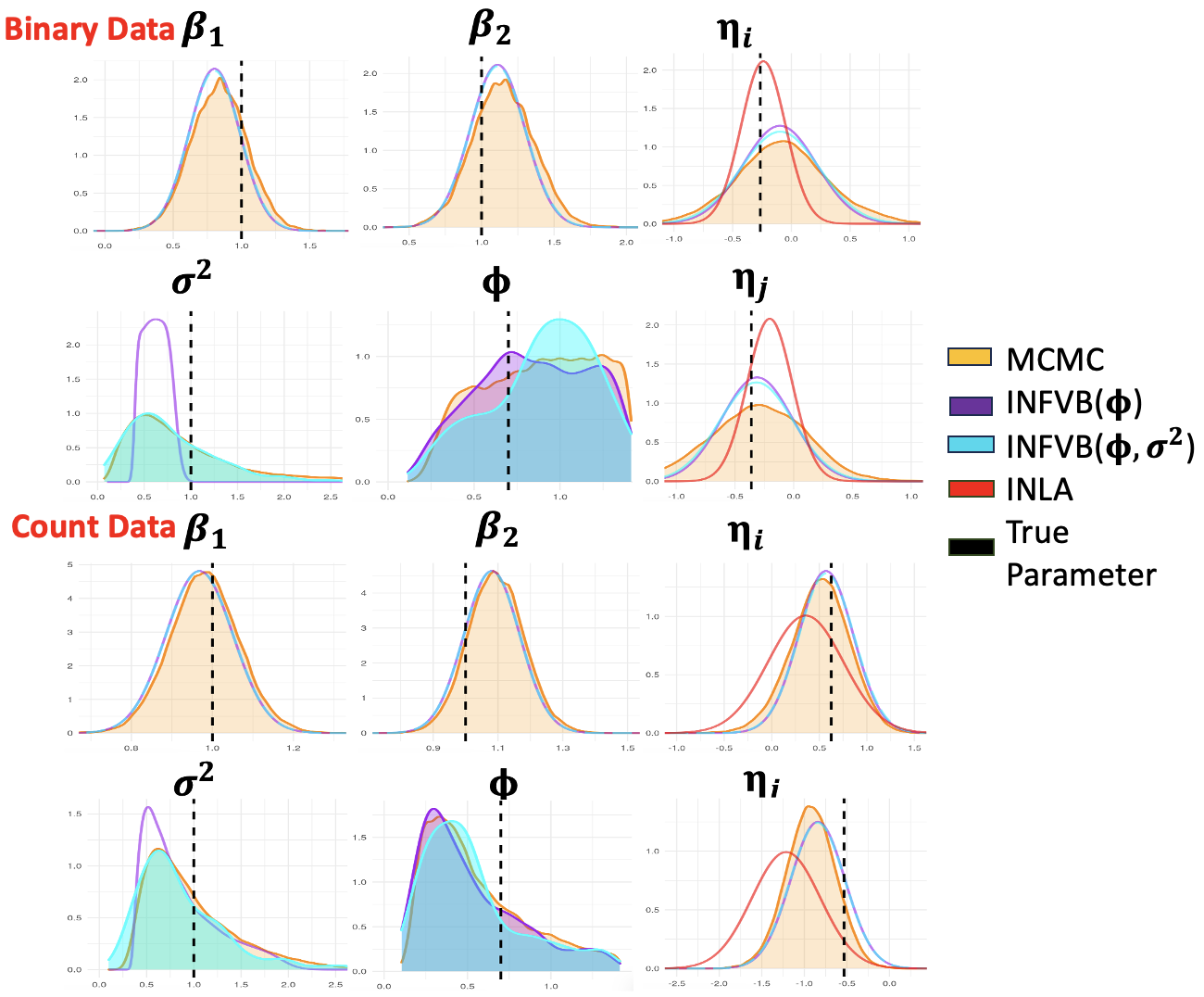}
\caption{Full-SGLMM: Comparison of posterior distributions for MCMC (orange), INFVB($\phi$) (purple), INFVB($\phi,\ssq$) (blue), and INLA (red) when $\phi=0.7$ and $N=500$. Results for the binary (top panel) and count datasets (bottom panel) are provided along with the true parameter values (black dashed lines). Posteriors are shown for all estimated parameters and two chosen linear predictors, $\eta_i$ and $\eta_j$. INLA is used only for estimating the linear predictor $\eta_i$, not the other parameters.} \label{Fig:INFVBn500binarycount}
\end{center}
\end{figure}

For all datasets, we find that the resulting posterior distributions are similar across the two INFVB methods, MCMC and HMC. This is notable because variational approaches have been critiqued for underestimating posterior variances. Figure~\ref{Fig:INFVBn500binarycount} includes comparisons of the posterior distributions for two cases - binary and count data using $\phi=0.7$. The figure also displays the posterior distributions of linear predictors $\eta_i$ and $\eta_j$ corresponding to randomly selected locations $i$ and $j$. Note that variational Bayes methods, namely MFVB, have been known to underestimate posterior variance \citep{han2013integrated, blei2017variational}. However, our results suggest that the INFVB method with two parameter discretizations, INFVB($\phi, \ssq$), provides accurate approximations of the true posterior distributions. While the single-parameter version INFVB($\phi$) provides comparable prediction results, it does underestimate posterior variances, particularly with the partial sill parameters $\ssq$. For practitioners, we recommend using the INFVB($\phi, \ssq$) as it better approximates posterior variances while preserving prediction accuracy. 

Since INLA models the latent surface using stochastic partial differential equations (SPDE), as opposed to a full Gaussian process, comparisons of $\beta_1$, $\beta_2$, $\ssq$, and $\phi$ with the other methods may not be reliable. However, we believe that a comparison of the linear predictors $\eta_i$'s is fair. Note that the marginal posterior distributions for $\eta_i$ and $\eta_j$ from INLA differ to those from the other methods.

\subsection{Case 2: Large Datasets with Basis-SGLMMs}\label{SubSec:simBasisSGLMM}
\paragraph{Simulation Design}
We randomly select 25,000 locations $\bs_i \in \mathcal D=[0,1]^2 \subset \mathbb R^2$ with $N_{\text{train}}=20,000$ locations for training and $N_{\text{test}}=5,000$ reserved for validation. The covariates ($X_1$ and $X_2$), model parameters ($\bbe,\ssq$), and spatial random effects ($\bomega$) are generated similarly as in the previous section. We employ the basis-SGLMM framework outlined in Section$~\ref{SubSec:BasisSGLMM}$ using the approximation $\bomega \approx\bgPhi \bdel$ where $\bgPhi$ is an $n \times m$ matrix where each column contains a pre-specified spatial eigenbasis function. For this particular implementation, the bases are made up of the $m$-leading eigenvectors of a Mat\'ern covariance function with smoothness $\nu=0.5$, partial sill $\ssq=1$, and the corresponding range parameters used to generate the data. We use the leading $m=\{20,50,100\}$ basis functions (eigenvectors) in this simulation study. 
Though basis function specification is an active area of research, incorporating basis selection lies beyond the scope of this study. We fit the basis-SGLMM models (Section$~\ref{SubSec:BasisSGLMM}$)
using four different methods - Hybrid MFVB (Algorithm~\ref*{alg:two}), INFVB($\ssq$) (Algorithm~\ref*{alg:twoB}), MCMC and INLA. Parameter models, stopping criteria for the VB methods, and MCMC implementation are similar to those in the previous section.

\paragraph{Results}
Out-of-sample prediction accuracy is comparable across all four model-fitting approaches, as shown in Table~\ref{Tab:BasisSGLMM} for the binary and count datasets using $50$-leading eigenvectors. However, the VI-based approaches show stark improvements in computational efficiency as evidenced by the large computational speedup factors (over MCMC). For the binary case, the speedup factor ranges from 3527-to-3649 for Hybrid MFVB and 47-to-50.91 for INFVB. For count datasets, the speedup factor is around 148-to-162 for Hybrid MFVB and 7-to-8 for INFVB. The count data cases require an embedded Laplace approximation, which explains the walltime differences between the binary and count cases. The computational cost is lower for Hybrid MFVB as it requires a single processor to run the Hybrid MFVB algorithm. However, the INFVB($\ssq$) approach entails running multiple ($J=1,000$) procedures across a limited number of processors (30 total), which detracts from the computational gains. This speedup would be more pronounced if more computational resources (processors) were available. INLA offers a similar speedup to INFVB($\ssq$) with comparable results, though it is slower than Hybrid MFVB.

For the binary data, MCMC, Hybrid MFVB, and INFVB($\ssq$) yield similar CRPS values, whereas INLA demonstrates a higher CRPS (Table~\ref*{Tab:BasisBinaryCountCRPSCImethod}). In contrast, for coverage, all approaches produce comparable results. For the count data, INLA also exhibits a higher CRPS, consistent with its performance on binary data. However, INLA achieves the best result in coverage, capturing 92$\%$ of the true values.

\begin{table}[ht]
\begin{center} 
\scalebox{0.9}{
\begin{tabular}{ccccccccc} \toprule 
Binary &\multicolumn{4}{c}{AUC (Walltime in seconds)} && \multicolumn{3}{c}{Speedup}\\
\cmidrule{2-5}\cmidrule{7-9}
&MCMC & Hybrid &INFVB &INLA &&Hybrid &INFVB &INLA\\
& & MFVB&($\ssq$) & && MFVB&($\ssq$)\\
\arrayrulecolor{black!30}\midrule
$\phi=0.1$&   0.75 (688.20) & 0.75 (0.19) & 0.75 (14.33) & 0.75 (14.21) & &3649.00 & 48.03 & 48.44 \\
$\phi=0.3$&  0.75 (695.70) & 0.75 (0.20) & 0.75 (14.56) & 0.75 (14.02) & &3527.19 & 47.78 & 49.63 \\
$\phi=0.5$&  0.74 (691.34) & 0.74 (0.19) & 0.74 (14.27) & 0.74 (14.33) & &3590.63 & 48.45 & 48.23 \\
$\phi=0.7$&  0.73 (685.51) & 0.73 (0.19) & 0.73 (13.46) & 0.73 (14.17) & &3641.29 & 50.91 & 48.37 \\
\toprule
\toprule
Count &\multicolumn{4}{c}{RMSPE (Walltime in seconds)} && \multicolumn{3}{c}{Speedup}\\
\cmidrule{2-5}\cmidrule{7-9}
&MCMC & Hybrid &INFVB &INLA &&Hybrid &INFVB &INLA\\
& & MFVB&($\ssq$) & && MFVB&($\ssq$)\\
\arrayrulecolor{black!30}\midrule
$\phi=0.1$&   2.06 (721.19) & 2.06 (4.44) & 2.06 (97.09) & 2.11 (18.56) & &162.26 & 7.43 & 38.85 \\
$\phi=0.3$&   1.37 (708.39) & 1.37 (4.74) & 1.37 (90.56) & 1.38 (15.50) & &149.61 & 7.82 & 45.71 \\
$\phi=0.5$&   1.16 (689.52) & 1.16 (4.64) & 1.16 (85.72) & 1.16 (13.99) & &148.75 & 8.04 & 49.30 \\
$\phi=0.7$&   1.08 (683.34) & 1.08 (4.34) & 1.08 (85.07) & 1.08 (13.67) & &157.37 & 8.03 & 49.98 \\
\bottomrule 
\end{tabular}
}
    \caption{Comparison of RMSPE (walltime in seconds) and average speedup for MCMC, Hybrid MFVB, INFVB($\ssq$), and INLA for the basis-SGLMM case when using 50 eigen Bases functions. Results for the binary (top) and count datasets (bottom) are provided. The reported values represent the averages across all replicate datasets.}\label{Tab:BasisSGLMM}
\end{center}
\end{table}

\begin{figure}[ht]
 \begin{center}
\includegraphics[width=1\linewidth]{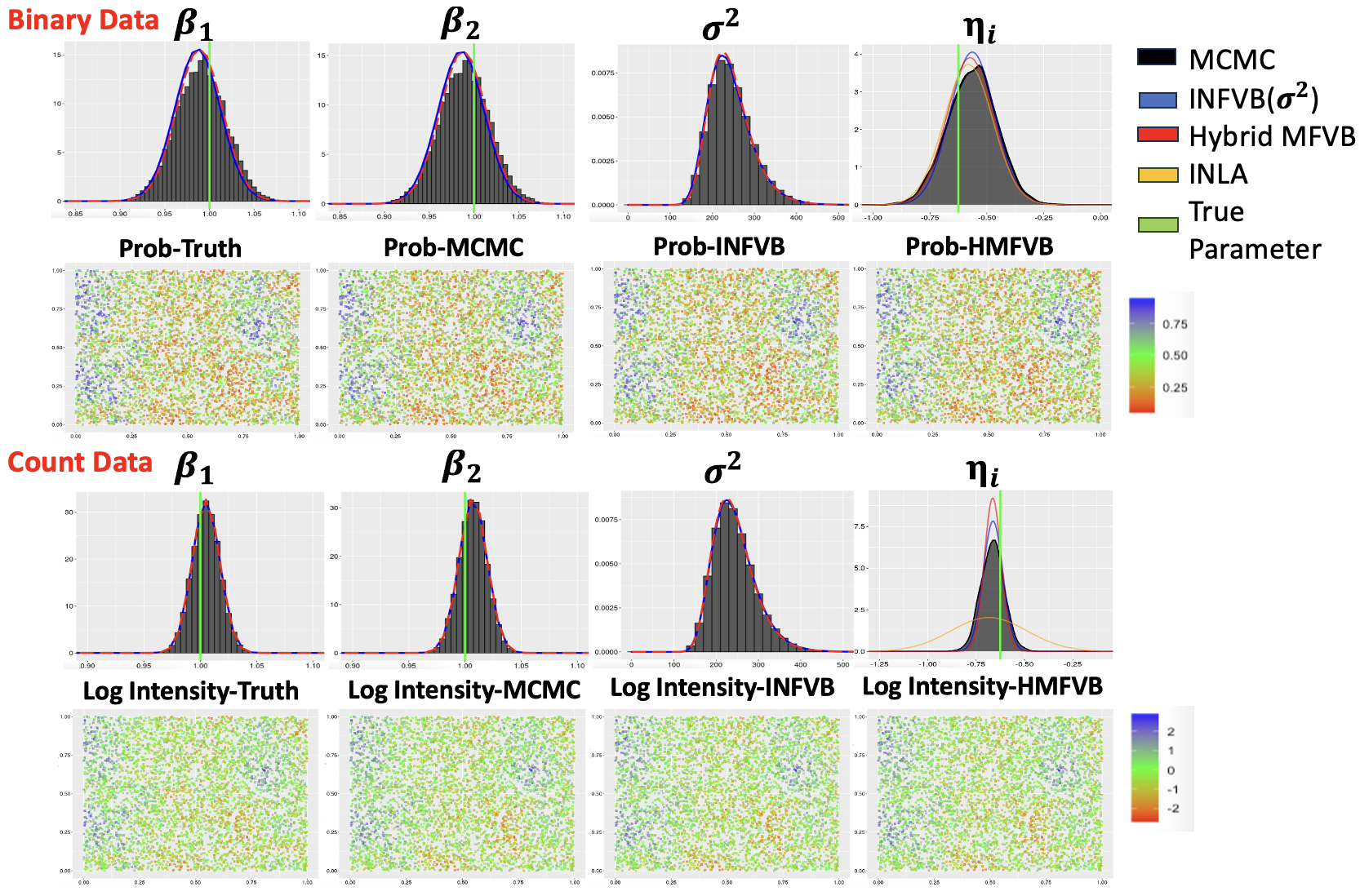}
\caption{Basis-SGLMM: Comparison of posterior distributions for MCMC (black), INFVB($\ssq$) (blue),  
Hybrid MFVB (red), and INLA (orange) for the case when $\phi=0.5$ and $N=25k$. Results for the binary (top panel) and count datasets (bottom panel) are provided. Posteriors shown for model parameters and linear predictor $\eta_i$.  INLA is used only for estimating the linear predictor $\eta_i$, not the other parameters. True and predicted latent probability (binary) and log intensity (count) are provided for all four methods.}\label{Fig:BasisSGLMMBinary}
\end{center}
\vspace{-0.5in}
\end{figure}

Figure~\ref{Fig:BasisSGLMMBinary} compares the relevant posterior distributions for two cases - binary and count data when $\phi=0.5$. The resulting posterior distributions are nearly identical across all methods, including MCMC, INFVB, Hybrid MFVB, and HMC. Since the predictive performances are similar, we suggest using the Hybrid MFVB approach for fitting basis-SGLMMs primarily due to its computational efficiency. We also employ INLA and four other methods to estimate the linear predictor $\eta$. The results appear comparable for binary data; however, for count data, INLA exhibits a larger variance.

\subsection{Case 3: Comparison with Scalable Methods (INLA and HMC)}
We evaluate the proposed VB methods to INLA and two MCMC algorithms (Metropolis---Hastings and HMC) through an abridged simulated example. The purpose of this separate analysis is to compare the VB approach (predictions and walltimes) to a more efficient MCMC sampler such as HMC. We generate data for both the full and basis-SGLMMs as in the previous subsections. The spatial random effects $\bomega$ come from a zero-mean Gaussian process with Mat\'ern covariance function parameterized by $\nu=0.5$, $\sigma^2=1$, and multiple spatial range parameters $\phi=\{0.1, 0.3, 0.5, 0.7 \}$. We use a total of 16 simulated datasets corresponding to the spatial model (full vs. basis-SGLMM), data type (count vs. binary), and spatial range parameters. 

\begin{table}[ht]
\begin{center} 
\scalebox{0.77}{
\begin{tabular}{ccccccccccc} \toprule 
Binary &\multicolumn{6}{c}{AUC (Walltime in seconds)} && \multicolumn{3}{c}{Speedup}\\
\cmidrule{2-9}\cmidrule{9-11}
&MCMC & INFVB &INFVB &INLA &HMC &&INFVB &INFVB &INLA &HMC\\
& &($\phi$) &($\phi,\ssq$) & && &($\phi$) &($\phi,\ssq$)\\
\arrayrulecolor{black!30}\midrule
$\phi=0.1$&   0.68 (7840.70) & 0.67 (24.80) & 0.67 (20.06) & 0.66 (2.09) & 0.69 (1561.60) & &316.11 & 390.78 & 3748.22 & 5.02 \\
$\phi=0.3$&  0.77 (7796.61) & 0.77 (18.61) & 0.77 (17.01) & 0.68 (2.08) &  0.79 (1279.33)& &418.88 & 458.49 & 3755.79 & 6.09 \\
$\phi=0.5$&  0.70 (8553.98) & 0.70 (29.63) & 0.70 (15.96) & 0.65 (2.17) & 0.70 (1263.89) & &288.69 & 535.93 & 3936.04 & 6.77 \\
$\phi=0.7$&  0.74 (8577.10) & 0.73 (27.14) & 0.73 (15.55) & 0.71 (2.07) & 0.75 (1109.72) & &316.05 & 551.65 & 4141.94 & 7.73 \\
\toprule
\toprule
Count &\multicolumn{6}{c}{RMSPE (Walltime in seconds)} && \multicolumn{3}{c}{Speedup}\\
\cmidrule{2-9}\cmidrule{9-11}
&MCMC & INFVB &INFVB &INLA &HMC &&INFVB &INFVB &INLA &HMC\\
& &($\phi$) &($\phi,\ssq$) & && &($\phi$) &($\phi,\ssq$)\\
\arrayrulecolor{black!30}\midrule
$\phi=0.1$&   1.97 (7468.93) & 1.96 (419.96) & 1.96 (436.53) & 2.25 (2.25) & 1.98 (63.02)& &17.78 & 17.11 & 3316.65 & 118.51 \\
$\phi=0.3$&  2.89 (7635.57) & 2.90 (406.64) & 2.90 (376.93) & 4.95 (2.48) &  2.90 (106.12)& &18.78 & 20.26 & 3083.21 & 71.95 \\
$\phi=0.5$&  0.99 (7566.05) & 0.99 (364.73) & 0.99 (371.47) & 1.25 (2.08) & 0.93 (86.57) & &20.74 & 20.37 & 3641.89 & 87.40 \\
$\phi=0.7$&  1.99 (7571.67) & 1.98 (477.54) & 1.98 (369.49) & 2.27 (2.18) & 2.03 (93.25) & &15.86 & 20.49 & 3473.50 & 81.20 \\
\bottomrule 
\end{tabular}
}
\caption{Comparison of RMSPE (walltime in seconds) and speedup for MCMC, INFVB($\phi$) INFVB($\phi, \ssq$), INLA and HMC for the full-SGLMM case. Results for the binary (top) and count datasets (bottom) are provided.}\label{Tab:ComparativeFullSGLMM}
\end{center}
\end{table}

\begin{table}[ht]
\begin{center} 
\scalebox{0.77}{
\begin{tabular}{ccccccccccc} \toprule 
Binary &\multicolumn{6}{c}{AUC (Walltime in seconds)} && \multicolumn{3}{c}{Speedup}\\
\cmidrule{2-9}\cmidrule{9-11}
&MCMC & Hybrid &INFVB &INLA &HMC &&Hybrid &INFVB &INLA &HMC\\
& &  MFVB&($\ssq$) & & && MFVB&($\ssq$) & &\\
\arrayrulecolor{black!30}\midrule
$\phi=0.1$&   0.74 (610.94) & 0.74 (0.20) & 0.74 (14.34) & 0.74 (14.19) & 0.74 (919.05)& &3133.04 & 42.60 & 43.05 &0.66 \\
$\phi=0.3$&  0.75 (606.96) & 0.75 (0.20) & 0.75 (14.87) & 0.75 (14.51) &  0.75 (837.21)& &2989.93 & 40.83 & 41.84 & 0.72 \\
$\phi=0.5$&  0.75 (601.01) & 0.75 (0.18) & 0.75 (13.65) & 0.75 (14.36) & 0.75 (1035.62) & &3357.60 & 44.04 & 41.85 & 0.57 \\
$\phi=0.7$&  0.73 (597.83) & 0.73 (0.20) & 0.73 (14.19) & 0.73 (13.65) & 0.73 (993.30) & &3034.69 & 42.12 & 43.78 & 0.60 \\
\toprule
\toprule
Count &\multicolumn{6}{c}{RMSPE (Walltime in seconds)} && \multicolumn{3}{c}{Speedup}\\
\cmidrule{2-9}\cmidrule{9-11}
&MCMC & Hybrid &INFVB &INLA &HMC &&Hybrid &INFVB &INLA &HMC\\
& &  MFVB&($\ssq$) & & && MFVB&($\ssq$) & &\\
\arrayrulecolor{black!30}\midrule
$\phi=0.1$&   2.13 (661.12) & 2.13 (4.36) & 2.13 (100.71) & 2.16 (18.28) & 2.13 (899.49)& &151.77 & 6.56 & 36.17 &0.73 \\
$\phi=0.3$&  1.34 (653.53) & 1.34 (4.68) & 1.34 (97.52) & 1.35 (15.55) &  1.34 (931.09)& &139.52 & 6.70 & 42.03 & 0.70 \\
$\phi=0.5$&  1.19 (655.06) & 1.19 (4.76) & 1.19 (87.16) & 1.19 (13.36) & 1.19 (932.24) & &137.65 & 7.52 & 49.02 & 0.70 \\
$\phi=0.7$&  1.09 (645.05) & 1.08 (4.42) & 1.08 (86.12) & 1.08 (14.76) & 1.08 (909.79) & &145.94 & 7.49 & 43.70 & 0.71 \\
\bottomrule 
\end{tabular}
}
\caption{Comparison of RMSPE (walltime in seconds) and speedup for MCMC, Hybrid MFVB, INFVB($\ssq$) and HMC for the basis-SGLMM case when using 50 basis functions. Results for the binary (top) and count datasets (bottom) are provided.}\label{Tab:Comparative50BasisSGLMM}
\end{center}
\end{table}

Results from this comparative analysis (Tables~\ref{Tab:ComparativeFullSGLMM} and~\ref{Tab:Comparative50BasisSGLMM}) show that VB-based methods obtain comparable prediction and inferential results as the MCMC-based methods, but at a fraction of the time. Although INLA is faster than our approaches in the full-SGLMM case, it results in lower AUC for binary data and higher RMSPE for count data. For basis-SGLMMs, INLA has comparable predictive performance to the VB-based approaches, but has longer model-fitting walltimes than the Hybrid MFVB case.  

HMC, implemented through  the \texttt{RStan} package \citep{carpenter2017stan}, produces rapidly mixing Markov chains with relatively large effective sample sizes with respect to the number of iterations. Hence, HMC requires significantly fewer samples than the Metropolis---Hastings algorithm \citep{betancourt2017conceptual} to obtain comparable effective sample sizes. HMC performs similarly to the Metropolis---Hastings and VB-based approaches in prediction, but has longer walltimes. HMC's slower walltimes can be attributed to its generation of near-independent samples with larger effective sample sizes. Though the Metropolis---Hastings algorithm has a faster per-iteration runtime compared to HMC, it does not generate nearly as many independent samples as evidenced by the smaller effective sample sizes. On the other hand, HMC's walltimes are considerably slower than those for the VB-based approaches and has much smaller effective sample sizes. Note that the VB-based methods generate independent samples from the variational distributions, which provides a clear advantage (e.g. higher effective sample sizes) over the auto-correlated samples from MCMC.

\vspace{-0.5cm}
\section{Application}\label{Sec:RealData}
In this study, we apply our approach to two large spatial environmental datasets. The first application showcases the dramatic scalability of our VI-based approach by modeling a massive spatial binary dataset ($N$=2.7 million) derived from remotely sensed satellite imagery. A second case that focuses on a large spatial count dataset is provided in the supplement~\ref*{SubSec:RealDataBlueJayBird}.

% \subsection{Binary Spatial Data: MODIS Cloud Mask Data}
As the flagship mission of the Earth Observing System, the National Aeronautics and Space Administration (NASA) launched the Terra Satellite in December 1999. Similar to past studies \citep{sengupta2013hierarchical,bradley2016comparison,lee2022picar}, we model the cloud mask data captured by the Moderate Resolution Imaging Spectroradiometer (MODIS) instrument onboard the Terra satellite. The spatial binary responses $\bZ(s)\in\{0,1\}$ represent the presence/absence of cloud mask at $1km$ spatial resolutions. The data consists of $n=2,748,620$ locations where 90$\%$ is used to fit the model and 10$\%$ reserved for validation. 

\begin{figure}[ht]
 \begin{center}
\includegraphics[width=0.7\linewidth]{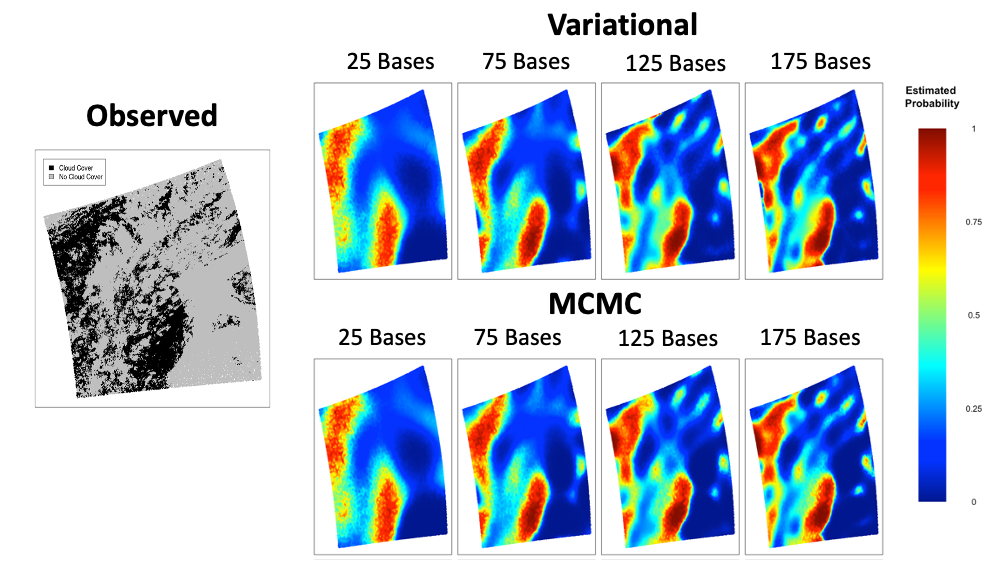}
\caption{Results for the MODIS cloud mask application. Binary cloud mask observations (left) and the estimated probability surfaces are shown for the Hybrid MFVB and MCMC methods. The probability surfaces are broken down with respect to the total number of basis functions used in the basis-SGLMM.}\label{Fig:MODIS}
\end{center}
\vspace{-0.6in}
\end{figure}

Since the cloud mask data is binary and massive in size, we employ the basis-SGLMM model (Section~\ref{SubSec:BasisSGLMM}) with a logit link function. The Moran's eigenbasis functions \citep{lee2022picar} are used for this application. As done in past studies \citep{sengupta2013hierarchical,bradley2016comparison,lee2022picar}, we use the vector $\mathbf{1}$ and the vector latitudes as covariates. We compare the Hybrid MFVB and MCMC-based approaches using maximum absolute error (MAE), area under the ROC curve (AUC), and the associated walltimes. We fit eight different basis-SGLMM models for various sets of PICAR basis functions. See \citep{lee2022picar} for additional details on constructing these basis functions. For each set, we chose the leading $m$ basis functions where $m\in \{25, 50, 75, 100, 125, 150, 175, 200\}$.  Table~\ref{Tab:MODIS} includes a summary of the results and Figure~\ref{Fig:MODIS} provides spatial maps of the predicted probability surfaces for a subset of the basis function sets. Results indicate that the Hybrid MFVB approach achieves near-identical predictive performance to the MCMC-based approach with respect to AUC and MAE, albeit at a mere fraction of the computational cost. As expected, the prediction accuracy is directly correlated with $m$, the number of basis functions, since including more basis functions can help capture the high-frequency behavior of the latent spatial surfaces. However, increasing $m$ also incurs larger computational costs resulting in longer walltimes.

\begin{table}[ht]
\centering
\begin{tabular}{|c|c|c|c|c|c|c|c|}
  \hline
  \hline
  &  \multicolumn{2}{c|}{\tb{AUC}} &  \multicolumn{2}{c|}{\tb{MAE}} &  \multicolumn{2}{c|}{\tb{Walltime (minutes)}}& \\
\hline
 \# Bases &  VB & MCMC & VB & MCMC & VB & MCMC & \tb{Speedup} \\ 
  \hline
  \hline
25&0.82&0.821&0.306&0.305&0.9&858.4&1008\\
\hline  
50&0.839&0.839&0.291&0.29&1.1&992.7&882\\
\hline  
75&0.853&0.854&0.28&0.279&2.6&1118.6&427\\
\hline  
100&0.862&0.862&0.271&0.271&3.7&1248.6&341\\
\hline  
125&0.867&0.866&0.267&0.267&7.5&1388.6&185\\
\hline  
150&0.873&0.873&0.262&0.261&10.9&1587.4&145\\
\hline  
175&0.876&0.875&0.259&0.26&11.5&1753.2&153\\
\hline  
200&0.879&0.878&0.256&0.257&15.5&1868.7&121\\
 \hline  
 \hline
\end{tabular}
\caption{Area under the ROC curve, mean absolute error, walltime (minutes), and speedup for the MODIS cloud mask example using VI and MCMC. Rows denote the number of basis functions used.} \label{Tab:MODIS}
\label{Table:MODIS}
\end{table}
The Hybrid MFVB approach exhibits exceptional portability as it enables practitioners to model millions of non-Gaussian spatial observations within a matter of minutes-to-hours, even on a laptop. Our proposed methodology delivers results that are on par with MCMC while offering a remarkable speed advantage of up to 1,008 times. These tools empower practitioners to explore a wide range of hierarchical spatial models without depleting their computational resources, particularly in modeling massive spatial datasets with millions of observations. 
\section{Discussion} \label{Sec:Discussion}
We introduce a variational Bayes approach for modeling Gaussian and discrete non-Gaussian (binary and count) spatial datasets in the continuous spatial domain. Our approach extends to SGLMMs where the spatial random effects are represented as stationary Gaussian processes (full-SGLMM) or basis expansions (basis-SGLMMs). For both cases, we incorporate the mean field (MFVB) and integrated non-factorized variational Bayes (INVB) method into the Bayesian spatial hierarchical modeling framework. For count and binary data, we employ a Laplace and quadratic approximation \citep{jaakkola1997variational} to ensure conjugacy and bypass costly expectation approximations via numerical methods. We demonstrate our approach through an extensive simulation study, a comparative study with state-of-the-art competitors, and two real-world environmental applications. The results indicate that our proposed approaches provide near-identical prediction and uncertainty quantification to the MCMC-based methods, albeit at a mere fraction of the computational walltime. For the vast majority of cases, the VB approaches outperform INLA in predictions. Due to the dramatic speedup and portability of code, Our proposed modeling framework enables practitioners to model massive non-Gaussian spatial datasets within practical timeframes.

Though INFVB methods better represent posterior variances \citep{han2013integrated,blei2017variational}, the quality of the approximations depends on user-specified tuning parameters, such as the number of discretized values used in INFVB. Future research would benefit from providing theoretically-grounded guidelines to determine the optimal range and discretization for INFVB. Next, the choice of spatial basis functions can impact the performance of INFVB and Hybrid MFVB. A comparative study examining multiple classes of spatial basis functions, like bi-square (radial) basis functions, empirical orthogonal functions, wavelets, and multiresolution basis functions (see Section \ref{SubSec:BasisSGLMM}), would be useful. Both the Hybrid MFVB and INFVB can be extended to the spatio-temporal setting using popular mechanisms such as integro-difference equations (IDE) \citep{dewar2008data}, vector autoregressive (VAR) \citep{lesage1999spatial} and non-separable spatio-temporal models \citep{prates2022non}.

\def\spacingset#1{\renewcommand{\baselinestretch}%
{#1}\small\normalsize} \spacingset{0}
\bibliographystyle{apalike}
\bibliography{references}

\newpage

\def\spacingset#1{\renewcommand{\baselinestretch}%
{#1}\small\normalsize} \spacingset{1}

  \title{ Supplemental Information for ``A Scalable Variational Bayes Approach to Fit High-dimensional Spatial Generalized Linear Mixed Models''}
 \author{}
\date{}
  \maketitle

\setcounter{page}{1}
\renewcommand{\thepage}{S.\arabic{page}}
\setcounter{section}{0}
\renewcommand{\thesection}{S.\arabic{section}}
\setcounter{equation}{0}
\renewcommand{\theequation}{S.\arabic{equation}}
\setcounter{table}{0}
\renewcommand{\thetable}{S.\arabic{table}}
\setcounter{figure}{0}
\renewcommand{\thefigure}{S.\arabic{figure}}
\setcounter{algorithm}{0}
\renewcommand{\thealgorithm}{S.\arabic{algorithm}}

\section{Derivation details of INFVB} \label{SubSec:Supple_INFVBderivation}
For INFVB, we partition the parameter vector into two disjoint blocks ($\btheta_c, \btheta_d$). Then, we represent $q_{\text{INFVB}}$ and the KL divergence as:
\begin{equation} \label{EQ:Supple_INFVB}
    q_{\text{INFVB}}(\btheta)=q(\btheta_c|{\pmb{\theta}_d}){q(\pmb{\theta}_d)}.
\end{equation}
\begin{equation}\label{EQ:Supple_KL}
\begin{aligned}
    \text{KL}(q||p(\btheta|\bZ)) &=\text{argmin}_{q \in Q}\int q(\btheta) \log \frac{q(\btheta)}{p(\btheta|\bZ)}d\btheta 
\end{aligned}
\end{equation}

To obtain the KL divergence using $q_{\text{INFVB}}(\btheta)$, we plug in (\ref{EQ:Supple_INFVB}) into (\ref{EQ:Supple_KL}) as follows:
\begin{equation}\label{EQ:Supple_INFVBderivation}
\begin{aligned}
    \text{KL}(q_{\text{INFVB}}||p(\btheta|\bZ)) &=\text{argmin}_{q_{\text{INFVB}}}\int \int q(\btheta_c|\btheta_d)q(\btheta_d) \log \frac{q(\btheta_c|\btheta_d)q(\btheta_d)}{p(\btheta_c,\btheta_d|\bZ)}d\btheta_cd\btheta_d \\
    &=\text{argmin}_{q_{\text{INFVB}}}\int q(\btheta_d) \Bigg[\int q(\btheta_c|\btheta_d) \log \frac{q(\btheta_c|\btheta_d) q(\btheta_d)}{p(\btheta_c,\btheta_d|\bZ)}d\btheta_c \Bigg] d\btheta_d\\
    &=\text{argmin}_{q_{\text{INFVB}}}\int q(\btheta_d) \Bigg[\int q(\btheta_c|\btheta_d) \log \frac{q(\btheta_c|\btheta_d)}{p(\btheta_c,\btheta_d|\bZ)}d\btheta_c \\
    & \qquad \qquad \qquad \qquad \qquad +\int q(\btheta_c|\btheta_d) \log q(\btheta_d) d\btheta_c \Bigg] d\btheta_d \\
    &= \text{argmin}_{q_{\text{INFVB}}}\int q(\btheta_d) \Bigg[\int q(\btheta_c|\btheta_d) \log \frac{q(\btheta_c|\btheta_d)}{p(\btheta_c,\btheta_d|\bZ)}d\btheta_c+ \log q(\btheta_d)\Bigg] d\btheta_d \\
\end{aligned}
\end{equation}

The objective in (\ref{EQ:Supple_INFVBderivation}) is easier to optimize than that in (\ref{EQ:Supple_KL}) due to the use of conditional distributions $q(\btheta_c|\btheta_d)$. Moreover, $q(\btheta_d)$ can be represented as a discretized empirical distribution. Past studies \citep{han2013integrated, wu2018fast} have proposed discretizing $q(\btheta_d)$ using a grid of $J$ points $\btheta_d=\{\btheta_d^{(1)},..,\btheta_d^{(J)}\}$ such that $q(\btheta_d)= \sum_{j=1}^{J}A_j 1(\btheta_d=\btheta_d^{(j)})$ with weights $A_j$. Note that this simplifies the optimization task to focus solely on the bracket within the integral. We highlight the following points for the INFVB approach:

\begin{enumerate}
    \item We derive the optimal $q(\btheta_c|\btheta_d^{(j)})$ by employing $J$ discretizations for $q(\btheta_d)$. This simplifies the optimization problem outlined in Equation \ref{EQ:Supple_INFVBderivation}.\\
     $q(\btheta_c|\btheta_d^{(j)})=\int q(\btheta_c|\btheta_d^{(j)}) \log \frac{q(\btheta_c|\btheta_d^{(j)})}{p(\btheta_c,\btheta_d^{(j)}|\bZ)}d\btheta_c$
     \item The optimal $q(\btheta_d^{(j)})$ can be calucalated for each $ q(\btheta_c|\btheta_d^{(j)})$ as follows \\
     $\log q(\btheta_d^{(j)})=\text{Constant}-\int q(\btheta_c|\btheta_d^{(j)}) \log \frac{q(\btheta_c|\btheta_d^{(j)})}{p(\btheta_c,\btheta_d^{(j)}|\bZ)}d\btheta_c$ \\
     $\approx \text{Constant}+\int q(\btheta_c|\btheta_d^{(j)}) \log \frac{p(\btheta_c,\btheta_d^{(j)}|\bZ)}{q(\btheta_c|\btheta_d^{(j)})}d\btheta_c $ \\
     $\approx \int q(\btheta_c|\btheta_d^{(j)}) \log \frac{p(\btheta_c,\btheta_d^{(j)}|\bZ)}{q(\btheta_c|\btheta_d^{(j)})}d\btheta_c $ which is ELBO$^{(j)}$ same as (\ref{EQ:ConditionalELBO}) 
     \item We can get the corresponding normalized weights are $A_j=\frac{ELBO^{(j)}}{\sum_{j=1}^{J}ELBO^{(j)}}$ with the j-th conditional $ELBO^{(j)}$
     \item Optimal approximate marginal posterior distribtuions for $q(\btheta_c)$ and $q(\btheta_d)$ are as follows \\
     $ q(\btheta_c)= \sum_{j=1}^{J}A_j q(\btheta_c|\btheta_d^{(j)}), \hspace{0.3cm} q(\btheta_d)= \sum_{j=1}^{J}A_j 1(\btheta_d=\btheta_d^{(j)}) \hspace{0.3cm} \text{where} \quad A_j=\frac{ELBO^{(j)}}{\sum_{j=1}^{J}ELBO^{(j)}}$
\end{enumerate}

\section{INFVB Workflow}
Figure~\ref{Fig:INFVB} in the main provides an overview of the INFVB workflow beginning from discretizing $\btheta_d$ to obtaining the variational functions. Steps 1 and 2 are parallelized and can be distributed across $J$ separate cores; thereby leading to a substantial computational speedup. Despite the advantages of INFVB, there are some important considerations for implementation. First, for the discretized parameter $\btheta_d$, it is important to specify a sensible discretization scheme $\{\btheta_d^{(1)},...,\btheta_d^{(J)}\}$. Selecting too many discretized values (large J) can increase computational costs, yet specifying fewer discretized values (small $J$) may adversely impact the accuracy of the final variational functions. Second, it is important to select which parameters to discretize (i.e., $\btheta_d$). Opting for lower-dimensional parameters minimizes the grid points required for $q(\btheta_d)$. However, it's important to note that lower-dimensional parameters might not always allow for a closed form of $q(\btheta_c|\btheta_d^{(j)})$. In such cases, a parametric-form assumption or a product form factorization may be necessary\citep{wu2018fast}.

\clearpage
\section{INFVB Algorithms}
\begin{algorithm}[ht]
\caption{INFVB Algorithm($\phi,\ssq$) for full-SGLMMs}\label{alg:oneB}
\SetKw{Initialize}{Initialize}
\Initialize{Set $\btheta_d= (\phi_d,\sigma_d^{2}) $ and discretize $\{\phi_d^{(1)},\phi_d^{(2)},...,\phi_d^{(J)}\}$ and 
$\{ \sigma_d^{2^{(1)}}, \sigma_d^{2^{(2)}} ,..., \sigma_d^{2^{(K)}} \}$ for user-specified $J$ and $K$.}\\
\For{j=1,...,J}{
\For{k=1,...,K}{
 \; \\ Update $ELBO^{(j,k)}$ and  $q_{0}(\bga|\phi_d^{(j)}, \sigma_d^{2^{(k)}})$ \; \\
$k \gets k + 1$\;
}
$j \gets j + 1$\;
}
\SetKw{Result}{Result}
\Result{Weighted Non-factorized Variational Functions}
\begin{multicols}{2}
\begin{enumerate}
\item $A_{j}=\frac{ELBO^{(j)}}{\sum_{i=1}^{J} ELBO^{(i)}}$
    \item $q(\boldsymbol{\bga})=\Sigma_{j=1}^{J}A_{j} q(\bga|\btheta_d^{(j)})$
        \item $q(\boldsymbol{\ssq})=\Sigma_{j=1}^{J}A_{j} 1(\btheta_d=\btheta_d^{(j)})$
            \item $q(\phi)=\Sigma_{j=1}^{J}A_{j} 1(\btheta_d=\btheta_d^{(j)})$
                
\end{enumerate}
\end{multicols}
\end{algorithm}
We provide the algorithms for the two-parameter discretized VB approach $INFVB(\phi,\ssq)$ for full-SGLMMs in Algorithm~\ref{alg:oneB}.

\RestyleAlgo{ruled}
\begin{algorithm}[ht]
\caption{Hybrid MFVB Algorithm for basis-SGLMMs}\label{alg:two}
\SetKw{Intialization}{Intialization}
\Intialization{Set $q_0(\gamma)$,$q_0(\ssq)$, $\epsilon_{*}$, $ELBO_0$, and $k=1$.}\\
\While{$|ELBO_{k}-ELBO_{k-1}|>\epsilon_{*}$}{
    Update $q_k(\gamma)$ \; \\
    Update $q_k(\ssq)$\; \\
     $k \gets k + 1$\;
}
\end{algorithm}

% \vspace{-0.7cm}

\RestyleAlgo{ruled}
\begin{algorithm}[ht]
\caption{INFVB($\ssq$) Algorithm for basis-SGLMMs}\label{alg:twoB}
\SetKw{Initialize}{Initialize}
\Initialize{Set $\epsilon_{*}$ and discretize $\{\sigma_d^{2^{(1)}},\sigma_d^{2^{(2)}},...,\sigma_d^{2^{(J)}}\}$ for user-specified $J$.}\\
\For{j=1,...,J}{
\Initialize{Set $ELBO^{(j)}_{0}$,  $q_{0}(\bga|\sigma_d^{2^{(j)}})$, and $k=1$}\; \\
\While{$|ELBO^{(j)}_{k}-ELBO^{(j)}_{k-1}|>\epsilon_{*}$}{
    Update $q_k(\bga|\sigma_d^{2^{(j)}})$\;
    Update $ELBO^{(j)}_{k}$\;
     $k \gets k + 1$\;
}
Set $q(\bga|\sigma_d^{2^{(j)}})=q_k(\bga|\sigma_d^{2^{(j)}}) $
}
\SetKw{Result}{Result}
\Result{Weighted Non-factorized Variational Functions}
\begin{multicols}{2}
\begin{enumerate}
\item $A_{j}=\frac{ELBO^{(j)}}{\sum_{i=1}^{J} ELBO^{(i)}}$
    \item $q(\boldsymbol{\bga})=\Sigma_{j=1}^{J}A_{j} q(\bga|\sigma_d^{2^{(j)}})$
        \item $q(\boldsymbol{\ssq})=\Sigma_{j=1}^{J}A_{j} 1(\btheta_d=\sigma_d^{2^{(j)}})$      
\end{enumerate}
\end{multicols}
\end{algorithm}

\clearpage
\section {Mat\'ern Covariance Function}
A popular class of stationary and isotropic covariance function is the Mat\'ern class \citep{stein1999interpolation} with covariance parameters $\Psi=\{\ssq, \phi, \nu\}$: 
\begin{equation}
C(s_i;s_j;\Psi) = \sigma^2 \frac{1}{\Gamma(\nu)2^{\nu-1}}\left(\sqrt{2\nu}\frac{h}{\phi}\right)^{\nu}K_{\nu}\left(\sqrt{2\nu}\frac{h}{\phi}\right)
\end{equation}
\noindent where $\ssq > 0$ is the partial sill, $\phi$ is the range parameter that summarizes the decay in spatial dependence with respect to distance, $\nu$ denotes the smoothness parameter, $K_{\nu}$ is the modified Bessel function of the second kind, and $h$ represents the distance (e.g., Euclidean) between locations, $\bs_i$ and $\bs_j$. 

\section{Approximations for Poisson and Bernoulli Data Models}
\paragraph{VB for Count SGLMMs}
We do not have a conjugate prior for $\bga$ when dealing with count data; hence, we approximate the variational function $q(\bga|\btheta_d^{(j)})$ with respect to the discretized parameter $\btheta_d^{(j)}$ using the Laplace approximation. 
 Referring to (\ref{EQ:MFVB}), we define \(f(\bga) = E_{\ssq}\Big[\log p(\theta_{\bga} | y, \theta_{{-\bga}})\Big]\). Subsequently, utilizing the Maximum a Posterior (MAP) point \(\hat{\btheta}\) of \(f(\btheta)\), we perform a second-order Taylor approximation of \(f(\btheta)\).
\begin{equation}
    f(\btheta_l) \approx f( \hat{\btheta_l})+ \triangledown f( \hat{\btheta_l})(\btheta_l-\hat{\btheta_l})+\frac{1}{2}(\btheta_l-\hat{\btheta_l})^{T}\triangledown^2f(\hat{\btheta_l}) (\btheta_l-\hat{\btheta_l})  
\end{equation}
\noindent where $\triangledown f(\hat{\btheta_l})=0 $ and $\triangledown^2f(\hat{\btheta_l}) $ is the Hessian matrix at MAP $\hat{\btheta_l} $ which leads to maximized $f(\btheta_l)$. Then the above equations simplifies to 

\begin{equation} \label{EQ:Laplacegamma}
    q_\gamma(\bga)  \propto \text{exp} \{f(\bga) \} \approx \text{exp}\{  f( \hat{\bga})+ \frac{1}{2}(\btheta_{\bga}-\hat{\btheta_{\bga}})^{T}\triangledown^2f(\hat{\btheta_{\bga}}) (\btheta_{\bga}-\hat{\btheta_{\bga}})  \} 
\end{equation}

\noindent It follows $q_{\bga}(\btheta_{\bga}) \approx \text{N}(\hat{\btheta_{\bga}}, - \triangledown^2 f(\hat{\btheta_{\bga}})^{-1} ) $. Laplace's approximation implies the variational distribution is a Gaussian distribution \citep{wu2018fast}.

The objective function and gradient are summarized as follows:
\begin{equation}
\begin{aligned}
\mbox{Objective Function:} \qquad f(\bga)&\propto E_{q(\ssq)}[\log p(\bZ, \bga, \ssq)]\\
& = \bZ'\tilde{\bX}\bga -\bOne' \Big(e^{\tilde{\bX}\bga}\Big)-\frac{1}{2}\bga'E[\bSig_{\gamma}^{-1}]\bga\\
\mbox{Gradient:} \qquad \nabla f(\bv)&= \tilde{\bX}'\bZ- \tilde{\bX}'Diag(e^{\tilde{\bX}\bga})\bOne'  - E[\bSig_{v}^{-1}]\bga
\end{aligned}
\end{equation}
where $\tilde{\bX}=[\bX, \bI]$ or $\tilde{\bX}=[\bX, \bgPhi]$, and $\bga$ is a multivariate normal random variable, representing a linear combination of $\bbe$ and $\bomega$ or $\delta $. Utilizing the gradient accelerates computations in comparison to relying solely on the objective function. In the end, Laplace approximation aids in obtaining the closed form of  $q_{\gamma}^{(k)}$. The resulting variational function takes the following form.
\begin{equation}
q_\gamma(\bga)= \mcN(\tbmu_\gamma, \tbC_{\gamma})    
\end{equation}
where $\tbmu_\gamma=\argmax_{\gamma}f(\bga)$ and $\tbC_\gamma= -(\bH)^{-1}$ where $\bH=\frac{\partial^2 f}{\partial \gamma^2}\Bigr|_{\substack{\gamma=\tbmu_\gamma}}$. More detailed calculations for the Laplace approximation for count data are provided in the later supplementary material.

\paragraph{VB for Binary SGLMMs}
When dealing with binary data, we have faced challenges of multivariate intractability arising from the term \((1+\exp\{-\mathbf{X}_{i}^\prime\mathbf{b} - W_i\})^{-1}\). We can address this by substituting the sigmoid equation with a quadratic approximation \citep{jaakkola1997variational}.

\begin{figure}[!ht]
 \begin{center}
\includegraphics[width=0.6\linewidth]{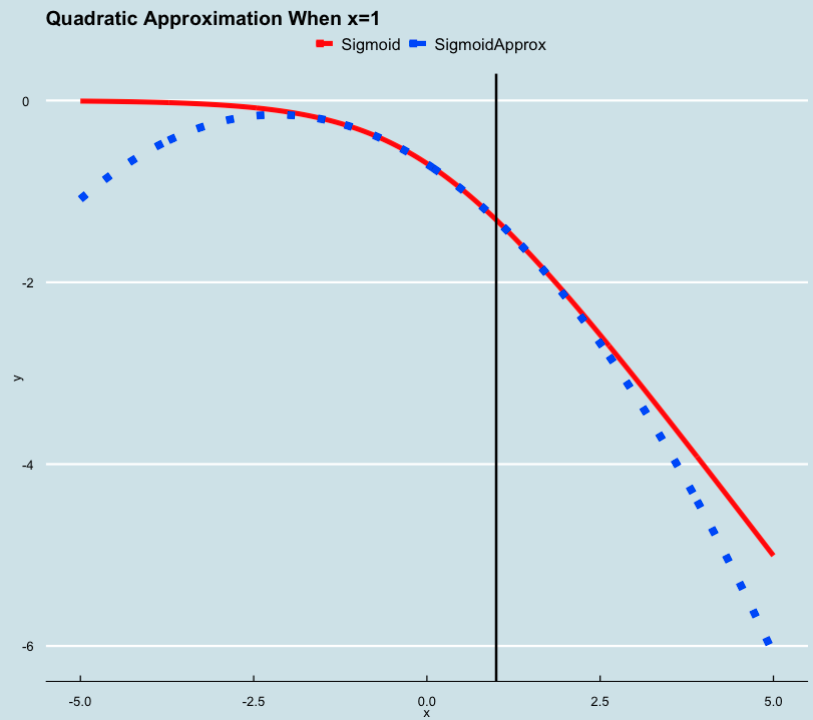}
\caption{Binary SGLMM : Quadratic Approximation When x=1. The red line represents the actual sigmoid values, while the blue line depicts the sigmoid approximation. The approximation closely aligns with the actual values, especially when x=1 \citep{jaakkola1997variational}.} \label{Fig:JaakkolaApproxi}
\end{center}
\end{figure}

Similar to the Count data, obtaining the Variational Distribution for Binary data $q_{\gamma}^{(k)}\propto E_{\ssq}[\log(p(\bga,\ssq, \bZ)]$ in closed form, a prerequisite for computing $q_{\sigma^2}(\ssq)$ and ELBO, poses a challenge. Additionally, employing Monte Carlo methods for the sigmoid function may be both costly and imprecise. For the Bernoulli data model, the log joint density is summarized as follows:
\begin{equation}
\begin{aligned}
    \log[p(\bZ, \bga , \ssq)]
    &\propto \bZ^{\prime}\bga +\bOne^\prime\Big[-\log (1+\exp\{\tilde{\bX}\bga\})\Big]-\frac{1}{2}\Big(\bga'\bSig_{\gamma}^{-1}\bga\Big)\\
    & -(\asi+1+\frac{n}{2}) \log\ssq -\frac{\bsi}{\ssq}
\end{aligned}
\end{equation}
where  $\tilde{\bX}=[\bX, \bI]$ or $\tilde{\bX}=[\bX, \bgPhi]$. When taking an expectation with respect to $q_{\gamma}$ to apply the MFVB approach, a challenge arises as $E_{q(\gamma)}\Big[-\log (1+\exp\{\tilde{\bX}\bga\})\Big]$ is not available in closed form. Our approach approximates $-\log (1+\exp\{\tilde{\bX}\bga\})$ as a quadratic function of $\bga$, leading to a Gaussian variational function. \\
\noindent \tb{Quadratic Approximation \citep{jaakkola1997variational}:}
\vspace{-0.5cm}
\begin{equation}
    {-\log(1+e^{x})}=\argmax_{\xi}\Big\{ \lambda(\xi)x^{2}-\frac{1}{2}x +\psi(\xi)\Big\}
\end{equation}
\noindent where $\lambda(\xi)=-\tanh(\xi/2)/(4\xi)$ and $\psi(\xi)=\xi/2-\log(1+e^{\xi})+\xi\tanh(\xi/2)/4$. To proceed with this approximation we need to introduce auxiliary variables $\pmb{\xi}$. \\
\noindent \tb{Optimal Value for the Auxiliary Variables} \citep{jaakkola1997variational}
\vspace{-0.5cm}
\begin{equation} \label{EQ:Supplement_OptimvalueJaakkola}
    \pmb{\xi}^{*}= \sqrt{\mbox{Diagonal}\Big\{\tilde{\bX}(\tbC_{\gamma}+\tbmu_{\gamma}\tbmu_{\gamma}')\tilde{\bX}'\Big\}} \qquad  \bf{D}=\mbox{diag}(\lambda(\pmb{\xi}^{*}))
\end{equation}
The matrix $\bf{D}$ is formed by placing $\pmb{\xi}$ on the diagonal. By updating $\pmb{\xi}$ at each iteration, we obtain the resulting Gaussian variational form.
\vspace{0.5cm}
\begin{equation}
    q_\gamma(\bga)= \mcN(\tbmu_\gamma, \tbC_{\gamma})
\end{equation}
where $\tbC_\gamma= (-2\tilde{\bX}'\mathbf{D}\tilde{\bX}+E[\bSig_{\gamma}^{-1}])^{-1}$ and $\tbmu=\tbC_\gamma(\bZ-\frac{1}{2}\bOne')\tilde{\bX}$.
full-SGLMM and Basis Representation include different components for $E[\bSig_{\gamma}^{-1}]$ as follows:
\vspace{0.5cm}
\begin{align*}
\mbox{Full SGLMM}\qquad & \qquad\qquad   \mbox{Basis Representation}\\
E[\bSig_{\gamma}^{-1}]=\begin{bmatrix}
\bSig_{\beta}^{-1}  & \bzero \\
\bzero & \frac{\tilde{\alpha}_\sigma}{ \tilde{\beta}_\sigma}\bR_{\phi}^{-1}
\end{bmatrix}, &
\qquad 
E[\bSig_{\gamma}^{-1}]=\begin{bmatrix}
\bSig_{\beta}^{-1}  & \bzero \\
\bzero & \frac{\tilde{\alpha}_\sigma}{ \tilde{\beta}_\sigma}\bSig_{\delta}^{-1}
\end{bmatrix} \qquad 
\end{align*}
(Figure~\ref{Fig:JaakkolaApproxi}) demonstrates a favorable approximation by using (\ref{EQ:Supplement_OptimvalueJaakkola}). The red line represents the actual sigmoid values, while the blue line depicts the sigmoid approximation. The approximation closely aligns with the actual values, particularly when x=1, demonstrating effective results.

\clearpage
\section{Estimating the linear Predictor}\label{Subsec:EstLinearPredic}
\begin{figure}[ht]
 \begin{center}
\includegraphics[width=0.8\linewidth]{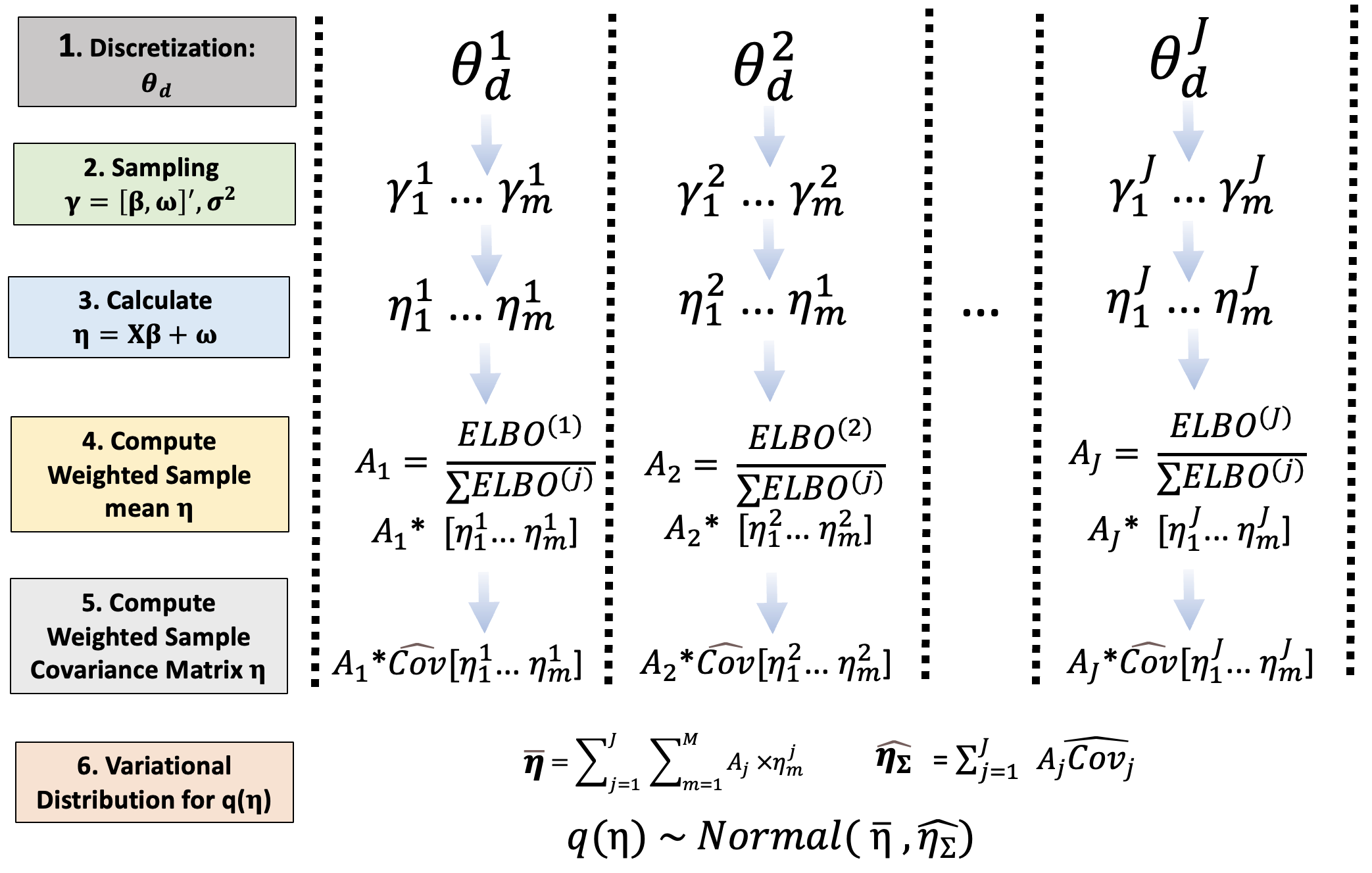}
\caption{Workflow of sampling-based approach for estimating  the linear predictor by using INFVB for full-SGLMM}\label{Fig:SamplingMethod_Full}
\end{center}
\end{figure}

\begin{figure}[ht]
 \begin{center}
\includegraphics[width=0.8\linewidth]{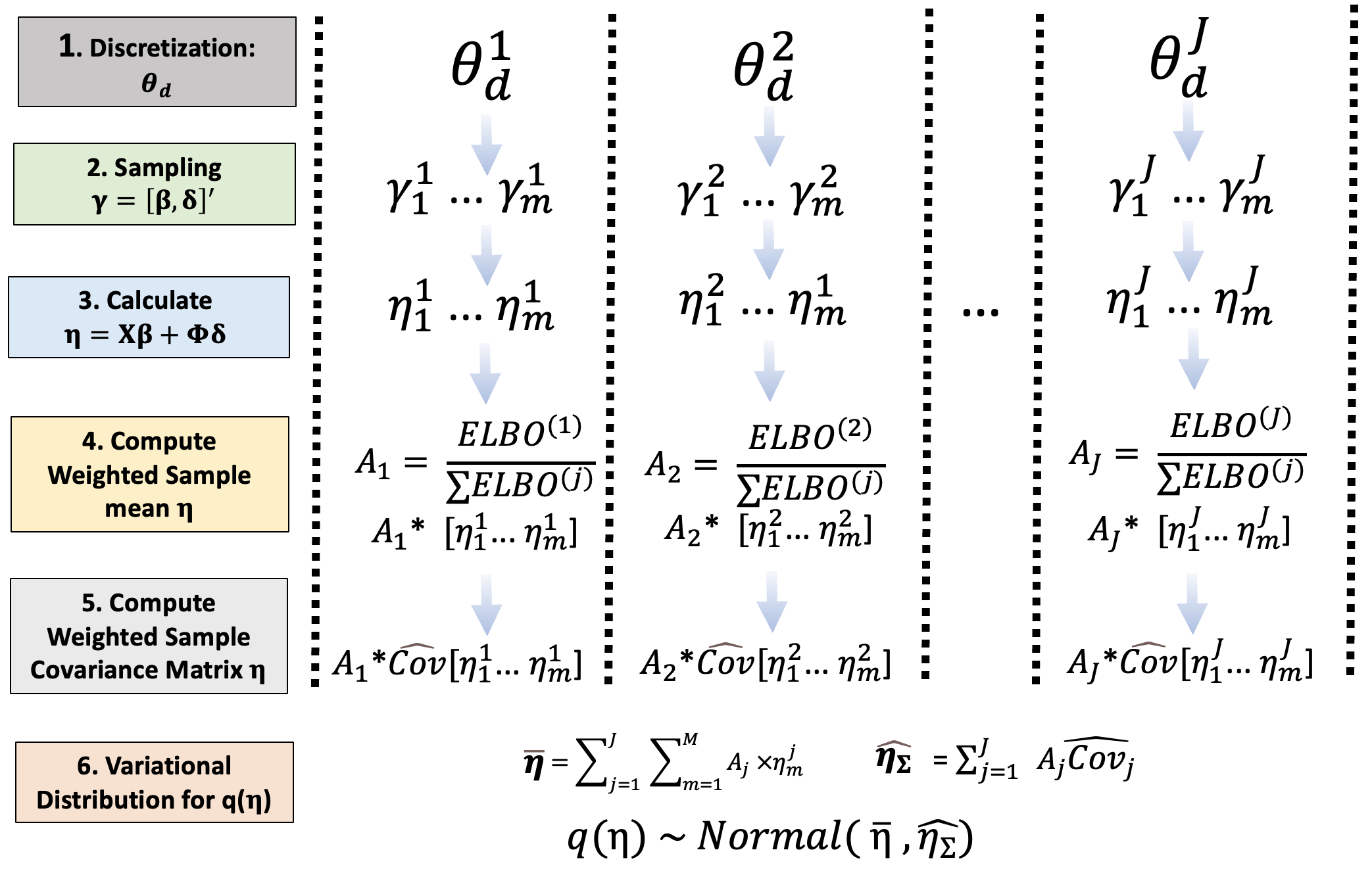}
\caption{Workflow of sampling-based approach for estimating  the linear predictor by using INFVB for Basis-SGLMM}\label{Fig:SamplingMethod_Basis}
\end{center}
\end{figure}

Estimating the linear predictor $\bfeta$ via Variational Bayes can challenging due to the tendency to underestimate the posterior variance \citep{han2013integrated}. To address this, we employed a sampling-based approach for estimating $\bfeta$. When using INFVB, we obtain \( m \) samples of \(\bga=(\bbe, \bdel)'\) and \(\ssq\) for each discretization step. We then calculate the weighted sample mean and covariance matrix of \(\bfeta\) by incorporating the ELBO ratio $A_j$, as shown in Figure~\ref{Fig:SamplingMethod_Full}. This allows us to derive the variational distribution \( q(\bfeta) \), which yields results comparable to other methods such as the Metropolis---Hastings Algorithm, INLA, and Hamiltonian Monte Carlo (HMC). The workflow for this sampling method in the Basis-SGLMM is in~\ref{Fig:SamplingMethod_Basis}.

\clearpage
\section{Tables and Figures(Gaussian, Count and Binary Data)} 
\begin{table}[ht]
\begin{center} 
\begin{tabular}{cccccc} \toprule 
Gaussian &\multicolumn{3}{c}{RMSPE (Walltime in seconds)} && \multicolumn{1}{c}{Speedup}\\
\cmidrule{2-4}\cmidrule{5-6}
&MCMC & INFVB && &INFVB \\
& & ($\phi$)  && &($\phi$) \\
\arrayrulecolor{black!30}\midrule
$\phi=0.1$&  0.524 (553.471) &  0.523 (0.758) &&  & 730.52 \\
$\phi=0.3$&  0.313 (596.148) &  0.313 (0.756) &&  & 788.68 \\
$\phi=0.5$&  0.245 (579.899) &  0.245 (0.753) &&  & 770.14 \\
$\phi=0.7$&  0.204 (568.686) &  0.204 (0.758) &&  & 750.11 \\
\bottomrule 
\end{tabular}
    \caption{Average Comparison of RMSPE (Walltime in seconds) and average Speedup for MCMC and INFVB($\phi$), for the full-SGLMMs case with normal dataset when $N=500$.} \label{Tab:GaussianFullsupplement}
\end{center}
\end{table}

\begin{table}[ht]
\begin{center} 
\begin{tabular}{ccccccc} \toprule 
Gaussian &\multicolumn{2}{c}{MCMC} && \multicolumn{2}{c}{INFVB($\phi$)}\\
\cmidrule{2-3}\cmidrule{4-6}
&CRPS & Coverage & CRPS &Coverage \\
\arrayrulecolor{black!30}\midrule
$\phi=0.1$&  0.39 & 0.89 &  0.39 & 0.82   \\
$\phi=0.3$&  0.23 & 0.94 &  0.23 & 0.92   \\
$\phi=0.5$&  0.18 & 0.96 &  0.18 & 0.95    \\
$\phi=0.7$&  0.15 & 0.94 &  0.15 & 0.94    \\
\bottomrule 
\end{tabular}
    \caption{Average Comparison of CRPS and coverage for Gaussian model across MCMC, and INFVB($\phi$) for the full-SGLMM.}
\label{Tab:GaussianCRPSCImethod}
\end{center}
\end{table}

\begin{figure}[!ht]
 \begin{center}
\includegraphics[width=1\linewidth]{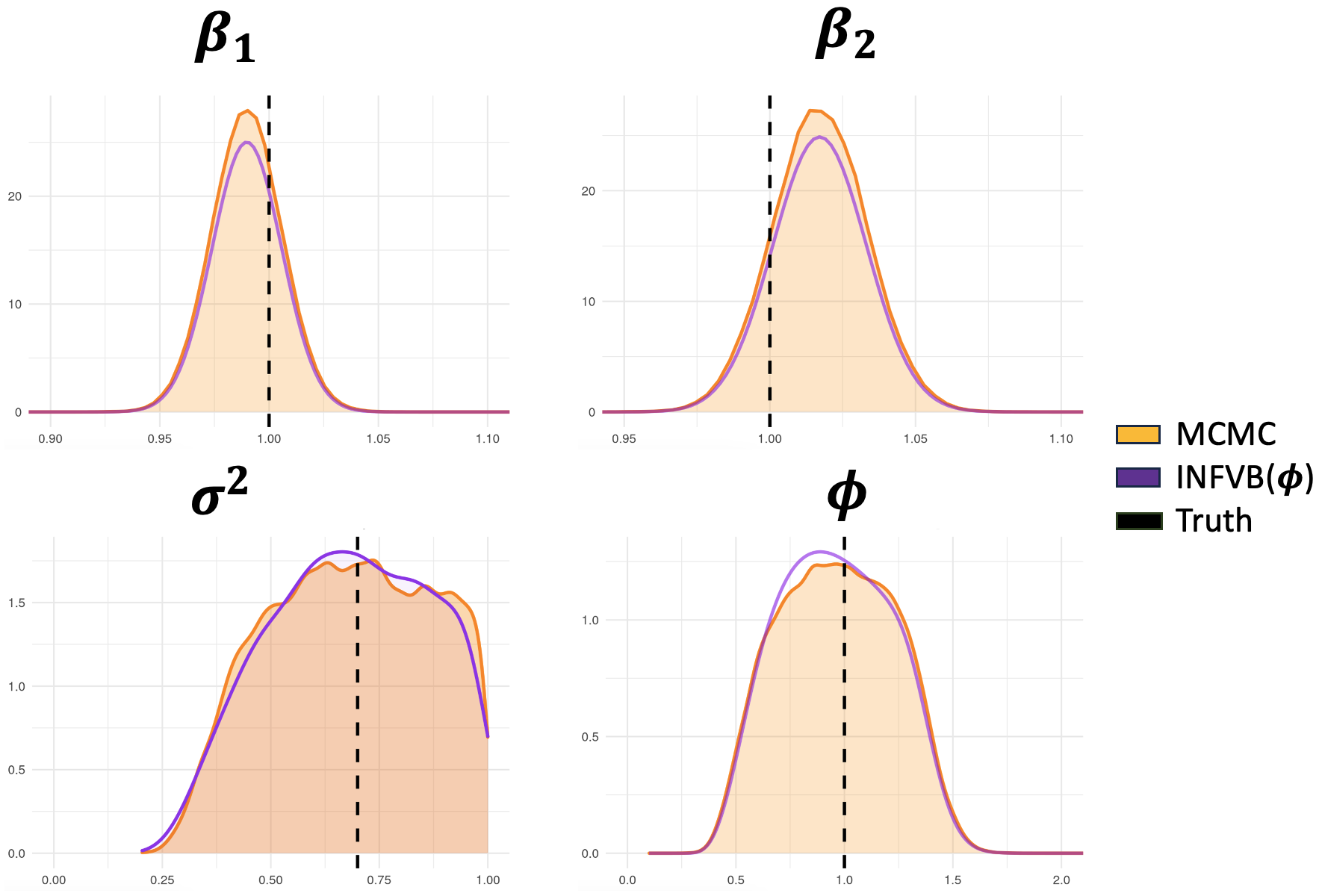}
\caption{Full-SGLMM: Comparison of posterior distributions for MCMC (orange) and INFVB($\phi$) (purple) when $\phi=0.7$ and $N=500$. Results for the normal datasets are provided along with the true parameter values (black dashed lines). Posteriors shown for all estimated parameters.}
\end{center}
\end{figure}

\clearpage

\begin{table}[ht]
\begin{center}
\scalebox{0.9}{
\begin{tabular}{cccccccccc}
\toprule
\multicolumn{2}{c}{Binary} & \multicolumn{2}{c}{MCMC} & \multicolumn{2}{c}{INFVB($\phi$)} & \multicolumn{2}{c}{INFVB($\phi,\ssq$)} & \multicolumn{2}{c}{INLA} \\
\cmidrule(r){3-4} \cmidrule(r){5-6} \cmidrule(r){7-8} \cmidrule(r){9-10}
 &  & CRPS & Coverage & CRPS & Coverage & CRPS & Coverage & CRPS & Coverage \\
\arrayrulecolor{black!30}\midrule
$\phi=0.1$ &  & 0.50 & 0.81 & 0.54 & 0.61 & 0.54 & 0.61 & 1.03 &0.30 \\
$\phi=0.3$ &  & 0.37 & 0.89 & 0.39 & 0.73 & 0.39 & 0.75 & 0.99 & 0.30 \\
$\phi=0.5$ &  & 0.33 & 0.90 & 0.35 & 0.75 & 0.34 & 0.77 & 0.97 & 0.31 \\
$\phi=0.7$ &  & 0.29 & 0.93 & 0.29 & 0.81 & 0.29 & 0.83 & 0.93 & 0.32 \\
\midrule
\midrule
\multicolumn{2}{c}{Count} & \multicolumn{2}{c}{MCMC} & \multicolumn{2}{c}{INFVB($\phi$)} & \multicolumn{2}{c}{INFVB($\phi,\ssq$)} & \multicolumn{2}{c}{INLA} \\
\cmidrule(r){3-4} \cmidrule(r){5-6} \cmidrule(r){7-8} \cmidrule(r){9-10}
 &  & CRPS & Coverage & CRPS & Coverage & CRPS & Coverage & CRPS & Coverage \\
\arrayrulecolor{black!30}\midrule
$\phi=0.1$ &  & 0.40 & 0.75 & 0.38 & 0.91 & 0.38 & 0.91 & 1.09 &0.57 \\
$\phi=0.3$ &  & 0.27 & 0.84 & 0.27 & 0.91 & 0.27 & 0.91 & 1.02 &0.55 \\
$\phi=0.5$ &  & 0.23 & 0.87 & 0.23 & 0.89 & 0.23 & 0.90 & 1.01 &0.49 \\
$\phi=0.7$ &  & 0.20 & 0.88 & 0.20 & 0.89 & 0.20 & 0.90 & 0.94 &0.47 \\
\bottomrule
\end{tabular}
}
\caption{Average Comparison of CRPS and coverage for Binary and Count models across MCMC, INFVB($\phi$), INFVB($\phi,\ssq$), and INLA approaches for the full-SGLMM.}
\label{Tab:BinaryCountCRPSCImethod}
\end{center}
\end{table}

\begin{table}[ht]
\begin{center} 
\scalebox{0.9}{
\begin{tabular}{ccccccccc} \toprule 
Binary &\multicolumn{4}{c}{AUC (Walltime in seconds)} && \multicolumn{3}{c}{Speedup}\\
\cmidrule{2-5}\cmidrule{7-9}
&MCMC & Hybrid &INFVB &INLA &&Hybrid &INFVB &INLA\\
& & MFVB &($\ssq$) & &&MFVB &($\ssq$) &\\
\arrayrulecolor{black!30}\midrule
$\phi=0.1$&   0.72 (718.54) & 0.72 (0.05) & 0.72 (5.85) & 0.72 (6.35) & &13136.03 & 122.83 & 113.09 \\
$\phi=0.3$&  0.73 (875.45) & 0.73 (0.06) & 0.73 (6.42) & 0.73 (6.34) &  & 15616.34 & 136.26 & 138.06 \\
$\phi=0.5$&  0.73 (706.27) & 0.73 (0.06) & 0.73 (6.37) & 0.73 (6.33) &  & 12403.76 & 110.91 & 111.56 \\
$\phi=0.7$&  0.73 (706.36) & 0.73 (0.06) & 0.73 (6.48) & 0.73 (6.36) &  & 12405.33 & 109.02 & 111.05 \\
\toprule
\toprule
Count &\multicolumn{4}{c}{RMSPE (Walltime in seconds)} && \multicolumn{3}{c}{Speedup}\\
\cmidrule{2-5}\cmidrule{7-9}
&MCMC & Hybrid &INFVB &INLA &&Hybrid &INFVB &INLA\\
& & MFVB &($\ssq$) & &&MFVB &($\ssq$) &\\
\arrayrulecolor{black!30}\midrule
$\phi=0.1$&   2.61 (561.14) & 2.61 (1.18) & 2.61 (18.38) & 2.66 (8.36) & &474.31 & 30.55 & 67.20 \\
$\phi=0.3$&   1.61 (571.49) & 1.61 (1.21) & 1.61 (18.23) & 1.63 (8.39) & &471.29 & 31.36 & 68.14 \\
$\phi=0.5$&   1.26 (570.03) & 1.26 (1.25) & 1.26 (17.59) & 1.26 (7.35) & &456.84 & 32.41 & 77.54 \\
$\phi=0.7$&   1.14 (558.26) & 1.14 (1.26) & 1.14 (17.34) & 1.15 (7.00) & &443.74 & 32.19 & 79.76 \\
\bottomrule 
\end{tabular}
}
\caption{Average Comparison of RMSPE (Walltime in seconds) and average speedup for MCMC, Hybrid MFVB and INFVB($\ssq$) for the basis-SGLMM case when using 20 Eigen Bases functions. Results for the binary (top) and count datasets (bottom) are provided.}\label{Tab:20BasisSGLMM}
\end{center}
\end{table}

\begin{table}[ht]
\begin{center}
\scalebox{0.9}{
\begin{tabular}{cccccccccc}
\toprule
\multicolumn{2}{c}{Binary} & \multicolumn{2}{c}{MCMC} & \multicolumn{2}{c}{Hybrid MFVB} & \multicolumn{2}{c}{INFVB($\ssq$)} & \multicolumn{2}{c}{INLA} \\
\cmidrule(r){3-4} \cmidrule(r){5-6} \cmidrule(r){7-8} \cmidrule(r){9-10}
 &  & CRPS & Coverage & CRPS & Coverage & CRPS & Coverage & CRPS & Coverage \\
\arrayrulecolor{black!30}\midrule
$\phi=0.1$ &  & 0.55 & 0.15 & 0.56 & 0.14 & 0.56 & 0.14 & 0.70 & 0.15 \\
$\phi=0.3$ &  & 0.36 & 0.23 & 0.36 & 0.21 & 0.36 & 0.21 & 0.63 & 0.23 \\
$\phi=0.5$ &  & 0.28 & 0.29 & 0.28 & 0.27 & 0.28 & 0.27 & 0.59 & 0.29 \\
$\phi=0.7$ &  & 0.23 & 0.33 & 0.23 & 0.31 & 0.23 & 0.31 & 0.56 & 0.34 \\
\midrule
\midrule
\multicolumn{2}{c}{Count} & \multicolumn{2}{c}{MCMC} & \multicolumn{2}{c}{Hybrid MFVB} & \multicolumn{2}{c}{INFVB($\ssq$)} & \multicolumn{2}{c}{INLA} \\

\cmidrule(r){3-4} \cmidrule(r){5-6} \cmidrule(r){7-8} \cmidrule(r){9-10}
 &  & CRPS & Coverage & CRPS & Coverage & CRPS & Coverage & CRPS & Coverage \\
\arrayrulecolor{black!30}\midrule
$\phi=0.1$ &  & 0.61 & 0.05 & 0.61 & 0.05 & 0.61 & 0.05 & 0.69 &0.94 \\
$\phi=0.3$ &  & 0.39 & 0.09 & 0.39 & 0.09 & 0.39 & 0.10 & 0.63 &0.93 \\
$\phi=0.5$ &  & 0.30 & 0.13 & 0.30 & 0.13 & 0.30 & 0.13 & 0.59 &0.93 \\
$\phi=0.7$ &  & 0.25 & 0.16 & 0.25 & 0.16 & 0.25 & 0.16 & 0.56 &0.93 \\
\bottomrule
\end{tabular}
}
\caption{Average Comparison of CRPS and coverage for Binary and Count models across MCMC, INFVB($\ssq$), INFVB($\ssq$), and INLA approaches for the basis-SGLMM when using 20 Eigen Bases functions.}
\label{Tab:20BasisBinaryCountCRPSCImethod}
\end{center}
\end{table}

\begin{table}[ht]
\begin{center} 
\scalebox{0.77}{
\begin{tabular}{ccccccccccc} \toprule 
Binary &\multicolumn{6}{c}{AUC (Walltime in seconds)} && \multicolumn{3}{c}{Speedup}\\
\cmidrule{2-9}\cmidrule{9-11}
&MCMC & Hybrid &INFVB &INLA &HMC &&Hybrid &INFVB &INLA &HMC\\
& &  MFVB&($\ssq$) & & &&MFVB &($\ssq$) & &\\
\arrayrulecolor{black!30}\midrule
$\phi=0.1$&   0.71 (583.04) & 0.71 (0.08) & 0.71 (5.97) & 0.71 (6.88) & 0.71 (760.65) & &7380.29 & 97.69 & 84.78 & 0.77 \\
$\phi=0.3$&  0.74 (527.75) & 0.74 (0.05) & 0.74 (6.59) & 0.74 (6.43) &  0.74 (733.49)& &9957.55 & 80.12 & 82.10 & 0.72 \\
$\phi=0.5$&  0.73 (529.90) & 0.74 (0.06) & 0.74 (6.53) & 0.74 (6.18) & 0.74 (737.83) & &9634.47 & 81.16 & 85.75 & 0.72 \\
$\phi=0.7$&  0.72 (548.26) & 0.72 (0.06) & 0.72 (5.62) & 0.72 (6.43) & 0.72 (666.94) & &9137.65 & 97.64 & 85.32 & 0.82 \\
\toprule
\toprule
Count &\multicolumn{6}{c}{RMSPE (Walltime in seconds)} && \multicolumn{3}{c}{Speedup}\\
\cmidrule{2-9}\cmidrule{9-11}
&MCMC & Hybrid &INFVB &INLA &HMC &&Hybrid &INFVB &INLA &HMC\\
& &  MFVB&($\ssq$) & & &&MFVB &($\ssq$) & &\\
\arrayrulecolor{black!30}\midrule
$\phi=0.1$&   2.75 (563.88) & 2.75 (1.26) & 2.75 (20.42) & 2.80 (7.39) & 2.75 (750.22)& &449.31 & 27.62 & 76.31 & 0.75 \\
$\phi=0.3$&  1.58 (863.34) & 1.58 (1.26) & 1.58 (18.32) & 1.59 (7.91) &  1.58 (797.96)& &687.37 & 47.13 & 109.21 & 1.08 \\
$\phi=0.5$&  1.33 (868.76) & 1.33 (1.41) & 1.33 (18.16) & 1.33 (7.44) & 1.33 (744.66) & &614.84 & 47.83 & 116.69 & 1.17 \\
$\phi=0.7$&  1.18 (563.89) & 1.18 (1.15) & 1.18 (16.83) & 1.18 (6.74) & 1.18 (753.45) & &489.49 & 33.50 & 83.72 & 0.75 \\
\bottomrule 
\end{tabular}
}
\caption{Comparison of RMSPE (walltime in seconds) and speedup for MCMC, Hybrid MFVB, INFVB($\ssq$) and HMC for the basis-SGLMM case when using 20 eigen Bases functions. Results for the binary (top) and count datasets (bottom) are provided.}\label{Tab:Comparative20BasisSGLMM}
\end{center}
\end{table}

\begin{table}[ht]
\begin{center} 
\scalebox{0.9}{
\begin{tabular}{ccccccccc} \toprule 
Binary &\multicolumn{4}{c}{AUC (Walltime in seconds)} && \multicolumn{3}{c}{Speedup}\\
\cmidrule{2-5}\cmidrule{7-9}
&MCMC & Hybrid &INFVB &INLA &&Hybrid &INFVB &INLA\\
& & MFVB &($\ssq$) & &&MFVB &($\ssq$) &\\
\arrayrulecolor{black!30}\midrule
$\phi=0.1$&   0.76 (1218.01) & 0.76 (0.68) & 0.76 (29.81) & 0.76 (47.48) & &1795.58 & 40.87 & 25.66 \\
$\phi=0.3$&  0.75 (1237.31) & 0.75 (0.66) & 0.75 (27.93) & 0.75 (47.12) &  & 1868.38 & 44.30 & 26.26 \\
$\phi=0.5$&  0.74 (1194.78) & 0.74 (0.63) & 0.74 (27.95) & 0.74 (47.38) &  & 1889.94 & 42.75 & 25.22 \\
$\phi=0.7$&  0.74 (1177.01) & 0.74 (0.62) & 0.74 (28.35) & 0.74 (48.07) &  & 1896.26 & 41.51 & 24.48 \\
\toprule
\toprule
Count &\multicolumn{4}{c}{RMSPE (Walltime in seconds)} && \multicolumn{3}{c}{Speedup}\\
\cmidrule{2-5}\cmidrule{7-9}
&MCMC & Hybrid &INFVB &INLA &&Hybrid &INFVB &INLA\\
& & MFVB &($\ssq$) & &&MFVB &($\ssq$) &\\
\arrayrulecolor{black!30}\midrule
$\phi=0.1$&   1.71 (1184.38) & 1.71 (23.18) & 1.71 (362.08) & 1.74 (52.17) & &51.09 & 3.27 & 22.70 \\
$\phi=0.3$&   1.27 (1248.65) & 1.27 (20.78) & 1.27 (346.51) & 1.27 (46.57) & &60.08 & 3.60 & 26.81 \\
$\phi=0.5$&   1.12 (1185.35) & 1.12 (15.43) & 1.12 (338.55) & 1.12 (45.61) & &76.80 & 3.50 & 25.99 \\
$\phi=0.7$&   1.05 (1190.44) & 1.05 (14.40) & 1.05 (333.10) & 1.05 (46.27) & &82.65 & 3.57 & 25.73 \\
\bottomrule 
\end{tabular}
}
\caption{Average Comparison of RMSPE (Walltime in seconds) and average speedup for MCMC, Hybrid MFVB and INFVB($\ssq$) for the basis-SGLMM case when using 100 Eigen Bases functions. Results for the binary (top) and count datasets (bottom) are provided.}\label{Tab:100BasisSGLMM}
\end{center}
\end{table}

\begin{table}[ht]
\begin{center}
\scalebox{0.9}{
\begin{tabular}{cccccccccc}
\toprule
\multicolumn{2}{c}{Binary} & \multicolumn{2}{c}{MCMC} & \multicolumn{2}{c}{Hybrid MFVB} & \multicolumn{2}{c}{INFVB($\ssq$)} & \multicolumn{2}{c}{INLA} \\
\cmidrule(r){3-4} \cmidrule(r){5-6} \cmidrule(r){7-8} \cmidrule(r){9-10}
 &  & CRPS & Coverage & CRPS & Coverage & CRPS & Coverage & CRPS & Coverage \\
\arrayrulecolor{black!30}\midrule
$\phi=0.1$ &  & 0.24 & 0.58 & 0.24 & 0.54 & 0.24 & 0.55 & 0.69 & 0.60 \\
$\phi=0.3$ &  & 0.16 & 0.76 & 0.16 & 0.72 & 0.16 & 0.72 & 0.63 & 0.77 \\
$\phi=0.5$ &  & 0.13 & 0.82 & 0.13 & 0.78 & 0.13 & 0.79 & 0.59 & 0.83 \\
$\phi=0.7$ &  & 0.12 & 0.85 & 0.12 & 0.82 & 0.12 & 0.83 & 0.56 & 0.86 \\
\midrule
\midrule
\multicolumn{2}{c}{Count} & \multicolumn{2}{c}{MCMC} & \multicolumn{2}{c}{Hybrid MFVB} & \multicolumn{2}{c}{INFVB($\ssq$)} & \multicolumn{2}{c}{INLA} \\

\cmidrule(r){3-4} \cmidrule(r){5-6} \cmidrule(r){7-8} \cmidrule(r){9-10}
 &  & CRPS & Coverage & CRPS & Coverage & CRPS & Coverage & CRPS & Coverage \\
\arrayrulecolor{black!30}\midrule
$\phi=0.1$ &  & 0.27 & 0.25 & 0.27 & 0.27 & 0.27 & 0.28 & 0.69 &0.90 \\
$\phi=0.3$ &  & 0.15 & 0.44 & 0.15 & 0.47 & 0.15 & 0.48 & 0.63 &0.90 \\
$\phi=0.5$ &  & 0.12 & 0.56 & 0.12 & 0.58 & 0.12 & 0.59 & 0.59 &0.84 \\
$\phi=0.7$ &  & 0.10 & 0.64 & 0.10 & 0.65 & 0.10 & 0.66 & 0.56 &0.77 \\
\bottomrule
\end{tabular}
}
\caption{Average Comparison of CRPS and coverage for Binary and Count models across MCMC, INFVB($\ssq$), INFVB($\ssq$), and INLA approaches for the basis-SGLMM when using 100 Eigen Bases functions.}
\label{Tab:100BasisBinaryCountCRPSCImethod}
\end{center}
\end{table}

\begin{table}[ht]
\begin{center} 
\scalebox{0.77}{
\begin{tabular}{ccccccccccc} \toprule 
Binary &\multicolumn{6}{c}{AUC (Walltime in seconds)} && \multicolumn{3}{c}{Speedup}\\
\cmidrule{2-9}\cmidrule{9-11}
&MCMC & Hybrid &INFVB &INLA &HMC &&Hybrid &INFVB &INLA &HMC\\
& &MFVB  &($\ssq$) && && MFVB&($\ssq$) & &\\
\arrayrulecolor{black!30}\midrule
$\phi=0.1$&   0.76 (864.33) & 0.76 (0.80) & 0.76 (31.00) & 0.76 (50.69) & 0.76 (1325.92) & &1076.37 & 27.89 & 17.05 & 0.65 \\
$\phi=0.3$&  0.75 (830.24) & 0.76 (0.65) & 0.76 (28.78) & 0.76 (47.42) &  0.76 (1087.19)& &1275.33 & 28.85 & 17.51 & 0.76 \\
$\phi=0.5$&  0.75 (819.75) & 0.75 (0.59) & 0.75 (27.14) & 0.75 (47.25) & 0.75 (802.84) & &1398.89 & 30.21 & 17.35 & 1.02 \\
$\phi=0.7$&  0.73 (837.84) & 0.73 (0.64) & 0.73 (28.13) & 0.73 (47.54) & 0.73 (1250.93) & &1307.07 & 29.78 & 17.63 & 0.67 \\
\toprule
\toprule
Count &\multicolumn{6}{c}{RMSPE (Walltime in seconds)} && \multicolumn{3}{c}{Speedup}\\
\cmidrule{2-9}\cmidrule{9-11}
&MCMC & Hybrid &INFVB &INLA &HMC &&Hybrid &INFVB &INLA &HMC\\
& &MFVB  &($\ssq$) && && MFVB&($\ssq$) & &\\
\arrayrulecolor{black!30}\midrule
$\phi=0.1$&   1.77 (944.22) & 1.77 (23.11) & 1.77 (392.71) & 1.77 (50.44) & 1.77 (1120.96)& &40.86 & 2.40 & 18.72 & 0.84 \\
$\phi=0.3$&  1.22 (938.40) & 1.22 (25.28) & 1.22 (347.12) & 1.22 (48.56) &  1.22 (1115.71)& &37.13 & 2.70 & 19.32 & 0.84 \\
$\phi=0.5$&  1.14 (871.51) & 1.14 (17.92) & 1.14 (330.20) & 1.14 (47.65) & 1.14 (1128.55) & &48.64 & 2.64 & 18.29 & 0.77 \\
$\phi=0.7$&  1.05 (950.00) & 1.05 (13.95) & 1.05 (343.39) & 1.05 (45.70) & 1.05 (1169.19) & &68.08 & 2.77 & 20.79 & 0.81 \\
\bottomrule 
\end{tabular}
}
\caption{Comparison of RMSPE (walltime in seconds) and speedup for MCMC, Hybrid MFVB, INFVB($\ssq$) and HMC for the basis-SGLMM case when using 100 eigen Bases functions. Results for the binary (top) and count datasets (bottom) are provided.}\label{Tab:Comparative100BasisSGLMM}
\end{center}
\end{table}

\begin{table}[ht]
\begin{center}
\scalebox{0.9}{
\begin{tabular}{cccccccccc}
\toprule
\multicolumn{2}{c}{Binary} & \multicolumn{2}{c}{MCMC} & \multicolumn{2}{c}{Hybrid MFVB} & \multicolumn{2}{c}{INFVB($\ssq$)} & \multicolumn{2}{c}{INLA} \\
\cmidrule(r){3-4} \cmidrule(r){5-6} \cmidrule(r){7-8} \cmidrule(r){9-10}
 &  & CRPS & Coverage & CRPS & Coverage & CRPS & Coverage & CRPS & Coverage \\
\arrayrulecolor{black!30}\midrule
$\phi=0.1$ &  & 0.37 & 0.32 & 0.37 & 0.30 & 0.37 & 0.30 & 0.69 & 0.33 \\
$\phi=0.3$ &  & 0.22 & 0.48 & 0.23 & 0.45 & 0.23 & 0.45 & 0.63 & 0.49 \\
$\phi=0.5$ &  & 0.17 & 0.57 & 0.18 & 0.54 & 0.18 & 0.54 & 0.59 & 0.58 \\
$\phi=0.7$ &  & 0.15 & 0.64 & 0.15 & 0.61 & 0.15 & 0.61 & 0.56 & 0.65 \\
\midrule
\midrule
\multicolumn{2}{c}{Count} & \multicolumn{2}{c}{MCMC} & \multicolumn{2}{c}{Hybrid MFVB} & \multicolumn{2}{c}{INFVB($\ssq$)} & \multicolumn{2}{c}{INLA} \\

\cmidrule(r){3-4} \cmidrule(r){5-6} \cmidrule(r){7-8} \cmidrule(r){9-10}
 &  & CRPS & Coverage & CRPS & Coverage & CRPS & Coverage & CRPS & Coverage \\
\arrayrulecolor{black!30}\midrule
$\phi=0.1$ &  & 0.42 & 0.12 & 0.42 & 0.12 & 0.42 & 0.13 & 0.69 &0.92 \\
$\phi=0.3$ &  & 0.24 & 0.22 & 0.24 & 0.23 & 0.24 & 0.24 & 0.63 &0.92 \\
$\phi=0.5$ &  & 0.18 & 0.30 & 0.18 & 0.31 & 0.18 & 0.31 & 0.59 &0.92 \\
$\phi=0.7$ &  & 0.15 & 0.36 & 0.15 & 0.37 & 0.15 & 0.37 & 0.56 &0.92 \\
\bottomrule
\end{tabular}
}
\caption{Comparison of CRPS and coverage for Binary and Count models across MCMC, INFVB($\ssq$), INFVB($\ssq$), and INLA approaches for the basis-SGLMM when using 50 Eigen Bases functions.}
\label{Tab:BasisBinaryCountCRPSCImethod}
\end{center}
\end{table}

\clearpage

\section{Count Spatial Data: Blue Jay Bird Data}\label{SubSec:RealDataBlueJayBird}
\begin{figure}[ht]
 \begin{center}
\includegraphics[width=1\linewidth]{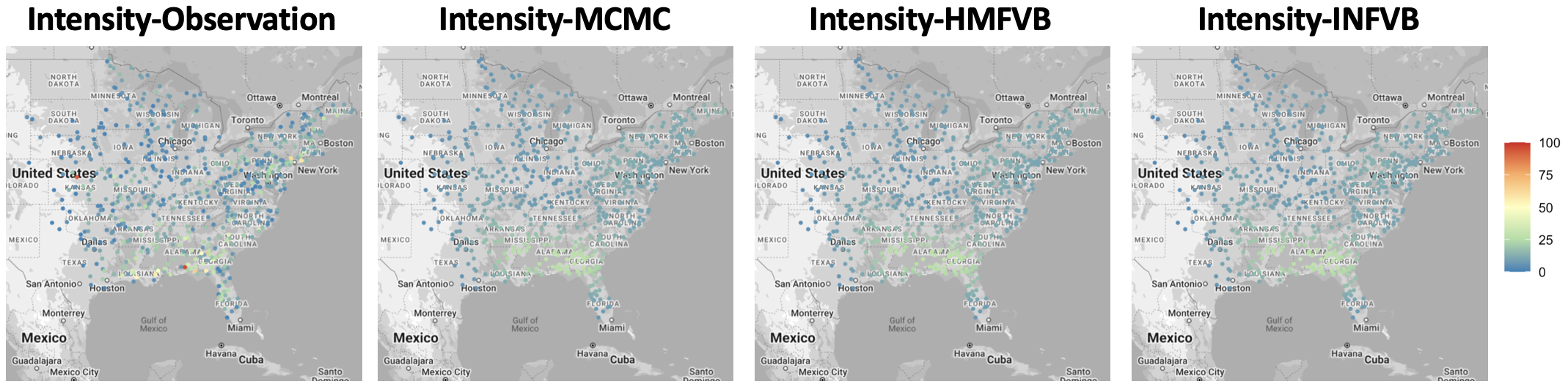}
\caption{True observations (first) and predicted intensity surfaces for the North American Blue Jay abundance dataset. The intensity surface is estimated using the basis-SGLMM model fit via MCMC (second), Hybrid MFVB (third), and INFVB (fourth).}\label{Fig:BlueJayFigure}
\end{center}
\end{figure}

The annual North American Breeding Bird Survey (BBS) \citep{BBS_2022} is a collaboration between the U.S. Geological Survey's Eastern Ecological Science Center and Environment Canada's Canadian Wildlife Service to monitor the abundance of bird populations across North America. The BBS includes population data for over 400 species, which is readily accessible to the public. We focus on counts of the Blue Jay (Cyanocitta cristata) species collected at $1,593$ locations along roadside routes in 2018 (Figure~\ref{Fig:BlueJayFigure}).

We randomly select 1,000 locations to train our models and reserve the remaining 593 locations for validation. We employ the basis-SGLMM model with the log link function (for count data) and embedded eigenvector basis functions for the MCMC, Hybrid MFVB, and INFVB($\ssq$) implementations. The eigenvector basis functions $\bgPhi$ consists of the leading 10 eigenvectors of a Mat\'ern correlation function with parameters for smoothness $\nu=0.5$ and range $\phi=0.5$ computed using all observed locations. The matrix of covariates $\bX$ includes the latitude and longitude of the locations. The prior distributions are  $\bbe \sim \mcN([0,0]',100 \cdot \bI_2)$ and $\ssq\sim \text{IG} (0.1, 0.1)$. The RMSPE and computational walltimes for MCMC are based on running the Metropolis---Hastings algorithm for $100k$ iterations. We assess convergence and set the stopping criterion similarly as in Section \ref{SubSec:simBasisSGLMM}.

\begin{table}[!ht]
\begin{center}
\begin{tabular}{c c c c}
&   &  &   \\
  & MCMC  & Hybrid MFVB  & INFVB($\ssq$)  \\
\hline
RMSPE          & 9.820  & 9.821 & 9.821 \\
Walltime (seconds)  & (182.882)  & (1.061)  & (12.783) \\
Computational Speedup & & 172.368 & 14.307 
\end{tabular}
\end{center}
\caption{Results for the blue jay abundance application. Predictive performance and model-fitting walltimes are provided for the MCMC, Hybrid MFVB, and INFVB implementations. The computational speedup is expressed as a ratio (e.g., VB walltime/MCMC walltime).}
\end{table}

Both the Hybrid MFVB and INFVB($\ssq$) perform comparably in predictive performance to MCMC, but at a fraction of the computational cost. Hybrid MFVB has a walltime of 1.061 seconds and a computational speedup factor of 172.368 compared to MCMC. INFVB($\ssq$) has a speedup of 14.307 over MCMC . All three methods accurately represent the latent intensity surfaces of Blue Jay abundance, as showcased in Figure~\ref{Fig:BlueJayFigure}.

\clearpage
\section{Full-SGLMM (Gaussian Data Model): Discretized \texorpdfstring{$\phi^{(j)}$}{Lg} }
\tb{Hierarchical Model :}
% \vspace{-0.3in}
\begin{align*}\vspace{-0.1in}
    \tb{Normal Data Model:} &\qquad  \bZ_i|\bbe,\ssq,\phi \sim \mbox{N}(\bX_{i}'\bbe,\ssq\bR_{\phi})\\
    \tb{Prior Model:}& \qquad \bbe \sim \mcN(0,\bSig_{\beta}), \qquad \ssq \sim IG(\alpha_\sigma, \beta_\sigma), \qquad \phi \sim \text{Unif}(0,\sqrt{2})
\end{align*}

\noindent \tb{Objective:} Obtain variational functions $q(\bbe)$ , $q(\ssq)$, and $q(\phi)$ via Mean Field Variational Bayes(MFVB) to approximate $p(\bbe|\cdot)$ , $p(\ssq|\cdot)$, and $p(\phi|\cdot)$  

\tb{Probability Density Functions : }
\begin{align*}
    \mbox{Joint:}& \qquad p(\bZ, \bbe, \phi, \sigma^2)= p(\bZ|\bbe)p(\bbe)p(\phi)p(\ssq)\\
    \mbox{Likelihood:}& \qquad p(\bZ|\bbe,\phi, \sigma^2)= \prod_{i=1}^{n}(2\pi)^{-\frac{n}{2}}|\ssq\bR_{\phi}|^{-1/2}\text{exp}\big(-\frac{1}{2\ssq}(\bZ_i-\bX_{i}'\bbe)'\bR_{\phi}^{-1}(\bZ_i-\bX_{i}'\bbe) \big)\\
    \mbox{Prior :}& \qquad p(\ssq)= \frac{\beta_0^{\alpha_0}}{\Gamma(\alpha_0)}(\ssq)^{-\alpha_0-1}\mbox{exp}(-\frac{\beta_0}{\ssq})\\
    & \qquad p(\phi) \sim \mbox{Unif}(0,\sqrt{2}) \\
    & \qquad p(\bbe) \sim 2\pi^{-\frac{p}{2}}|\bSig_{\beta}|^{-\frac{1}{2}}\mbox{exp}\big(-\frac{1}{2}\bbe'\bSig_{\beta}^{-1}\bbe \big) \\
\end{align*}

\noindent \tb{Log Joint posterior density}:
\begin{align*}
    \log[p(\bZ, \bbe, \phi, \ssq)]& = \log[p(\bZ|\bbe)]+\log[p(\bbe)]+\log[p(\phi)]+\log[p(\ssq)]\\
    &\propto -\frac{n}{2}\log2\pi+\log|\ssq\bR_{\phi}|^{-\frac{1}{2}}-\frac{1}{2\ssq} \sum_{i=1}^{n}(\bZ_i-\bX_{i}'\bbe)'\bR_{\phi}^{-1}(\bZ_i-\bX_{i}'\bbe)\\
    &\qquad -\frac{p}{2}\log2\pi-\frac{1}{2}\log|\bSig_{\beta}|-\frac{1}{2}\bbe'\bSig_{\beta}^{-1}\bbe+\log p(\phi) \\
    &\qquad +\alpha_0\log \beta_0-\log\Gamma(\alpha_0)-(\alpha_0+1)\log\ssq-\frac{\beta_0}{\ssq}\\
    &\propto -\frac{n}{2}\log\ssq-\frac{1}{2}\log|\bR_{\phi}|-\frac{1}{2\ssq} \sum_{i=1}^{n}(\bZ_i-\bX_{i}'\bbe)'\bR_{\phi}^{-1}(\bZ_i-\bX_{i}'\bbe)\\
    &\qquad -\frac{1}{2}\log|\bSig_{\beta}|-\frac{1}{2}\bbe'\bSig_{\beta}^{-1}\bbe+\log p(\phi) \\
    &\qquad +\alpha_0\log \beta_0-\log\Gamma(\alpha_0)-(\alpha_0+1)\log\ssq-\frac{\beta_0}{\ssq} \\
    &\propto -\frac{n}{2}\log\ssq-\frac{1}{2}\log|\bR_{\phi}|-\frac{1}{2\ssq} (\bZ-\bX\bbe)'\bR_{\phi}^{-1}(\bZ-\bX\bbe)\\
    &\qquad -\frac{1}{2}\log|\bSig_{\beta}|-\frac{1}{2}\bbe'\bSig_{\beta}^{-1}\bbe+\log p(\phi) \\
    &\qquad +\alpha_0\log \beta_0-\log\Gamma(\alpha_0)-(\alpha_0+1)\log\ssq-\frac{\beta_0}{\ssq}
\end{align*}

\subsection{Variational Function for \texorpdfstring{$\bbe$ and $\ssq$}{Lg} } 

\noindent \tb{Computing the Variational Function $q(\bbe)$:} We implement Mean Field Variational Bayes (MFVB) using the Kernel of a Normal Distribution. The objective function is as follows:

\begin{align*}
    q(\bbe)&\propto \exp\big[E_{q(-\bbe)}[\log p(\bZ, \bbe, \phi, \ssq)]\big]\\
    &= \exp\big[E_{q(\ssq,\phi)}[\log p(\bZ, \bbe, \phi, \ssq)]\big]\\
    &\propto E_{q(\ssq,\phi)}[\exp\big( -\frac{1}{2\ssq} (\bZ-\bX\bbe)'\bR_{\phi}^{-1}(\bZ-\bX\bbe)-\frac{1}{2}\bbe'\bSig_{\beta}^{-1}\bbe  \big)] \\
    &\propto E_{q(\ssq,\phi)}[\exp\big( -\frac{1}{2}[\frac{1}{\ssq}(\bZ-\bX\bbe)'\bR_{\phi}^{-1}(\bZ-\bX\bbe)+\bbe'\bSig_{\beta}^{-1}\bbe] \big)] \\
    &\propto E_{q(\ssq,\phi)}[\exp\big( -\frac{1}{2}[\bbe'(\frac{1}{\ssq}\bX'\bR_{\phi}^{-1}\bX+\bSig_{\beta}^{-1})\bbe-2\frac{1}{\ssq} \bZ'\bR_{\phi}^{-1}\bX\bbe] \big)]\\
    &\propto \exp\big( -\frac{1}{2}[\bbe'(\frac{\alpha}{\beta}\bX'\bR_{\phi}^{-1}\bX+\bSig_{\beta}^{-1})\bbe-2\frac{\alpha}{\beta} \bZ'\bR_{\phi}^{-1}\bX\bbe] 
\end{align*}

\tb{Key Components:}
\begin{enumerate}
    \item Variational Function for $\bbe$: $q(\bbe)= \mcN(\tbmu_{\beta}, \tbC_{\beta})$
    \item $E[\frac{1}{\ssq}] = \frac{\alpha}{\beta}$
\end{enumerate}
\tb{Resulting Variational Function:} 
$$q(\bbe)= \mcN(\tbmu_{\beta}, \tbC_{\beta})$$ where $\tbC_{\beta}=\big[\frac{\alpha}{\beta}\bX'\bR_{\phi}^{-1}\bX+\bSig_{\beta}^{-1}\big]$ and $\tbmu_{\beta}=\tbC_{\beta} \big[ \frac{\alpha}{\beta}\bX'\bR_{\phi}^{-1}\bZ \big]$

\subsection{Variational Function for \texorpdfstring{$\ssq$}{Lg}}
Note that there is a conjugacy for $\ssq$. 
\begin{align*}
    q(\ssq)&\propto \exp\big[E_{q(-\ssq)}[\log p(\bZ, \bbe, \phi, \ssq)]\big]\\
    &\propto \exp\big[E_{q(\bbe,\phi)}[\log p(\bZ, \bbe, \phi, \ssq)]\big]\\
    &\propto \exp \big[E_{q(\bbe,\phi)}[-\frac{n}{2}\log\ssq -\frac{1}{2\ssq} (\bZ-\bX\bbe)'\bR_{\phi}^{-1}(\bZ-\bX\bbe)  \\
    &\qquad +\alpha_0\log \beta_0-\log\Gamma(\alpha_0)-(\alpha_0+1)\log\ssq-\frac{\beta_0}{\ssq} ]\big]\\
    &\propto E_{q(\bbe,\phi)}[(\ssq)^{-(\frac{n}{2}+\alpha_0)-1} \exp[-\frac{1}{\ssq}(\bbe_0+\frac{1}{2} (\bZ-\bX\bbe)'\bR_{\phi}^{-1}(\bZ-\bX\bbe))]]\\
    &\propto (\ssq)^{-(\frac{n}{2}+\alpha_0)-1} \exp[-\frac{1}{\ssq}(\bbe_0+\frac{1}{2} (\bZ-\bX\tbmu_{\beta})'\bR_{\phi}^{-1}(\bZ-\bX\tbmu_{\beta}) \\
    &\qquad +\frac{1}{2}\mbox{tr}[\bR_{\phi}^{-1}\bX\tbC_{\beta}\bX'])]
\end{align*}

\tb{Key Components:}
\begin{enumerate}
    \item Variational Function for $\ssq$: $q(\ssq)= IG(\tilde{\alpha}, \tilde{\beta})$
    \item Expectation of Quadratic Forms: $E_{q(\bga)}\Big[\bga'\bSig_{\gamma}^{-1}\bga \Big]=\tbmu_\gamma'\bSig_{\gamma}^{-1}\tbmu_\gamma+tr[\bSig_{\gamma}^{-1}\tbC_\gamma]$
    \item $E[\log\frac{1}{\ssq}] \approx \psi(\alpha)-\log\beta$
\end{enumerate}
\tb{Resulting Variational Function:}  $q(\ssq)=IG(\tilde{\alpha},\tilde{\beta})$
\begin{align*}
  &\widetilde{\alpha}=\alpha_0+\frac{n}{2} \\
  &\widetilde{\beta}=\beta_0+\frac{1}{2} (\bZ-\bX\tbmu_{\beta})'\bR_{\phi}^{-1}(\bZ-\bX\tbmu_{\beta})+\frac{1}{2}\mbox{tr}[\bR_{\phi}^{-1}\bX\tbC_{\beta}\bX']
\end{align*}

\subsection{Evidence Lower Bound Calculation}
The evidence lower bound (ELBO) is a critical component for the stopping criterion ($\epsilon$) in the iteration as well as the variational weights $A_j=\frac{ELBO^{(j)}}{\sum_{i=1}^{J} ELBO^{(i)}}$  The ELBO is computed as follows:
\begin{align*}
    ELBO&=E_q\Bigg[\log \frac{p(\bZ, \bbe, \ssq, \phi)}{q(\bbe, \ssq, \phi)}\Bigg]\\
    & = \tblue{E_q[\log p(\bZ, \bbe, \ssq, \phi)]} - \tred{E_q[\log q(\bbe, \ssq, \phi)]}
\end{align*}

\begin{center}
$\theta_c=\{ \bbe, \ssq\},\qquad \btheta_d=\{ \phi \}$    
\end{center}

\tblue{\tb{First Term:}}
\begin{align*}
\tblue{E_q[\log p(\bZ, \bbe, \ssq, \phi)]}&=E_q\Big[ -\frac{n}{2}\log\ssq-\frac{1}{2}\log|\bR_{\phi}|-\frac{1}{2\ssq} (\bZ-\bX\bbe)'\bR_{\phi}^{-1}(\bZ-\bX\bbe) \\
&\qquad -\frac{1}{2}\log|\bSig_{\beta}|-\frac{1}{2}\bbe'\bSig_{\beta}^{-1}\bbe+\log p(\phi)\\
&\qquad +\alpha_0\log \beta_0-\log\Gamma(\alpha_0)-(\alpha_0+1)\log\ssq-\frac{\beta_0}{\ssq} \Big]\\
&\propto E_q\Big[ -\frac{1}{2}\log|\bR_{\phi}|-\frac{1}{2\ssq} (\bZ-\bX\bbe)'\bR_{\phi}^{-1}(\bZ-\bX\bbe) \\
& \qquad -\frac{1}{2}\log|\bSig_{\beta}|-\frac{1}{2}\bbe'\bSig_{\beta}^{-1}\bbe\\
& \qquad +\alpha_0\log \beta_0-\log\Gamma(\alpha_0)-(\alpha_0+\frac{n}{2}+1)\log\ssq-\frac{\beta_0}{\ssq} \Big]\\
&\propto  -\frac{1}{2}\log|\bR_{\phi}|-E[\frac{1}{\ssq}][\frac{1}{2}(\bZ-\bX\tbmu_{\beta})'\bR_{\phi}^{-1}(\bZ-\bX\tbmu_{\beta})+\frac{1}{2}\mbox{tr}[\bR_{\phi}^{-1}\bX\tbC_{\beta}\bX']] \\
& \qquad -\frac{1}{2}\log|\bSig_{\beta}|-\frac{1}{2}\tbmu_{\beta}'\bSig_{\beta}^{-1}\tbmu_{\beta}-\frac{1}{2}\mbox{tr}[\bSig_{\beta}^{-1}\tbC_{\beta} ]\\
& \qquad +\alpha_0\log \beta_0-\log\Gamma(\alpha_0)+(\alpha_0+\frac{n}{2}+1)E[\log\frac{1}{\ssq}]-\beta_0E[\frac{1}{\ssq}] \\
&\propto  -\frac{1}{2}\log|\bR_{\phi}|-\frac{\tilde{\alpha}}{\tilde{\beta}}[\frac{1}{2}(\bZ-\bX\tbmu_{\beta})'\bR_{\phi}^{-1}(\bZ-\bX\tbmu_{\beta})+\frac{1}{2}\mbox{tr}[\bR_{\phi}^{-1}\bX\tbC_{\beta}\bX']] \\
& \qquad -\frac{1}{2}\log|\bSig_{\beta}|-\frac{1}{2}\tbmu_{\beta}'\bSig_{\beta}^{-1}\tbmu_{\beta}-\frac{1}{2}\mbox{tr}[\bSig_{\beta}^{-1}\tbC_{\beta} ]\\
& \qquad +\alpha_0\log \beta_0-\log\Gamma(\alpha_0)+(\alpha_0+\frac{n}{2}+1)(\psi(\tilde{\alpha})-\log\tilde{\beta})-\beta_0\frac{\tilde{\alpha}}{\tilde{\beta}} \\
\end{align*}

\tred{\tb{Second Term:}}
\begin{align*}
\tred{E_q[\log p(\bbe, \ssq, \phi)]}&=E_q\Big[-\frac{p}{2}\log2\pi-\frac{1}{2}\log|\tbC_{\beta}|-\frac{1}{2}(\bbe-\tbmu_{\beta})'\tbC_{\beta}^{-1}(\bbe-\tbmu_{\beta})  \\
& \qquad \tilde{\alpha}\log\tilde{\beta}-\log\Gamma(\tilde{\alpha})+(\tilde{\alpha}+1)\log\frac{1}{\ssq}-\tilde{\beta}\frac{1}{\ssq} \Big]\\
&\propto -\frac{1}{2}\log|\tbC_{\beta}|-\frac{1}{2}\mbox{tr}(\tbC_{\beta}^{-1}\tbC_{\beta}) \\
& \qquad \tilde{\alpha}\log\tilde{\beta}-\log\Gamma(\tilde{\alpha})+(\tilde{\alpha}+1)E[\log\frac{1}{\ssq}]-\tilde{\beta}E[\frac{1}{\ssq}]\\
&\propto -\frac{1}{2}\log|\tbC_{\beta}|-\frac{p}{2} \\
& \qquad \tilde{\alpha}\log\tilde{\beta}-\log\Gamma(\tilde{\alpha})+(\tilde{\alpha}+1)(\psi(\tilde{\alpha})-\log\tilde{\beta})-\tilde{\beta}\frac{\tilde{\alpha}}{\tilde{\beta}}\\
\end{align*}

\noindent \tb{Evidence Lower Bound Calculation}
\begin{align*}
    ELBO&=E_q\Bigg[\log \frac{p(\bZ, \bbe, \ssq, \phi)}{q(\bbe, \ssq, \phi)}\Bigg]\\
    & = \tblue{E_q[\log p(\bZ, \bbe, \ssq, \phi)]} - \tred{E_q[\log q(\bbe, \ssq, \phi)]} \\
    &\propto  \tblue{-\frac{1}{2}\log|\bR_{\phi}|-\frac{\tilde{\alpha}}{\tilde{\beta}}[\frac{1}{2}(\bZ-\bX\tbmu_{\beta})'\bR_{\phi}^{-1}(\bZ-\bX\tbmu_{\beta})+\frac{1}{2}\mbox{tr}[\bR_{\phi}^{-1}\bX\tbC_{\beta}\bX']]} \\
    & \qquad \tblue{-\frac{1}{2}\log|\bSig_{\beta}|-\frac{1}{2}\tbmu_{\beta}'\bSig_{\beta}^{-1}\tbmu_{\beta}-\frac{1}{2}\mbox{tr}[\bSig_{\beta}^{-1}\tbC_{\beta}]}\\
    & \qquad \tblue{+\alpha_0\log \beta_0-\log\Gamma(\alpha_0)+(\alpha_0+\frac{n}{2}+1)(\psi(\tilde{\alpha})-\log\tilde{\beta})-\beta_0\frac{\tilde{\alpha}}{\tilde{\beta}}}\\
    &\qquad \tred{+\frac{1}{2}\log|\tbC_{\beta}|+\frac{p}{2}-\tilde{\alpha}\log\tilde{\beta}+\log\Gamma(\tilde{\alpha})-(\tilde{\alpha}+1)(\psi(\tilde{\alpha})-\log\tilde{\beta})+\tilde{\beta}\frac{\tilde{\alpha}}{\tilde{\beta}}}\\
    &\propto \tblue{-\frac{1}{2}\log|\bR_{\phi}|-\frac{1}{2}\log|\bSig_{\beta}|-\frac{1}{2}\tbmu_{\beta}'\bSig_{\beta}^{-1}\tbmu_{\beta}-\frac{1}{2}\mbox{tr}[\bSig_{\beta}^{-1}\tbC_{\beta}]+\alpha_0\log \beta_0-\log\Gamma(\alpha_0)} \\
    &\qquad \tred{+\frac{1}{2}\log|\tbC_{\beta}|+\frac{p}{2}-\tilde{\alpha}\log\tilde{\beta}+\log\Gamma(\tilde{\alpha})} \\
    &\propto \tblue{-\frac{1}{2}\log|\bR_{\phi}|-\frac{1}{2}\log|\bSig_{\beta}|-\frac{1}{2}\tbmu_{\beta}'\bSig_{\beta}^{-1}\tbmu_{\beta}-\frac{1}{2}\mbox{tr}[\bSig_{\beta}^{-1}\tbC_{\beta}]}\tred{+\frac{1}{2}\log|\tbC_{\beta}|-\tilde{\alpha}\log\tilde{\beta}}
\end{align*}

\tb{Key Components:}
\begin{enumerate}
    \item Variational Function for $\ssq$: $q(\ssq)= IG(\tilde{\alpha}, \tilde{\beta})$
    \item $\widetilde{\alpha}=\alpha_0+\frac{n}{2}$ 
    \item $\widetilde{\beta}=\beta_0+\frac{1}{2} (\bZ-\bX\tbmu_{\beta})'\bR_{\phi}^{-1}(\bZ-\bX\tbmu_{\beta})+\frac{1}{2}\mbox{tr}[\bR_{\phi}^{-1}\bX\tbC_{\beta}\bX']$
\end{enumerate}

\newpage
\section{Full-SGLMM (Poisson Data Model): Discretized \texorpdfstring{$\phi^{(j)}$}{Lg} }
\tb{Hierarchical Model (Original):}
% \vspace{-0.3in}
\begin{align*}\vspace{-0.1in}
    \tb{Poisson Data Model:} &\qquad  Z_i|\lambda_i \sim \mbox{Pois}(\lambda_i)\\
    & \qquad \lambda_i = \exp\{\bX_{i}^\prime\bbe +W_i\}\\
    \tb{Process Model:}& \qquad \bW|\ssq,\phi \sim \mcN(\bzero,\ssq\bR_{\phi}), \qquad \bW=(W_1,...,W_n)'\\
    \tb{Prior Model:}& \qquad \bbe \sim \mcN(\bzero,\bSig_{\beta}), \qquad \ssq \sim IG(\alpha_\sigma, \beta_\sigma), \qquad \phi \sim \text{Unif}(0,\sqrt{2})
\end{align*}
\textbf{Hierarchical Model (Modified):}
% \vspace{-0.3in}
\begin{align*}\vspace{-0.1in}
       \textbf{\mbox{Poisson Data Model:}} &\quad  Z_i|\lambda_i \sim \mbox{Pois}(\lambda_i) \quad \text{where} \hspace{0.1cm} \lambda_i=\exp\{\textcolor{black}{\widetilde{\bX_{i}}'} \textcolor{black}{\boldsymbol{\gamma}} \} \\
    & \quad \textcolor{black}{\widetilde{\bX}}=[\bX \quad \textcolor{black}{I_{n}}]\mbox{ and } \textcolor{black}{\boldsymbol{\gamma}}=\pr{(\bbe , \bW)}\\
    \textbf{\mbox{Process Model:}} &\\
        & \quad  \textcolor{black}{\boldsymbol{\gamma}|\ssq, \phi  \sim \mcN \bigg(\begin{bmatrix}\mu_{\beta} \\ 0 \end{bmatrix},  \begin{bmatrix}
\Sigma_{\beta} & 0  \\
0 & \sigma^2R_{\phi}  
\end{bmatrix} \bigg)} \\ 
    \textbf{\mbox{Prior Model:}}& \\ 
    & \quad \sigma^2 \sim \mbox{IG}(\alpha_{0},\beta_{0}) \\
    & \quad \phi \sim \mbox{Unif}(0,\sqrt{2}) \\
\end{align*}

\noindent \tb{Objective:} Obtain variational functions $q(\bga)$ , $q(\ssq)$, and $q(\phi)$ via Mean Field Variational Bayes(MFVB) to approximate $p(\bga|\cdot)$ , $p(\ssq|\cdot)$, and $p(\phi|\cdot)$

\tb{Probability Density Functions}
\begin{align*}
    \mbox{Joint:}& \qquad p(\bZ, \bga, \phi, \sigma^2)= p(\bZ|\bga)p(\bga|\phi,\ssq)p(\phi)p(\ssq)\\
    \mbox{Likelihood:}& \qquad p(\bZ|\bga)= \prod_{i=1}^{n}\frac{\lambda_i^{Z_i}e^{-\lambda_i}}{Z_i!}, \qquad \mbox{where } \lambda_i=\exp\{\widetilde{\bX_i'}\bga\}\\
    \mbox{Process :}&\qquad p(\bga|\phi,\ssq) = (2\pi)^{-(n+p)/2}|\bSig_{\gamma}|^{-1/2}\exp\{-\frac{1}{2}\bga'\bSig_{\gamma}^{-1}\bga\}\\
    \mbox{Prior :}& \qquad p(\ssq)= \frac{\beta_0^{\alpha_0}}{\Gamma(\alpha_0)}(\ssq)^{-\alpha_0-1}\mbox{exp}(-\frac{\beta_0}{\ssq})\\
    & \qquad p(\phi) \sim \mbox{Unif}(0,\sqrt{2}) \\
    \mbox{Proposal :}& \qquad q(\bga)=2\pi^{\frac{-(n+p)}{2}} |\tbC_{\bga}|^{\frac{-1}{2}} \exp(-\frac{1}{2}(\bga-\tbmu_{\bga})'\tbC_{\bga}^{-1}(\bga-\tbmu_{\bga}))
\end{align*}

\noindent \tb{Log Joint posterior density}:
\begin{align*}
    \log[p(\bZ, \bga, \phi, \ssq)]& = \log[p(\bZ|\bga)]+\log[p(\bga|\phi,\ssq)]+\log[p(\phi)]+\log[p(\ssq)]\\
    &\propto \sum_{i=1}^{n}Z_i\log\lambda_i-\sum_{i=1}^{n}\lambda_i -\sum_{i=1}^{n}\log Z_i!-\frac{1}{2}\log|\bSig_\gamma|-\frac{1}{2}\bga'\bSig_{\gamma}^{-1}\bga+\log p(\phi) \\
    &\qquad+\alpha_0\log \beta_0-\log\Gamma(\alpha_0)-(\alpha_0+1)\log\ssq-\frac{\beta_0}{\ssq}\\
    &\propto \sum_{i=1}^{n}Z_i(\widetilde{\bX_i'}\bga) -\sum_{i=1}^{n}\exp\{\widetilde{\bX_i'}\bga\}-\frac{1}{2}(\log|\bSig_\gamma|+\bga'\bSig_{\gamma}^{-1}\bga)+\log p(\phi) \\
    &\qquad+\alpha_0\log \beta_0-\log\Gamma(\alpha_0)-(\alpha_0+1)\log\ssq-\frac{\beta_0}{\ssq}\\
    &\propto \bZ'(\widetilde{\bX}\bga) -\bOne_n' \Big(e^{\widetilde{\bX}\bga}\Big)-\frac{1}{2}(\log|\bSig_\gamma|+\bga'\bSig_{\gamma}^{-1}\bga)+\log p(\phi) \\
    &\qquad+\alpha_0\log \beta_0-\log\Gamma(\alpha_0)-(\alpha_0+1)\log\ssq-\frac{\beta_0}{\ssq}
\end{align*}

\subsection{Variational Function for \texorpdfstring{$\bbe$ and $\bW$}{Lg} } 
We represent $\boldsymbol{\gamma}=\pr{(\bbe , \bW)}$ to preserve dependence between $\bbe$ and $\bW$.\\ 

\noindent \tb{Computing the Variational Function $q(\bga)$:} We implement Mean Field Variational Bayes (MFVB) and include a Laplace approximation (2nd order Taylor Expansion). The objective function is as follows:
\begin{align*}
    f(\bga)&= E_{q(-\bga)}[\log p(\bZ, \bga, \phi, \ssq)]\\
    &= E_{q(\ssq)}[\log p(\bZ, \bga, \phi, \ssq)] \\
    &= E_{q(\ssq)}[\bZ'(\widetilde{\bX}\bga) -\bOne_n' \Big(e^{\widetilde{\bX}\bga}\Big)-\frac{1}{2}(\log|\bSig_\gamma|+\bga'\bSig_{\gamma}^{-1}\bga)\\
    & \qquad \qquad -(\alpha_0+1)\log\ssq -\frac{\beta_0}{\ssq} ] \\
    &= \bZ'(\widetilde{\bX}\bga) -\bOne_n' \Big(e^{\widetilde{\bX}\bga}\Big)-\frac{1}{2}(\log|\bSig_\gamma|+\bga'\bSig_{\gamma}^{-1}\bga)\\
    & \qquad \qquad +(\alpha_0+1)E[\log\frac{1}{\ssq}] -\beta_0E[\frac{1}{\ssq}] 
\end{align*}
\tb{Key Components:}
\begin{enumerate}
    \item Variational Function for $\bga$: $q_\gamma(\bga)= \mcN(\tbmu_\gamma, \tbC_{\gamma})$
    \item Expectation of a lognormal RV: $E_{q(\bga)}[e^{\bX\bga}]= \exp\{\bX\tbmu_{\gamma}+\frac{1}{2}\mbox{diag}(\bX\tbC_{\gamma}\bX')\}$ 
    \item Expectation of Quadratic Forms: $E_{q(\bga)}\Big[\bga'\bSig_{\gamma}^{-1}\bga \Big]=\tbmu_\gamma'\bSig_{\gamma}^{-1}\tbmu_\gamma+tr[\bSig_{\gamma}^{-1}\tbC_\gamma]$
    \item $E[\log\frac{1}{\ssq}] \approx \psi(\alpha)-\log\beta$
\end{enumerate}
\tb{Resulting Variational Function:} 
$$q_\gamma(\bga)= \mcN(\tbmu_\gamma, \tbC_{\gamma})$$ where $\tbmu_\gamma=\argmax_{\gamma}f(\bga)$ and $\tbC_\gamma= -(\bH)^{-1}$ where $\bH=\frac{\partial^2 f}{\partial \gamma^2}\Bigr|_{\substack{\gamma=\tbmu_\gamma}}$

\subsection{Variational Function for \texorpdfstring{$\ssq$}{Lg}}
Note that there is a conjugacy for $\ssq$. 
\begin{align*}
    p(\ssq| \cdot) &\propto p(\bZ|\bga, \phi, \ssq)p(\ssq)=p(\bZ,\bga, \phi, \ssq)  \\
    &\propto (2\pi)^{\frac{-n}{2}}|\ssq R_{\phi}|^{-\frac{1}{2}} \cdot \mbox{exp}(\frac{-1}{2\ssq}\bW'R_{\phi}^{-1}\bW) \cdot \frac{\beta^{\alpha_0}}{\Gamma(\alpha_0)}(\ssq)^{-\alpha_0-1}\mbox{exp}(-\frac{\beta_0}{\ssq}) \\
    &\propto (\ssq)^{\frac{-n}{2}-\alpha_0-1} |R_{\phi}|^{-\frac{1}{2}} \cdot \mbox{exp}[\frac{-1}{\ssq}(\beta_0+\frac{1}{2}\bW'R_{\phi}^{-1}\bW )]
\end{align*}
To obtain the variational function $q(\ssq)$, we require $\tilde{\mu}_W$ and $\tilde{C}_W$ from the previous section. 
\begin{equation*}
    \tbmu_\gamma=\pr{(\tbmu_\beta, \tbmu_W)},\qquad  \tbC_{\gamma}=\begin{bmatrix}
        \tbC_{\beta} & \tbC_{\beta,W} \\
        \tbC_{W,\beta} & \tbC_{W} \\
    \end{bmatrix}
\end{equation*}

\noindent \tb{Computing the Variational Function $q(\ssq)$} We obtain the variational function via MFVB by taking expectation of the log joint probability distribution with respect to $\bga$.
\begin{align*}
    q(\ssq) &\propto \mbox{exp}\big[E_{q(-\ssq)}[\log p(\bZ, \bga, \phi, \ssq)]\big] \\
    &\propto \mbox{exp}\big[E_{q(\bga)}[\log p(\bZ, \bga, \phi, \ssq)]\big] \\
    &\propto \mbox{exp}\big[E_{q(\bga)}[(-(\frac{n}{2}+\alpha_0)-1)\log \ssq -\frac{1}{2}\log |R_{\phi}|-\frac{1}{\ssq}(\beta_0+\frac{1}{2}\bW'R_{\phi}^{-1}\bW)] \big] \\
    &\propto \mbox{exp}\big[(-(\frac{n}{2}+\alpha_0)-1)\log \ssq -\frac{1}{2}\log |R_{\phi}|-\frac{1}{\ssq}(\beta_0+\frac{1}{2}\tbmu_w'R_{\phi}^{-1}\tbmu_w+\frac{1}{2}\mbox{tr}(R_{\phi}^{-1}\tbC_{\bw})) \big]
\end{align*}
\tb{Resulting Variational Function:}  $q(\ssq)=IG(\tilde{\alpha},\tilde{\beta})$
\begin{align*}
  &\widetilde{\alpha}=\alpha_0+\frac{n}{2} \\
  &\widetilde{\beta}=\beta_0+\frac{1}{2}(\tbmu_w'R_{\phi}^{-1}\tbmu_w+\mbox{tr}(R_{\phi}^{-1}\tbC_{\bw}))
\end{align*}

\subsection{Evidence Lower Bound Calculation}
The evidence lower bound (ELBO) is a critical component for the stopping criterion ($\epsilon$) in the iteration as well as the variational weights $A_j=\frac{ELBO^{(j)}}{\sum_{i=1}^{J} ELBO^{(i)}}$  The ELBO is computed as follows:
\begin{align*}
    ELBO&=E_q\Bigg[\log \frac{p(\bZ, \bga, \ssq, \phi)}{q(\bga, \ssq, \phi)}\Bigg]\\
    & = \tblue{E_q[\log p(\bZ, \bga, \ssq, \phi)]} - \tred{E_q[\log q(\bga, \ssq, \phi)]}
\end{align*}

\begin{center}
$\theta_c=\{ \bga, \ssq\},\qquad \btheta_d=\{ \phi \}$    
\end{center}

\tblue{\tb{First Term:}}
\begin{align*}
\tblue{E_q[\log p(\bZ, \bga, \ssq, \phi)]}&=E_q\big[\bZ'(\widetilde{\bX}\bga) -\bOne_n' \Big(e^{\widetilde{\bX}\bga}\Big)-\frac{1}{2}(\log|\bSig_\gamma|+\bga'\bSig_{\gamma}^{-1}\bga)+\log p(\phi) \\
& \qquad+\alpha_0\log \beta_0-\log\Gamma(\alpha_0)-(\alpha_0+1)\log\ssq-\frac{\beta_0}{\ssq}\big] \\
& \propto \bZ'(\widetilde{\bX}\tbmu_\gamma) -\bOne_n' \Big(e^{\widetilde{\bX}\tbmu_\gamma+\frac{1}{2}\mbox{diag}(\widetilde{\bX}\tbC_{\gamma}\widetilde{\bX'} ) }\Big)-\frac{1}{2}(\log|\bSig_\gamma|+\tbmu_\gamma'\bSig_{\gamma}^{-1}\tbmu_\gamma+\mbox{tr}(\bSig_\gamma^{-1}\tbC_\gamma)) \\
& \qquad+\alpha_0\log \beta_0-\log\Gamma(\alpha_0)+(\alpha_0+1)E[\log\frac{1}{\ssq}]-\beta_0E[\frac{1}{\ssq}]\\
& \propto \bZ'(\widetilde{\bX}\tbmu_\gamma) -\bOne_n' \Big(e^{\widetilde{\bX}\tbmu_\gamma+\frac{1}{2}\mbox{diag}(\widetilde{\bX}\tbC_{\gamma}\widetilde{\bX'} ) }\Big)-\frac{1}{2}(\log|\bSig_\gamma|+\tbmu_\gamma'\bSig_{\gamma}^{-1}\tbmu_\gamma+\mbox{tr}(\bSig_\gamma^{-1}\tbC_\gamma)) \\
& \qquad+(\alpha_0+1)(\psi(\tilde{\alpha})-\log\tilde{\beta})-\beta_0 \frac{\tilde{\alpha}}{\tilde{\beta}}
\end{align*}

\tred{\tb{Second Term:}}
\begin{align*}
&\tred{E_q[\log q(\bga, \ssq, \phi)]} \\
&=E_q\big[\log \big[(2\pi)^{\frac{-(n+p)}{2}}|\tbC_\gamma|^{-\frac{1}{2}} \cdot \mbox{exp}(-\frac{1}{2}(\bga-\tbmu_\gamma)'\tbC_\gamma^{-1}(\bga-\tbmu_\gamma) ) \cdot \frac{\widetilde{\beta}^{\widetilde{\alpha}}}{\Gamma(\widetilde{\alpha})}(\ssq)^{-\widetilde{\alpha}-1}\mbox{exp}(-\frac{\widetilde{\beta}}{\ssq})\big]\big] \\
& \propto -\frac{1}{2} \log |\tbC_\gamma| -\frac{1}{2}\mbox{tr}(\tbC_\gamma^{-1} \tbC_\gamma)+\widetilde{\alpha} \log \widetilde{\beta}-\log\Gamma(\widetilde{\alpha})+(\widetilde{\alpha}+1)E[\log\frac{1}{\ssq}]-\widetilde{\beta}E[\frac{1}{\ssq}] \\
& \propto -\frac{1}{2} \log |\tbC_\gamma| -\frac{1}{2}(n+p)+\widetilde{\alpha} \log \widetilde{\beta}-\log\Gamma(\widetilde{\alpha})+(\widetilde{\alpha}+1)E[\log\frac{1}{\ssq}]-\widetilde{\beta}E[\frac{1}{\ssq}]\\
& \propto -\frac{1}{2} \log |\tbC_\gamma| +\widetilde{\alpha} \log \widetilde{\beta}-\log\Gamma(\widetilde{\alpha})+(\widetilde{\alpha}+1)(\psi(\tilde{\alpha})-\log\tilde{\beta})-\tilde{\alpha}
\end{align*}

\noindent \tb{Evidence Lower Bound Calculation}
\begin{align*}
    ELBO&=E_q\Bigg[\log \frac{p(\bZ, \bga, \ssq, \phi)}{q(\bga, \ssq, \phi)}\Bigg]\\
    & = \tblue{E_q[\log p(\bZ, \bga, \ssq, \phi)]} - \tred{E_q[\log q(\bga, \ssq, \phi)]} \\
    &\propto \bZ'(\widetilde{\bX}\tbmu_\gamma) -\bOne_n' \Big(e^{\widetilde{\bX}\tbmu_\gamma+\frac{1}{2}\mbox{diag}(\widetilde{\bX}\tbC_{\gamma}\widetilde{\bX'} ) }\Big)-\frac{1}{2}(\log|\bSig_\gamma|+\tbmu_\gamma'\bSig_{\gamma}^{-1}\tbmu_\gamma+\mbox{tr}(\bSig_\gamma^{-1}\tbC_\gamma)) \\
    &\qquad +\frac{1}{2}\log|\tbC_\gamma|-\widetilde{\alpha}\log\widetilde{\beta}+(\alpha_0-\widetilde{\alpha})(\psi(\tilde{\alpha})-\log\tilde{\beta}) +(\widetilde{\beta}-\beta_0)\frac{\tilde{\alpha}}{\tilde{\beta}}
\end{align*}

\newpage
\section{Full-SGLMM (Poisson Data Model): Discretized \texorpdfstring{$\phi^{(j)}$ and $\sigma^{2(j)}$}{Lg} }
\textbf{Hierarchical Model (Original):}
% \vspace{-0.3in}
\begin{align*}\vspace{-0.1in}
    \tb{Poisson Data Model:} &\qquad  Z_i|\lambda_i \sim \mbox{Pois}(\lambda_i)\\
    & \qquad \lambda_i = \exp\{\bX_{i}^\prime\bbe +W_i\}\\
    \tb{Process Model:}& \qquad \bW|\ssq,\phi \sim \mcN(\bzero,\ssq\bR_{\phi}), \qquad \bW=(W_1,...,W_n)'\\
    \tb{Prior Model:}& \qquad \bbe \sim \mcN(\bzero,\bSig_{\beta}), \qquad \ssq \sim IG(\alpha_\sigma, \beta_\sigma), \qquad \phi \sim \text{Unif}(0,\sqrt{2})
\end{align*}
\textbf{Hierarchical Model (Modified):}
% \vspace{-0.3in}
\begin{align*}\vspace{-0.1in}
       \textbf{\mbox{Poisson Data Model:}} &\quad  Z_i|\lambda_i \sim \mbox{Pois}(\lambda_i) \quad \text{where} \hspace{0.1cm} \lambda_i=\exp\{\textcolor{black}{\widetilde{\bX_{i}}'} \textcolor{black}{\boldsymbol{\gamma}} \} \\
    & \quad \textcolor{black}{\widetilde{\bX}}=[\bX \quad \textcolor{black}{I_{n}}]\mbox{ and } \textcolor{black}{\boldsymbol{\gamma}}=\pr{(\bbe , \bW)}\\
    \textbf{\mbox{Process Model:}} &\\
        & \quad  \textcolor{black}{\boldsymbol{\gamma}|\ssq, \phi  \sim \mcN \bigg(\begin{bmatrix}\mu_{\beta} \\ 0 \end{bmatrix},  \begin{bmatrix}
\Sigma_{\beta} & 0  \\
0 & \sigma^2R_{\phi}  
\end{bmatrix} \bigg)} \\ 
    \textbf{\mbox{Prior Model:}}& \\ 
    & \quad \sigma^2 \sim \mbox{IG}(\alpha_{0},\beta_{0}) \\
    & \quad \phi \sim \mbox{Unif}(0,\sqrt{2}) \\
\end{align*}
\noindent \tb{Objective:} Obtain variational functions $q(\bga)$ , $q(\ssq)$, and $q(\phi)$ via Mean Field Variational Bayes(MFVB) to approximate $p(\bga|\cdot)$ , $p(\ssq|\cdot)$, and $p(\phi|\cdot)$

\tb{Probability Density Functions}
\begin{align*}
    \mbox{Joint:}& \qquad p(\bZ, \bga, \phi, \sigma^2)= p(\bZ|\bga)p(\bga|\phi,\ssq)p(\phi)p(\ssq)\\
    \mbox{Likelihood:}& \qquad p(\bZ|\bga)= \prod_{i=1}^{n}\frac{\lambda_i^{Z_i}e^{-\lambda_i}}{Z_i!}, \qquad \mbox{where } \lambda_i=\exp\{\widetilde{\bX_i'}\bga\}\\
    \mbox{Process :}&\qquad p(\bga|\phi,\ssq) = (2\pi)^{-(n+p)/2}|\bSig_{\gamma}|^{-1/2}\exp\{-\frac{1}{2}\bga'\bSig_{\gamma}^{-1}\bga\}\\
    \mbox{Prior :}& \qquad p(\ssq)= \frac{\beta_0^{\alpha_0}}{\Gamma(\alpha_0)}(\ssq)^{-\alpha_0-1}\mbox{exp}(-\frac{\beta_0}{\ssq})\\
    & \qquad p(\phi) \sim \mbox{Unif}(0,\sqrt{2}) \\
    \mbox{Proposal :}& \qquad q(\bga)=2\pi^{\frac{-(n+p)}{2}} |\tbC_{\bga}|^{\frac{-1}{2}} \exp(-\frac{1}{2}(\bga-\tbmu_{\bga})'\tbC_{\bga}^{-1}(\bga-\tbmu_{\bga}))
\end{align*}

\noindent \tb{Log Joint posterior density}:
\begin{align*}
    \log[p(\bZ, \bga, \phi, \ssq)]& = \log[p(\bZ|\bga)]+\log[p(\bga|\phi,\ssq)]+\log[p(\phi)]+\log[p(\ssq)]\\
    &\propto \sum_{i=1}^{n}Z_i\log\lambda_i-\sum_{i=1}^{n}\lambda_i -\sum_{i=1}^{n}\log Z_i!-\frac{1}{2}\log|\bSig_\gamma|-\frac{1}{2}\bga'\bSig_{\gamma}^{-1}\bga+\log p(\phi) \\
    &\qquad+\alpha_0\log \beta_0-\log\Gamma(\alpha_0)-(\alpha_0+1)\log\ssq-\frac{\beta_0}{\ssq}\\
    &\propto \sum_{i=1}^{n}Z_i(\widetilde{\bX_i'}\bga) -\sum_{i=1}^{n}\exp\{\widetilde{\bX_i'}\bga\}-\frac{1}{2}(\log|\bSig_\gamma|+\bga'\bSig_{\gamma}^{-1}\bga)+\log p(\phi) \\
    &\qquad+\alpha_0\log \beta_0-\log\Gamma(\alpha_0)-(\alpha_0+1)\log\ssq-\frac{\beta_0}{\ssq}\\
    &\propto \bZ'(\widetilde{\bX}\bga) -\bOne_n' \Big(e^{\widetilde{\bX}\bga}\Big)-\frac{1}{2}(\log|\bSig_\gamma|+\bga'\bSig_{\gamma}^{-1}\bga)+\log p(\phi) \\
    &\qquad+\alpha_0\log \beta_0-\log\Gamma(\alpha_0)-(\alpha_0+1)\log\ssq-\frac{\beta_0}{\ssq}
\end{align*}

\subsection{Variational Function for \texorpdfstring{$\bbe$ and $\bW$}{Lg} }
We represent $\boldsymbol{\gamma}=\pr{(\bbe , \bW)}$ to preserve dependence between $\bbe$ and $\bW$.\\
\medskip   
\noindent \tb{Computing the Variational Function $q(\bga)$:} We implement Mean Field Variational Bayes (MFVB) and include a Laplace approximation (2nd order Taylor Expansion). The objective function is as follows:
\begin{align*}
    f(\bga)&= E_{q(-\bga)}[\log p(\bZ, \bga, \phi, \ssq)]\\
    &= E_{q(-\bga)}[\log p(\bZ, \bga, \phi, \ssq)] \\
    &= E_{q(-\bga)}\Bigg[\bZ'(\widetilde{\bX}\bga) -\bOne_n' \Big(e^{\widetilde{\bX}\bga}\Big)-\frac{1}{2}(\log|\bSig_\gamma|+\bga'\bSig_{\gamma}^{-1}\bga)\\
    & \qquad \qquad -(\alpha_0+1)\log\ssq -\frac{\beta_0}{\ssq} \Bigg] \\
    &= \bZ'(\widetilde{\bX}\bga) -\bOne_n' \Big(e^{\widetilde{\bX}\bga}\Big)-\frac{1}{2}(\log|\bSig_\gamma|+\bga'\bSig_{\gamma}^{-1}\bga)\\
    & \qquad \qquad -(\alpha_0+1)\log\ssq -\frac{\beta_0}{\ssq}  
\end{align*}
\tb{Key Components:}
\begin{enumerate}
    \item Variational Function for $\bga$: $q_\gamma(\bga)= \mcN(\tbmu_\gamma, \tbC_{\gamma})$
    \item Expectation of a lognormal RV: $E_{q(\bga)}[e^{\bX\bga}]= \exp\{\bX\tbmu_{\gamma}+\frac{1}{2}\mbox{diag}(\bX\tbC_{\gamma}\bX')\}$ 
    \item Expectation of Quadratic Forms: $E_{q(\bga)}\Big[\bga'\bSig_{\gamma}^{-1}\bga \Big]=\tbmu_\gamma'\bSig_{\gamma}^{-1}\tbmu_\gamma+tr[\bSig_{\gamma}^{-1}\tbC_\gamma]$
    \item $E[\log\frac{1}{\ssq}] \approx \psi(\alpha)-\log\beta$
\end{enumerate}
\tb{Resulting Variational Function:} 
$$q_\gamma(\bga)= \mcN(\tbmu_\gamma, \tbC_{\gamma})$$ where $\tbmu_\gamma=\argmax_{\gamma}f(\bga)$ and $\tbC_\gamma= -(\bH)^{-1}$ where $\bH=\frac{\partial^2 f}{\partial \gamma^2}\Bigr|_{\substack{\gamma=\tbmu_\gamma}}$

\subsection{Evidence Lower Bound Calculation}
The evidence lower bound (ELBO) is a critical component for the stopping criterion ($\epsilon$) in the iteration as well as the variational weights $A_j=\frac{ELBO^{(j)}}{\sum_{i=1}^{J} ELBO^{(i)}}$  The ELBO is computed as follows:
\begin{align*}
    ELBO&=E_q\Bigg[\log \frac{p(\bZ, \bga, \ssq, \phi)}{q(\bga, \ssq, \phi)}\Bigg]\\
    & = \tblue{E_q[\log p(\bZ, \bga, \ssq, \phi)]} - \tred{E_q[\log q(\bga, \ssq, \phi)]}
\end{align*}

\begin{center}
$\theta_c=\{ \bga\}, \qquad \btheta_d=\{ \phi,\ssq \}$    
\end{center}

\tblue{\tb{First Term:}}
\begin{align*}
\tblue{E_q[\log p(\bZ, \bga, \ssq, \phi)]}&=E_q\big[\bZ'(\widetilde{\bX}\bga) -\bOne_n' \Big(e^{\widetilde{\bX}\bga}\Big)-\frac{1}{2}(\log|\bSig_\gamma|+\bga'\bSig_{\gamma}^{-1}\bga)+\log p(\phi) \\
& \qquad+\alpha_0\log \beta_0-\log\Gamma(\alpha_0)-(\alpha_0+1)\log\ssq-\frac{\beta_0}{\ssq}\big] \\
& \propto \bZ'(\widetilde{\bX}\tbmu_\gamma) -\bOne_n' \Big(e^{\widetilde{\bX}\tbmu_\gamma+\frac{1}{2}\mbox{diag}(\widetilde{\bX}\tbC_{\gamma}\widetilde{\bX'} ) }\Big)-\frac{1}{2}(\log|\bSig_\gamma|+\tbmu_\gamma'\bSig_{\gamma}^{-1}\tbmu_\gamma+\mbox{tr}(\bSig_\gamma^{-1}\tbC_\gamma)) \\
& \qquad+\alpha_0\log \beta_0-\log\Gamma(\alpha_0)+(\alpha_0+1)\log\frac{1}{\ssq}-\beta_0\frac{1}{\ssq}
\end{align*}

\tred{\tb{Second Term:}}
\begin{align*}
&\tred{E_q[\log q(\bga, \ssq, \phi)]} \\
&=E_q\big[\log (2\pi)^{\frac{-(n+p)}{2}}|\tbC_\gamma|^{-\frac{1}{2}} \cdot \mbox{exp}(-\frac{1}{2}(\bga-\tbmu_\gamma)'\tbC_\gamma^{-1}(\bga-\tbmu_\gamma) ) \\
& \propto -\frac{1}{2} \log |\tbC_\gamma| -\frac{1}{2}\mbox{tr}(\tbC_\gamma^{-1} \tbC_\gamma)\\
& \propto -\frac{1}{2} \log |\tbC_\gamma| -\frac{1}{2}(n+p)\\
& \propto -\frac{1}{2} \log |\tbC_\gamma|
\end{align*}

\noindent \tb{Evidence Lower Bound Calculation}
\begin{align*}
    ELBO&=E_q\Bigg[\log \frac{p(\bZ, \bga, \ssq, \phi)}{q(\bga, \ssq, \phi)}\Bigg]\\
    & = \tblue{E_q[\log p(\bZ, \bga, \ssq, \phi)]} - \tred{E_q[\log q(\bga, \ssq, \phi)]} \\
    &\propto \bZ'(\widetilde{\bX}\tbmu_\gamma) -\bOne_n' \Big(e^{\widetilde{\bX}\tbmu_\gamma+\frac{1}{2}\mbox{diag}(\widetilde{\bX}\tbC_{\gamma}\widetilde{\bX'} ) }\Big)-\frac{1}{2}(\log|\bSig_\gamma|+\tbmu_\gamma'\bSig_{\gamma}^{-1}\tbmu_\gamma+\mbox{tr}(\bSig_\gamma^{-1}\tbC_\gamma)) \\
    &\qquad +\frac{1}{2}\log|\tbC_\gamma|+\alpha_0\log \beta_0-\log\Gamma(\alpha_0)+(\alpha_0+1)\log\frac{1}{\ssq}-\beta_0\frac{1}{\ssq}
\end{align*}

\newpage
\section{Full-SGLMM (Bernoulli Data Model): Discretized \texorpdfstring{$\phi^{(j)}$}{Lg} }
\tb{Hierarchical Model (Original):}
% \vspace{-0.3in}
\begin{align*}\vspace{-0.1in}
    \tb{Binary Data Model:} &\qquad  Z_i|p_i \sim \mbox{Bern}(p_i)\\
    & \qquad p_i = (1+\exp\{-\bX_{i}^\prime\bbe -W_i\})^{-1}\\
    \tb{Process Model:}& \qquad \bW|\ssq,\phi \sim \mcN(\bzero,\ssq\bR_{\phi}), \qquad \bW=(W_1,...,W_n)'\\
    \tb{Prior Model:}& \qquad \bbe \sim \mcN(\bzero,\bSig_{\beta}), \qquad \ssq \sim IG(\alpha_\sigma, \beta_\sigma), \qquad \phi \sim \text{Unif}(0,\sqrt{2})
\end{align*}

\textbf{Hierarchical Model (Modified):}
% \vspace{-0.3in}
\begin{align*}\vspace{-0.1in}
       \textbf{\mbox{Binary Data Model:}} &\quad  Z_i|p_i \sim \mbox{Bern}(p_i) \quad \text{where} \hspace{0.1cm} p_i = (1+\exp\{-{\widetilde{\bX_{i}}'}\bv \})^{-1} \\
    & \quad \textcolor{black}{\widetilde{\bX}}=[\bX \quad \textcolor{black}{I_{n}}]\mbox{ and } \textcolor{black}{\bv}=\pr{(\bbe , \bW)}\\
    \textbf{\mbox{Process Model:}} &\\
        & \quad  \textcolor{black}{\bv|\ssq, \phi  \sim \mcN \bigg(\begin{bmatrix}\mu_{\beta} \\ 0 \end{bmatrix},  \begin{bmatrix}
\Sigma_{\beta} & 0  \\
0 & \sigma^2R_{\phi}  
\end{bmatrix} \bigg)} \\ 
    \textbf{\mbox{Prior Model:}}& \\ 
    & \quad \sigma^2 \sim \mbox{IG}(\alpha_{0},\beta_{0}) \\
    & \quad \phi \sim \mbox{Unif}(0,\sqrt{2}) \\
\end{align*}
\noindent \tb{Objective:} Obtain variational functions $q(\bv)$ , $q(\ssq)$, and $q(\phi)$ via Mean Field Variational Bayes(MFVB) to approximate $p(\bv|\cdot)$ , $p(\ssq|\cdot)$, and $p(\phi|\cdot)$  \\ 

\tb{Probability Density Functions}
% \vspace{-0.3in}
\begin{align*}
    \mbox{Joint :}& \qquad p(\bZ, \bbe, \bW,\ssq)= p(\bZ|\bbe, \bW)p(\bbe)p(\bW|\ssq,\phi)p(\ssq)p(\phi)\\
    \mbox{Likelihood :}& \qquad p(\bZ|\bbe, \bW)= \prod_{i=1}^{n}p_{i}^{z_i}(1-p_{i})^{1-z_i}, \qquad \mbox{where } p_i=(1+\exp\{-\bX_{i}^\prime\bbe -W_i\})^{-1}\\
    \mbox{Process :}&\qquad p(\bW|\ssq, \phi) = (2\pi)^{-n/2}(\ssq)^{-n/2}|\bR_{\phi}|^{-1/2}\exp\{-\frac{1}{2\ssq}\bW'\bR_{\phi}^{-1}\bW\}\\
    \mbox{Prior :}& \qquad p(\bbe) = (2\pi)^{-p/2}|\bSig_{\beta}|^{-1/2}\exp\{-\frac{1}{2}\bbe'\bSig_{\beta}^{-1}\bbe\}\\
    & \qquad  p(\ssq)= \frac{\bsi^{\asi}}{\Gamma(\asi)}(\ssq)^{-\asi-1}\exp\{-\frac{\bsi}{\ssq}\}\\
    &\qquad  p(\phi) \sim \text{Unif}(0,\sqrt{2})    \\
    \mbox{Proposal :}& \qquad q(v)=2\pi^{\frac{-(n+p)}{2}} |\tbC_\bv|^{\frac{-1}{2}} \exp(-\frac{1}{2}(\bv-\tbmu_{\bv})'\tbC_\bv^{-1}(\bv-\tbmu_{\bv}))
\end{align*}

\noindent \tb{Log joint posterior density (Original)}:
\begin{align*}
    \log[p(\bZ, \bbe, \bW , \ssq, \phi)]& = \log[p(\bZ|\bbe, \bW)]+\log[p(\bbe)]+\log[p(\bW|\ssq,\phi)]+\log[p(\ssq)]+\log[p(\phi)]\\
    &= \bZ^{\prime}\bX\bbe+\bZ^{\prime}\bW -\bOne^\prime \log (1+\exp\{\bX\bbe+\bW\})\\
    & \qquad -\frac{1}{2}\Big(\bbe'\bSig_{\beta}^{-1}\bbe+n\log\ssq+\log|\bR_{\phi}|+(n+p)\log(2\pi)+\log|\bSig_{\beta}| +\frac{1}{\ssq}\bW'\bR_{\phi}^{-1}\bW\Big)\\
    & \qquad +\asi\log\bsi-\log\Gamma(\asi) -(\asi+1) \log\ssq -\frac{\bsi}{\ssq}
\end{align*}
\noindent \tb{Quadratic Approximation (Jaakola and Jordan, 1997):}
$$
-\log(1+e^{x})=\argmax_{\xi}\Big\{ \lambda(\xi)x^{2}-\frac{1}{2}x +\psi(\xi)\Big\}
$$
where $\lambda(\xi)=-\tanh(\xi/2)/(4\xi)$ and $\psi(\xi)=\xi/2-\log(1+e^{\xi})+\xi\tanh(\xi/2)/4$
\\
\\
\noindent \tb{Computing the optimal Auxiliary Variables} (Jaakola and Jordan, 1997)

$$
\pmb{\xi}= \sqrt{\mbox{Diagonal}\Big\{\tilde{X}(\tbC_{\bv}+\tbmu_{\bv}\tbmu_{\bv}')\tilde{X}'\Big\}}
$$
\\
\noindent \tb{Log joint posterior density (Modified)}:\\
Let $\bv=(\bbe, \bW)^\prime$, $\tilde{X}=[\bX\quad \bI]$, and $\bSig_{\bv}=\begin{bmatrix}
\bSig_{\beta}  & \bzero \\
\bzero & \ssq\bR_{\phi}
\end{bmatrix}$, then 
\begin{align*}
    \log[p(\bZ, \bbe, \bW , \ssq, \phi)]& = \bZ^{\prime}\tilde{X}\bv+ \bOne'\Big[\bv'\tilde{X}'\bD\tilde{X}\bv-\frac{1}{2}\tilde{X}\bv+\psi(\xi)\Big]\\
    & \qquad -\frac{1}{2}\Big(\bbe'\bSig_{\beta}^{-1}\bbe+n\log\ssq+\log|\bR_{\phi}| +(n+p)\log(2\pi)+\log|\bSig_{\beta}| +\frac{1}{\ssq}\bW'\bR_{\phi}^{-1}\bW\Big)\\
    & \qquad +\asi\log\bsi-\log\Gamma(\asi)-(\asi+1) \log\ssq -\frac{\bsi}{\ssq}\\
    &= \bv'\tilde{X}'\bD\tilde{X}\bv - \frac{1}{2}\bOne'\tilde{X}\bv+\bOne'\psi(\xi)+\bZ'\tilde{X}\bv-\frac{1}{2}\bv'\bSig_{\bv}^{-1}\bv \\
    & \qquad -\frac{1}{2}\Big(\log{|\bSig_{\bv}|}+(n+p)\log(2\pi) \Big) \\
    & \qquad +\asi\log\bsi-\log\Gamma(\asi)-(\alpha_\sigma+1)\log\ssq-\frac{\bsi}{\ssq}\\
    & = -\frac{1}{2}\Big(\bv'(-2\tilde{X}'\bD\tilde{X}+\bSig_{\bv}^{-1})\bv-2(\bZ'-\frac{1}{2}\bOne')\tilde{X}\bv\Big)+\bOne'\psi(\xi)\\
    & \qquad -\frac{1}{2}\Big(\log{|\bSig_{\bv}|}+(n+p)\log(2\pi) \Big) \\
    & \qquad +\asi\log\bsi-\log\Gamma(\asi)-(\alpha_\sigma+1)\log\ssq-\frac{\bsi}{\ssq} 
\end{align*}
where $\bf{D}=\mbox{diag}(\lambda(\pmb{\xi}))$. 

\subsection{Variational Function for \texorpdfstring{$\bbe$ and $\bW$}{Lg}}
We represent $\bv=(\bbe,\bW)'$ to preserve dependece between $\bbe$ and $\bW$. The distribution that minimizes the KL divergence is $q(\bv)\propto \exp\{E_{-\bv}[ \log p(\bZ, \bbe, \bW , \ssq, \phi)]\}$

$$q(\bv)\sim \mcN(\tbmu_{\bv}, \tbC_{\bv})$$
where $\tbC_{\bv}= (-2\tilde{X}'\bf{D}\tilde{X}+E[\bSig_{\bv}^{-1}])^{-1}$ and $\tbmu_{\bv}=\tbC_{\bv}\tilde{X}'(\bZ-\frac{1}{2}\bOne')$ and 

$$
E[\bSig_{\bv}^{-1}]=\begin{bmatrix}
\bSig_{\beta}^{-1}  & \bzero \\
\bzero & \frac{\tilde{\alpha}_\sigma}{ \tilde{\beta}_\sigma}\bR_{\phi}^{-1}
\end{bmatrix}
$$

Note that the covariance matrix and mean vector are split as follows:
$$
\tbC_{\bv}=\begin{bmatrix}
\tbC_\beta  & \tbC_{\beta,W} \\
\tbC_{W,\beta} & \tbC_{W}
\end{bmatrix}, \qquad \tbmu_{\bv} = (\tbmu_\beta , \tbmu_W)'
$$
\subsection{Variational Distribution for \texorpdfstring{$\sigma^2$}{Lg} }
The distribution that minimizes the KL divergence is $q(\ssq)\propto \exp\{E_{-\ssq}[ \log p(\bZ, \bbe, \bW , \ssq, \phi)]\}$
\begin{align*}
    \log [p(\bZ, \bbe, \bW , \ssq, \phi)]& \propto -(\alpha_\sigma +\frac{n}{2}+1)\log(\ssq)- \frac{1}{\ssq}[\beta_\sigma+\frac{1}{2}E[\bW'\bR_{\phi}^{-1}\bW]]\\
    & \qquad \propto -(\alpha_\sigma +\frac{n}{2}+1)\log(\ssq)- \frac{1}{\ssq}[\beta_\sigma+\frac{1}{2}(tr(\bR_\phi^{-1}\tbC_{W})+\bmu_w'\bR_\phi^{-1}\bmu_w)]
\end{align*}
It follows that $q(\ssq)= IG(\tilde{\alpha},\tilde{\beta})$ where $\tilde{\alpha}= \alpha_\sigma+n/2$ and $\tilde{\beta}= \beta_\sigma+\frac{1}{2}(tr(\bR_\phi^{-1}\tbC_{W})+\bmu_w'\bR_\phi^{-1}\bmu_w)$.

\subsection{Evidence Lower Bound}
\begin{align*}
    ELBO&=E_q\Bigg[\log \frac{p(\bZ, \bbe, \bW ,\ssq, \phi)}{q(\bbe,\bW,\ssq, \phi)}\Bigg]\\
    & = \tblue{E_q[\log p(\bZ, \bbe, \bW,\ssq, \phi)]} - \tred{E_q[log(q(\bbe,\bW,\ssq, \phi))]}
\end{align*}

Here, we have to decompose by partitioning the parameter space $\btheta=(\pmb{\theta}_c,{\pmb{\theta}_d})'$. We estimate $\pmb{\theta}_c$ but fix $ \pmb{\theta}_d$
$$\pmb{\theta}_c={ \{\bv, \ssq \} }, \qquad \pmb{\theta}_d={ \{\phi \} } $$

\tblue{Part 1:} $\tblue{E_q[\log p(\bZ, \bbe, \bW,\ssq, \phi)]}$
\begin{align*}
    E_q[\log p(\bZ, \bbe, \bW,\ssq, \phi)]& = E_q\Big[-\frac{1}{2}\Big(\bv'(-2\tilde{X}'\bD\tilde{X}+\bSig_{\bv}^{-1})\bv-2(\bZ'-\frac{1}{2}\bOne')\tilde{X}\bv\Big)+\bOne'\psi(\xi)\\
    & \qquad -\frac{1}{2}\Big(\log{|\bSig_{\bv}|}+(n+p)\log(2\pi) \Big) \\
    & \qquad +\asi\log\bsi-\log\Gamma(\asi)-(\alpha_\sigma+1)\log\ssq-\frac{\bsi}{\ssq}\Big] \\
    & =
    E_{q}[\bv'\tilde{X}'\bD\tilde{X}\bv]-\frac{1}{2}E_{q}[\bv'\bSig_{\bv}^{-1}\bv]+E_{q}[(\bZ'-\frac{1}{2}\bOne')\tilde{X}\bv]+\bOne'\psi(\xi) \\
    & \qquad -\frac{1}{2}\Big(\log{|\bSig_{\bv}|}+(n+p)\log(2\pi) \Big) \\
    & \qquad +\asi\log\bsi-\log\Gamma(\asi)+(\alpha_\sigma+1)
    E_{q}[\log\frac{1}{\ssq}]-\bsi E_{q}[\frac{1}{\ssq}] \\
    & =
    \tbmu_{\bv}'\tilde{X}'\bD\tilde{X}\tbmu_{\bv}+tr[\tilde{X}'\bD\tilde{X} \tbC_{\bv}]-\frac{1}{2}\tbmu_{\bv}'\bSig_{\bv}^{-1}\tbmu_{\bv}-\frac{1}{2} tr[\bSig_{\bv}^{-1}\tbC_{\bv}] +\bOne'\psi(\xi)\\
    & \qquad +(\bZ'-\frac{1}{2}\bOne')\tilde{X}\tbmu_{\bv} -\frac{1}{2}\Big(\log{|\bSig_{\bv}|}+(n+p)\log(2\pi) \Big) \\
    & \qquad +\asi\log\bsi-\log\Gamma(\asi)+(\alpha_\sigma+1)
    E_{q}[\log\frac{1}{\ssq}]-\bsi E_{q}[\frac{1}{\ssq}]\\
    & = \tbmu_{\bv}'\tilde{X}'\bD\tilde{X}\tbmu_{\bv}+tr[\tilde{X}'\bD\tilde{X} \tbC_{\bv}]-\frac{1}{2}\tbmu_{\bv}'\bSig_{\bv}^{-1}\tbmu_{\bv}-\frac{1}{2} tr[\bSig_{\bv}^{-1}\tbC_{\bv}] +\bOne'\psi(\xi)\\
    & \qquad +(\bZ'-\frac{1}{2}\bOne')\tilde{X}\tbmu_{\bv} -\frac{1}{2}\Big(\log{|\bSig_{\bv}|}+(n+p)\log(2\pi) \Big) \\
    & \qquad +\asi\log\bsi-\log\Gamma(\asi)+(\alpha_\sigma+1)
    (\psi(\tilde{\alpha})-\log\tilde{\beta}) -\bsi \frac{\tilde{\alpha}}{\tilde{\beta}}
\end{align*}

\tb{Key Components:}
\begin{enumerate}
    \item Expectation of a log Gamma RV: $E[\text{log} \frac{1}{\ssq}] \approx \psi(\tilde{\alpha})-\log\tilde{\beta} $
    \item Expectation of Quadratic Forms: $E_{q(\bv)}\Big[\bv'\bSig_{\bv}^{-1}\bv\Big]=\tbmu_\bv'\bSig_{\bv}^{-1}\tbmu_\bv+tr[\bSig_{\bv}^{-1}\tbC_\bv]$
\end{enumerate}

\tred{Part 2:} $\tred{E_q[\log q(\bbe, \bW,\ssq, \phi)]}$
\begin{align*}
    E_q[\log q(\bbe, \bW,\ssq, \phi)]& \propto E_q\Big[\log[q(v)]+\log[q(\ssq)]\Big] \\
    & = E_q\Big[\log[2\pi^{\frac{-(n+p)}{2}} |\tbC_\bv|^{\frac{-1}{2}} \exp(-\frac{1}{2}(\bv-\tbmu_{\bv})'\tbC_\bv^{-1}(\bv-\tbmu_{\bv}) ) \\  
    & \qquad +\log [\frac{\tilde{\beta}^{\tilde{\alpha}}}{\Gamma(\tilde{\alpha})} (\ssq)^{-\tilde{\alpha}-1} \cdot \exp(\frac{-\tilde{\beta}}{\ssq}) ]\Big] \\
    & = 
    E_{q}\Big[-\frac{(n+p)}{2}\log2\pi-\frac{1}{2} \log|\tbC_\bv|-\frac{1}{2}(\bv-\tbmu_{\bv})'\tbC_\bv^{-1}(\bv-\tbmu_{\bv})+\tilde{\alpha}\log\tilde{\beta} \\
    & \qquad -\log\Gamma(\tilde{\alpha}) + (\tilde{\alpha}+1)\log\frac{1}{\ssq}-\tilde{\beta}\frac{1}{\ssq} \Big]\\
    & =
    -\frac{(n+p)}{2}\log2\pi-\frac{1}{2} \log|\tbC_\bv| -\frac{1}{2} tr[\tbC_\bv^{-1}\tbC_\bv]+\tilde{\alpha}\log\tilde{\beta}-\log\Gamma(\tilde{\alpha}) \\
    &\qquad +(\tilde{\alpha}+1)E_q[\log \frac{1}{\ssq}]-\tilde{\beta}E_q[\frac{1}{\ssq}] \\
    & =
    -\frac{(n+p)}{2}\log2\pi-\frac{1}{2} \log|\tbC_\bv| -\frac{1}{2} (n+p)+\tilde{\alpha}\log\tilde{\beta} -\log\Gamma(\tilde{\alpha})\\
    &\qquad +(\tilde{\alpha}+1)(\psi(\tilde{\alpha})-\log\tilde{\beta}) -\tilde{\beta}\frac{\tilde{\alpha}}{\tilde{\beta}}
\end{align*}

\noindent \tb{Evidence Lower Bound:}
\begin{align*}
    ELBO&=E_q\Bigg[\log \frac{p(\bZ, \bbe, \bW ,\ssq, \phi)}{q(\bbe,\bW,\ssq, \phi)}\Bigg]\\
    & = \tblue{E_q[\log p(\bZ, \bbe, \bW,\ssq, \phi)]} - \tred{E_q[log(q(\bbe,\bW,\ssq, \phi))]} \\
       & = \tblue{\tbmu_{\bv}'\tilde{X}'\bD\tilde{X}\tbmu_{\bv}+tr[\tilde{X}'\bD\tilde{X} \tbC_{\bv}]-\frac{1}{2}\tbmu_{\bv}'\bSig_{\bv}^{-1}\tbmu_{\bv}-\frac{1}{2} tr[\bSig_{\bv}^{-1}\tbC_{\bv}] +\bOne'\psi(\xi)}\\
    & \qquad \tblue{+(\bZ'-\frac{1}{2}\bOne')\tilde{X}\tbmu_{\bv} -\frac{1}{2}\Big(\log{|\bSig_{\bv}|}+(n+p)\log(2\pi) \Big)} \\
    & \qquad \tblue{+\asi\log\bsi-\log\Gamma(\asi)+(\alpha_\sigma+1)
    (\psi(\tilde{\alpha})-\log\tilde{\beta}) -\bsi \frac{\tilde{\alpha}}{\tilde{\beta}}} \\
    & \qquad
    \tred{-\Bigg[-\frac{(n+p)}{2}\log2\pi-\frac{1}{2} \log|\tbC_\bv| -\frac{1}{2} (n+p)+\tilde{\alpha}\log\tilde{\beta} -\log\Gamma(\tilde{\alpha})}\\
    &\qquad \tred{+(\tilde{\alpha}+1)(\psi(\tilde{\alpha})-\log\tilde{\beta}) -\tilde{\beta}\frac{\tilde{\alpha}}{\tilde{\beta}} \Bigg]} \\
    &=
    \tbmu_{\bv}'\tilde{X}'\bD\tilde{X}\tbmu_{\bv}+tr[\tilde{X}'\bD\tilde{X} \tbC_{\bv}]-\frac{1}{2}\tbmu_{\bv}'\bSig_{\bv}^{-1}\tbmu_{\bv}-\frac{1}{2} tr[\bSig_{\bv}^{-1}\tbC_{\bv}]+\bOne'\psi(\xi)-\frac{1}{2}\log{|\bSig_{\bv}|}\\
    & \qquad +(\bZ'-\frac{1}{2}\bOne')\tilde{X}\tbmu_{\bv}+\frac{1}{2} \log|\tbC_\bv|+\asi\log\bsi-\log\Gamma(\asi)-\tilde{\alpha}\log\tilde{\beta} \\
    & \qquad +(\asi-\tilde{\alpha})(\psi(\tilde{\alpha})-\log\tilde{\beta}) +(\tilde{\beta}-\bsi)\frac{\tilde{\alpha}}{\tilde{\beta}}
    \end{align*}

\newpage
\section{Full-SGLMM (Bernoulli Data Model): Discretized \texorpdfstring{$\phi^{(j)}$ and $\sigma^{2(j)}$}{Lg} }
\tb{Hierarchical Model (Original):}
% \vspace{-0.3in}
\begin{align*}\vspace{-0.1in}
    \tb{Binary Data Model:} &\qquad  Z_i|p_i \sim \mbox{Bern}(p_i)\\
    & \qquad p_i = (1+\exp\{-\bX_{i}^\prime\bbe -W_i\})^{-1}\\
    \tb{Process Model:}& \qquad \bW|\ssq,\phi \sim \mcN(\bzero,\ssq\bR_{\phi}), \qquad \bW=(W_1,...,W_n)'\\
    \tb{Prior Model:}& \qquad \bbe \sim \mcN(\bzero,\bSig_{\beta}), \qquad \ssq \sim IG(\alpha_\sigma, \beta_\sigma), \qquad \phi \sim \text{Unif}(0,\sqrt{2})
\end{align*}
\textbf{Hierarchical Model (Modified):}
% \vspace{-0.3in}
\begin{align*}\vspace{-0.1in}
       \textbf{\mbox{Binary Data Model:}} &\quad  Z_i|p_i \sim \mbox{Bern}(p_i) \quad \text{where} \hspace{0.1cm} p_i = (1+\exp\{-{\widetilde{\bX_{i}}'}\bv \})^{-1} \\
    & \quad \textcolor{black}{\widetilde{\bX}}=[\bX \quad \textcolor{black}{I_{n}}]\mbox{ and } \textcolor{black}{\bv}=\pr{(\bbe , \bW)}\\
    \textbf{\mbox{Process Model:}} &\\
        & \quad  \textcolor{black}{\bv|\ssq, \phi  \sim \mcN \bigg(\begin{bmatrix}\mu_{\beta} \\ 0 \end{bmatrix},  \begin{bmatrix}
\Sigma_{\beta} & 0  \\
0 & \sigma^2R_{\phi}  
\end{bmatrix} \bigg)} \\ 
    \textbf{\mbox{Prior Model:}}& \\ 
    & \quad \sigma^2 \sim \mbox{IG}(\alpha_{0},\beta_{0}) \\
    & \quad \phi \sim \mbox{Unif}(0,\sqrt{2}) \\
\end{align*}
\noindent \tb{Objective:} Obtain variational functions $q(\bv)$ , $q(\ssq)$, and $q(\phi)$ via Mean Field Variational Bayes(MFVB) to approximate $p(\bv|\cdot)$ , $p(\ssq|\cdot)$, and $p(\phi|\cdot)$  \\ 

\tb{Probability Density Functions}
% \vspace{-0.3in}
\begin{align*}
    \mbox{Joint :}& \qquad p(\bZ, \bbe, \bW,\ssq)= p(\bZ|\bbe, \bW)p(\bbe)p(\bW|\ssq,\phi)p(\ssq)p(\phi)\\
    \mbox{Likelihood :}& \qquad p(\bZ|\bbe, \bW)= \prod_{i=1}^{n}p_{i}^{z_i}(1-p_{i})^{1-z_i}, \qquad \mbox{where } p_i=(1+\exp\{-\bX_{i}^\prime\bbe -W_i\})^{-1}\\
    \mbox{Process :}&\qquad p(\bW|\ssq, \phi) = (2\pi)^{-n/2}(\ssq)^{-n/2}|\bR_{\phi}|^{-1/2}\exp\{-\frac{1}{2\ssq}\bW'\bR_{\phi}^{-1}\bW\}\\
    \mbox{Prior :}& \qquad p(\bbe) = (2\pi)^{-p/2}|\bSig_{\beta}|^{-1/2}\exp\{-\frac{1}{2}\bbe'\bSig_{\beta}^{-1}\bbe\}\\
    & \qquad  p(\ssq)= \frac{\bsi^{\asi}}{\Gamma(\asi)}(\ssq)^{-\asi-1}\exp\{-\frac{\bsi}{\ssq}\}\\
    &\qquad  p(\phi) \sim \text{Unif}(0,\sqrt{2}) \\    
    \mbox{Proposal :}& \qquad q(v)=2\pi^{\frac{-(n+p)}{2}} |\tbC_\bv|^{\frac{-1}{2}} \exp(-\frac{1}{2}(\bv-\tbmu_{\bv})'\tbC_\bv^{-1}(\bv-\tbmu_{\bv}) )
\end{align*}

\noindent \tb{Log joint posterior density Fix $\ssq$ and $\phi$ (Original) }:
\begin{align*}
    \log[p(\bZ, \bbe, \bW , \ssq, \phi)]& = \log[p(\bZ|\bbe, \bW)]+\log[p(\bbe)]+\log[p(\bW|\ssq,\phi)]+\log[p(\ssq)]+\log[p(\phi)]\\
    &= \bZ^{\prime}\bX\bbe+\bZ^{\prime}\bW -\bOne^\prime \log (1+\exp\{\bX\bbe+\bW\})\\
    & \qquad -\frac{1}{2}\Big(\bbe'\bSig_{\beta}^{-1}\bbe+n\log\ssq+\log|\bR_{\phi}|+(n+p)\log(2\pi)+\log|\bSig_{\beta}| +\frac{1}{\ssq}\bW'\bR_{\phi}^{-1}\bW\Big)\\
    & \qquad +\asi\log\bsi-\log\Gamma(\asi) -(\asi+1) \log\ssq -\frac{\bsi}{\ssq}
\end{align*}

\noindent \tb{Quadratic Approximation (Jaakola and Jordan, 1997):}
$$
-\log(1+e^{x})=\argmax_{\xi}\Big\{ \lambda(\xi)x^{2}-\frac{1}{2}x +\psi(\xi)\Big\}
$$
where $\lambda(\xi)=-\tanh(\xi/2)/(4\xi)$ and $\psi(\xi)=\xi/2-\log(1+e^{\xi})+\xi\tanh(\xi/2)/4$
\\
\\
\noindent \tb{Computing the optimal Auxiliary Variables} (Jaakola and Jordan, 1997)

$$
\pmb{\xi}= \sqrt{\mbox{Diagonal}\Big\{\tilde{X}(\tbC_{\bv}+\tbmu_{\bv}\tbmu_{\bv}')\tilde{X}'\Big\}}
$$
\\
\noindent \tb{Log joint posterior density Fix $\ssq$ and $\phi$ (Modified)}:\\
Let $\bv=(\bbe, \bW)^\prime$, $\tilde{X}=[\bX\quad \bI]$, and $\bSig_{\bv}=\begin{bmatrix}
\bSig_{\beta}  & \bzero \\
\bzero & \ssq\bR_{\phi}
\end{bmatrix}$, then 
\begin{align*}
    \log[p(\bZ, \bbe, \bW , \ssq, \phi)]& = \bZ^{\prime}\tilde{X}\bv + \bOne'\Big[\bv'\tilde{X}'\bD\tilde{X}\bv-\frac{1}{2}\tilde{X}\bv+\psi(\xi)\Big]\\
    & \qquad -\frac{1}{2}\Big(\bbe'\bSig_{\beta}^{-1}\bbe+n\log\ssq+\log|\bR_{\phi}|+(n+p)\log(2\pi)+\log|\bSig_{\beta}| +\frac{1}{\ssq}\bW'\bR_{\phi}^{-1}\bW\Big)\\
    & \qquad +\asi\log\bsi-\log\Gamma(\asi)-(\asi+1) \log\ssq -\frac{\bsi}{\ssq}\\
    &= \bv'\tilde{X}'\bD\tilde{X}\bv - \frac{1}{2}\bOne'\tilde{X}\bv+\bOne'\psi(\xi)+\bZ'\tilde{X}\bv-\frac{1}{2}\bv'\bSig_{\bv}^{-1}\bv \\
    & \qquad -\frac{1}{2}\Big(\log{|\bSig_{\bv}|}+(n+p)\log(2\pi) \Big) \\
    & \qquad +\asi\log\bsi-\log\Gamma(\asi)-(\alpha_\sigma+1)\log\ssq-\frac{\bsi}{\ssq}\\
    & = -\frac{1}{2}\Big(\bv'(-2\tilde{X}'\bD\tilde{X}+\bSig_{\bv}^{-1})\bv-2(\bZ'-\frac{1}{2}\bOne')\tilde{X}\bv\Big)+\bOne'\psi(\xi)\\
    & \qquad -\frac{1}{2}\Big(\log{|\bSig_{\bv}|}+(n+p)\log(2\pi) \Big) \\
    & \qquad +\asi\log\bsi-\log\Gamma(\asi)-(\alpha_\sigma+1)\log\ssq-\frac{\bsi}{\ssq}
\end{align*}
where $\bf{D}=\mbox{diag}(\lambda(\pmb{\xi}))$.

\subsection{Variational Function for \texorpdfstring{$\bbe$ and $\bW$}{Lg}}
We represent $\bv=(\bbe,\bW)'$ to preserve dependece between $\bbe$ and $\bW$. The distribution that minimizes the KL divergence is $q(\bv)\propto \exp\{E_{-\bv}[ \log p(\bZ, \bbe, \bW , \ssq, \phi)]\}$

$$q(\bv)\sim \mcN(\tbmu_{\bv}, \tbC_{\bv})$$
where $\tbC_{\bv}= (-2\tilde{\bX}'\bf{D}\tilde{\bX}+\bSig_{\bv}^{-1})^{-1}$ and $\tbmu_{\bv}=\tbC_{\bv}\tilde{X}'(\bZ-\frac{1}{2}\bOne')$ and 

$$
\bSig_{\bv}^{-1}=\begin{bmatrix}
\bSig_{\beta}^{-1}  & \bzero \\
\bzero & \frac{1}{ \ssq}\bR_{\phi}^{-1}
\end{bmatrix}
$$

Note that the covariance matrix and mean vector are split as follows:
$$
\tbC_{\bv}=\begin{bmatrix}
\tbC_\beta  & \tbC_{\beta,W} \\
\tbC_{W,\beta} & \tbC_{W}
\end{bmatrix}, \qquad \tbmu_{\bv} = (\tbmu_\beta , \tbmu_W)'
$$

\subsection{Evidence Lower Bound}
\begin{align*}
    ELBO&=E_q\Bigg[\log \frac{p(\bZ, \bbe, \bW ,\ssq, \phi)}{q(\bbe,\bW,\ssq, \phi)}\Bigg]\\
    & = \tblue{E_q[\log p(\bZ, \bbe, \bW,\ssq, \phi)]} - \tred{E_q[logq(\bbe,\bW,\ssq, \phi)]}
\end{align*}

Here, we have to decompose by partitioning the parameter space $\btheta=(\pmb{\theta}_c,{\pmb{\theta}_d})'$. We estimate $\pmb{\theta}_c$ but fix $ \pmb{\theta}_d$
$$\pmb{\theta}_c={ \{\bv \} }, \qquad \pmb{\theta}_d={ \{\phi,\ssq \} } $$

\tblue{Part 1:} $\tblue{E_q[\log p(\bZ, \bbe, \bW,\ssq, \phi)]}$
\begin{align*}
    E_q[\log p(\bZ, \bbe, \bW,\ssq, \phi)]& = E_q\Big[-\frac{1}{2}\Big(\bv'(-2\tilde{X}'\bD\tilde{X}+\bSig_{\bv}^{-1})\bv-2(\bZ'-\frac{1}{2}\bOne')\tilde{X}\bv\Big)+\bOne'\psi(\xi) \\
    & \qquad -\frac{1}{2}\Big(\log{|\bSig_{\bv}|}+(n+p)\log(2\pi)\Big)  \\
    & \qquad +\asi\log\bsi-\log\Gamma(\asi)-(\alpha_\sigma+1)\log\ssq-\frac{\bsi}{\ssq}\Big] \\
    & =
    E_{q}[\bv'\tilde{X}'\bD\tilde{X}\bv]-\frac{1}{2}E_{q}[\bv'\bSig_{\bv}^{-1}\bv]+E_{q}[(\bZ'-\frac{1}{2}\bOne')\tilde{X}\bv]+\bOne'\psi(\xi) \\
    & \qquad -\frac{1}{2}\Big(\log{|\bSig_{\bv}|}+(n+p)\log(2\pi)\Big)\\
    & \qquad +\asi\log\bsi-\log\Gamma(\asi)-(\alpha_\sigma+1)\log\ssq-\frac{\bsi}{\ssq} \\
    & =
    \tbmu_{\bv}'\tilde{X}'\bD\tilde{X}\tbmu_{\bv}+tr[\tilde{X}'\bD\tilde{X} \tbC_{\bv}]-\frac{1}{2}\tbmu_{\bv}'\bSig_{\bv}^{-1}\tbmu_{\bv}-\frac{1}{2} tr[\bSig_{\bv}^{-1}\tbC_{\bv}] \\
    & \qquad +(\bZ'-\frac{1}{2}\bOne')\tilde{X}\tbmu_{\bv} +\bOne'\psi(\xi) \\
    & \qquad -\frac{1}{2}\Big(\log{|\bSig_{\bv}|}+(n+p)\log(2\pi)\Big) \\
    & \qquad +\asi\log\bsi-\log\Gamma(\asi)-(\alpha_\sigma+1)\log\ssq-\frac{\bsi}{\ssq}
\end{align*}

\tb{Key Components:}
\begin{enumerate}
    \item Expectation of a log Gamma RV: $E[\text{log} \frac{1}{\ssq}] \approx \psi(\tilde{\alpha})-\log\tilde{\beta} $
    \item Expectation of Quadratic Forms: $E_{q(\bv)}\Big[\bv'\bSig_{\bv}^{-1}\bv\Big]=\tbmu_\bv'\bSig_{\bv}^{-1}\tbmu_\bv+tr[\bSig_{\bv}^{-1}\tbC_\bv]$
\end{enumerate}

\tred{Part 2:} $\tred{E_q[\log q(\bbe, \bW,\ssq, \phi)]}$
\begin{align*}
    E_q[\log q(\bbe, \bW,\ssq, \phi)]& \propto E_q[\log[q(v)]] \\
    & = E_q[\log[2\pi^{\frac{-(n+p)}{2}} |\tbC_\bv|^{\frac{-1}{2}} \exp(-\frac{1}{2}(\bv-\tbmu_{\bv})'\tbC_\bv^{-1}(\bv-\tbmu_{\bv}) ) \\  
    & = 
    E_{q}[ -\frac{(n+p)}{2}\log(2\pi) -\frac{1}{2} \log|\tbC_\bv|-\frac{1}{2}(\bv-\tbmu_{\bv})'\tbC_\bv^{-1}(\bv-\tbmu_{\bv})] \\
    & =
    -\frac{(n+p)}{2}\log(2\pi)-\frac{1}{2} \log|\tbC_\bv| -\frac{1}{2} tr[\tbC_\bv^{-1}\tbC_\bv] \\
    & =
    -\frac{(n+p)}{2}\log(2\pi)-\frac{1}{2} \log|\tbC_\bv| -\frac{1}{2} (n+p)\\
\end{align*}

\noindent \tb{Evidence Lower Bound:}
\begin{align*}
    ELBO&=E_q\Bigg[\log \frac{p(\bZ, \bbe, \bW ,\ssq, \phi)}{q(\bbe,\bW,\ssq, \phi)}\Bigg]\\
    & = \tblue{E_q[\log p(\bZ, \bbe, \bW,\ssq, \phi)]} - \tred{E_q[log(q(\bbe,\bW,\ssq, \phi))]} \\
    &= \tblue{\tbmu_{\bv}'\tilde{X}'\bD\tilde{X}\tbmu_{\bv}+tr[\tilde{X}'\bD\tilde{X} \tbC_{\bv}]-\frac{1}{2}\tbmu_{\bv}'\bSig_{\bv}^{-1}\tbmu_{\bv}-\frac{1}{2} tr[\bSig_{\bv}^{-1}\tbC_{\bv}]} \\
    & \qquad \tblue{+(\bZ'-\frac{1}{2}\bOne')\tilde{X}\tbmu_{\bv} +\bOne'\psi(\xi)} \\
    & \qquad \tblue{-\frac{1}{2}\Big(\log{|\bSig_{\bv}|}+(n+p)\log(2\pi)\Big)} \\
    & \tblue{\qquad +\asi\log\bsi-\log\Gamma(\asi)-(\alpha_\sigma+1)\log\ssq-\frac{\bsi}{\ssq}} \\
    & \qquad \tred{-\Bigg[-\frac{(n+p)}{2}\log(2\pi)-\frac{1}{2} \log|\tbC_\bv| -\frac{1}{2} (n+p) \Bigg] } \\
    &=
    \tbmu_{\bv}'\tilde{X}'\bD\tilde{X}\tbmu_{\bv}+tr[\tilde{X}'\bD\tilde{X} \tbC_{\bv}]-\frac{1}{2}\tbmu_{\bv}'\bSig_{\bv}^{-1}\tbmu_{\bv}-\frac{1}{2} tr[\bSig_{\bv}^{-1}\tbC_{\bv}]\\
    & \qquad +(\bZ'-\frac{1}{2}\bOne')\tilde{X}\tbmu_{\bv}+\bOne'\psi(\xi)-\frac{1}{2}\log{|\bSig_{\bv}|}\\
    & \qquad +\frac{1}{2} \Big(\log|\tbC_\bv| +(n+p)\Big)\\
    & \qquad +\asi\log\bsi-\log\Gamma(\asi)-(\alpha_\sigma+1)\log\ssq-\frac{\bsi}{\ssq}
\end{align*}

\newpage
\section{Basis-SGLMM (Poisson Data Model): MFVB}
\tb{Hierarchical Model:}
% \vspace{-0.3in}
\begin{align*}\vspace{-0.1in}
         \textbf{\mbox{Data Model:}} &\quad  Z_i|\lambda_i \sim \mbox{Pois}(\lambda_i)\\
    & \qquad \lambda_i=\exp\{\bX_{i}^\prime\bbe +\bgPhi_{i}^\prime\bdel\}\\
    \tb{Process Model:}& \qquad \bdel|\ssq \sim \mcN(\bzero,\ssq\bI)\\
    \tb{Prior Model:}& \qquad \bbe \sim \mcN(\bzero,\bSig_{\beta}), \qquad \ssq \sim IG(\alpha_\sigma, \beta_\sigma)
\end{align*}

\tb{Probability Density Functions}
% \vspace{-0.3in}
\begin{align*}
    \mbox{Joint :}& \qquad p(\bZ, \bbe, \bdel,\ssq)= p(\bZ|\bbe, \bdel)p(\bbe)p(\bdel|\ssq)p(\ssq)\\
    \mbox{Likelihood :}& \qquad p(\bZ|\bga)= \prod_{i=1}^{n}\frac{\lambda_i^{Z_i}e^{-\lambda_i}}{Z_i!}, \qquad \mbox{where } \lambda_i=\exp\{\bX_{i}^\prime\bbe +\bgPhi_{i}^\prime\bdel\}\\
    \mbox{Process :}&\qquad p(\bdel|\ssq) = (2\pi)^{-m/2}(\ssq)^{-m/2}\exp\{-\frac{1}{2\ssq}\bdel'\bdel\}\\
    \mbox{Prior :}& \qquad p(\bbe) = (2\pi)^{-p/2}|\bSig_{\beta}|^{-1/2}\exp\{-\frac{1}{2}\bbe'\bSig_{\beta}^{-1}\bbe\}\\
    & \qquad  p(\ssq)= \frac{\bsi^{\asi}}{\Gamma(\asi)}(\ssq)^{-\asi-1}\exp\{-\frac{\bsi}{\ssq}\}
\end{align*}

\noindent \tb{Log joint posterior density (Original)}:
\begin{align*}
    \log[p(\bZ, \bbe, \bdel , \ssq)]& = \log[p(\bZ|\bbe, \bdel)]+\log[p(\bbe)]+\log[p(\bdel|\ssq)]+\log[p(\ssq)]\\
    & = \bZ'(\bX\bbe+\bgPhi\bdel) -\bOne' \Big(e^{\bX\bbe+\bgPhi\bdel}\Big)- \bOne' \log\bZ!\\
    & \qquad -\frac{1}{2}\Bigg((m+p)\log(2\pi)+m\log\ssq+\log|\bSig_{\beta}|+\frac{1}{\ssq}\bdel'\bdel+\bbe'\bSig_{\beta}^{-1}\bbe
    \Bigg)\\
    & \qquad +\asi\log\bsi-\log\Gamma(\asi) -(\asi+1) \log\ssq -\frac{\bsi}{\ssq}
\end{align*}

\newpage
\noindent \tb{Log joint posterior density (Modified)}:\\
Let $\bv=(\bbe, \bdel)^\prime$, $\tilde{\bX}=[\bX\quad \bgPhi]$, and $\bSig_{\bv}=\begin{bmatrix}
\bSig_{\beta}  & \bzero \\
\bzero & \ssq\bI
\end{bmatrix}$, then 
\begin{align*}
    \log[p(\bZ, \bbe, \bdel , \ssq,)]& =\bZ'\tilde{\bX}\bv -\bOne' \Big(e^{\tilde{\bX}\bv}\Big)-\bOne' \log\bZ!\\
    & \qquad -\frac{1}{2}\Big((m+p)\log(2\pi)+m\log \ssq +\log |\bSig_{\beta}|+\bv'\bSig_{v}^{-1}\bv\Big)\\
    & \qquad +\asi\log\bsi-\log\Gamma(\asi)-(\asi+1) \log\ssq -\frac{\bsi}{\ssq}
\end{align*}

\subsection{Variational Function for \texorpdfstring{$\bbe$ and $\bW$}{Lg} }
\noindent \tb{Variational Distribution for $\bv=(\bbe,\bdel)'$:} The distribution that minimizes the KL divergence is $q(\bv)\propto \exp\{E_{-\bv}[ \log p(\bZ, \bv , \ssq)]\}$. For the Poisson case, this distribution is not available in closed form; hence, we provide a Gaussian approximation via Laplace approximation (2nd order Taylor Expansion). The objective function is as follows:

\begin{align*}
    f(\bv)&= E_{q(-\bv)}[\log p(\bZ, \bv, \ssq)] = E_{q(\ssq)}[\log p(\bZ, \bv, \ssq)] \\
    &= E_{q(\ssq)}\Big[\bZ'\tilde{\bX}\bv -\bOne' \Big(e^{\tilde{\bX}\bv}\Big)-\bOne' \log\bZ!\\
& \qquad \qquad -\frac{1}{2}\Big((m+p)\log(2\pi)+m\log \ssq +\log |\bSig_{\beta}|+\bv'\bSig_{v}^{-1}\bv\Big)\\
    & \qquad \qquad +\asi\log\bsi-\log\Gamma(\asi)-(\asi+1) \log\ssq -\frac{\bsi}{\ssq} \Big] \\    
     &= \bZ'\tilde{\bX}\bv -\bOne' \Big(e^{\tilde{\bX}\bv}\Big)-\bOne' \log\bZ! -\frac{(m+p)}{2}\log(2\pi)-\frac{1}{2}\log|\bSig_{\beta}|-\frac{1}{2}\bv'E[\bSig_{v}^{-1}]\bv\\
    & \qquad \qquad +\asi\log\bsi-\log\Gamma(\asi)+(\asi+\frac{m}{2}+1) E[\log\frac{1}{\ssq}] -\bsi E[\frac{1}{\ssq}]   \\
         &= \bZ'\tilde{\bX}\bv -\bOne' \Big(e^{\tilde{\bX}\bv}\Big)-\bOne' \log\bZ! -\frac{(m+p)}{2}\log(2\pi)\\
         & \qquad \qquad +\frac{1}{2}\log |\bSig_\beta^{-1}|-\frac{1}{2}\bv'E[\bSig_{v}^{-1}]\bv \\
    & \qquad \qquad +\asi\log\bsi-\log\Gamma(\asi)+(\asi+\frac{m}{2}+1) (\psi(\talpha)-\log\tibeta) -\bsi \frac{\talpha}{\tibeta} 
\end{align*}

\tb{Key Components:}
\begin{enumerate}
    \item Expectation of a lognormal RV: $E_{q(\bv)}[e^{\tbX\bv}]= \exp\{\tbX\tbmu_{v}+\frac{1}{2}\mbox{diag}(\tbX\tbC_{v}\tbX')\}$ 
    \item Expectation of Quadratic Forms: $E_{q(\bv)}\Big[\bv'\bSig_{v}^{-1}\bv \Big]=\tbmu_v'\bSig_{v}^{-1}\tbmu_v+tr[\bSig_{v}^{-1}\tbC_v]$
    \item $E[\log\frac{1}{\ssq}] = \psi(\talpha)-\log\tibeta$ 
    \item $E[\frac{1}{\ssq}] = \frac{\talpha}{\tibeta}$
    \item $E[\bSig_{\bv}^{-1}]=\begin{bmatrix}
\bSig_{\beta}^{-1}  & \bzero \\
\bzero & \frac{\tilde{\alpha}_\sigma}{ \tilde{\beta}_\sigma}\bI
\end{bmatrix}$
\item 
$-\frac{1}{2}E[\log{|\bSig_{\bv}|}]=\frac{1}{2}E[\log |\bSig_\beta^{-1}|]+ \frac{m}{2}E[\log \frac{1}{\ssq}]= \frac{1}{2}\log |\bSig_\beta^{-1}|+ \frac{m}{2}(\psi(\tilde{\alpha})-\log\tilde{\beta})$
\end{enumerate}

\tb{Variational Function for $\bv$:} For objective function and corresponding gradient
$$
f(\bv)\propto \bZ'\tilde{\bX}\bv -\bOne' \Big(e^{\tilde{\bX}\bv}\Big)-\frac{1}{2}\bv'E[\bSig_{v}^{-1}]\bv
$$
$$\nabla f(\bv)= \tilde{\bX}'\bZ- \tilde{\bX}'Diag(e^{\tilde{\bX}\bv})\bOne'  - E[\bSig_{v}^{-1}]\bv$$
we have the resulting variational function 
$$q_v(\bv)= \mcN(\tbmu_v, \tbC_{v})$$ where $\tbmu_\bv=\argmax_{\bv}f(\bv)$ and $\tbC_\bv= -(\bH)^{-1}$ where $\bH=\frac{\partial^2 f}{\partial \bv^2}\Bigr|_{\substack{\bv=\tbmu_\bv}}$.\\ 
Note that the mean vector and covariance matrix are partitioned as follows:

$$
\tbmu_{\bv} = (\tbmu_\beta , \tbmu_\delta)', \qquad \tbC_{\bv}=\begin{bmatrix}
\tbC_\beta  & \tbC_{\beta,\delta} \\
\tbC_{\delta,\beta} & \tbC_{\delta}
\end{bmatrix}, 
$$

\subsection{Variational Distribution for \texorpdfstring{$\sigma^2$}{Lg}}
The distribution that minimizes the KL divergence is $q(\ssq)\propto \exp\{E_{-\ssq}[ \log p(\bZ, \bv , \ssq)]\}$
\begin{align*}
    \log [p(\bZ, \bv , \ssq)]& \propto -(\alpha_\sigma +\frac{m}{2}+1)\log(\ssq)- \frac{1}{\ssq}[\beta_\sigma+\frac{1}{2}E[\bdel^{\prime}\bdel]]\\
    & \qquad \propto -(\alpha_\sigma +\frac{m}{2}+1)\log(\ssq)- \frac{1}{\ssq}[\beta_\sigma+\frac{1}{2}(tr(\tbC_{\delta})+\tbmu_\delta'\tbmu_\delta)]
\end{align*}
It follows that $q(\ssq)= IG(\tilde{\alpha},\tilde{\beta})$ where $\tilde{\alpha}= \alpha_\sigma+m/2$ and $\tilde{\beta}= \beta_\sigma+\frac{1}{2}(tr(\tbC_{\delta})+\tbmu_\delta'\tbmu_\delta)$.

\subsection{Evidence Lower Bound}
\begin{align*}
    ELBO&=E_q\Bigg[\log \frac{p(\bZ, \bv ,\ssq)}{q(\bv,\ssq)}\Bigg]\\
    & = \tblue{E_q[\log p(\bZ, \bv,\ssq)]} - \tred{E_q[log(q(\bv,\ssq))]}
\end{align*}

\tblue{Part 1:} $\tblue{E_q[\log p(\bZ, \bv,\ssq)]}$
\begin{align*}
    E_q[\log p(\bZ, \bv,\ssq)]& = E_q\Big[\bZ'\tilde{\bX}\bv -\bOne' \Big(e^{\tilde{\bX}\bv}\Big)-\bOne' \log\bZ! -\frac{(m+p)}{2}\log(2\pi)\\
         & \qquad \qquad +\frac{1}{2}\log |\bSig_\beta^{-1}|-\frac{1}{2}\bv'\bSig_{v}^{-1}\bv \\
    & \qquad \qquad +\asi\log\bsi-\log\Gamma(\asi)+(\asi+\frac{m}{2}+1) (\psi(\talpha)-\log\tibeta) -\bsi \frac{\talpha}{\tibeta}\Big] \\
& = \bZ'\tilde{\bX}\tbmu_v -\bOne' E[e^{\tilde{\bX}\bv}]-\bOne' \log\bZ! -\frac{(m+p)}{2}\log(2\pi)\\
         & \qquad \qquad +\frac{1}{2}\log |\bSig_\beta^{-1}|-\frac{1}{2}E[\bv'E[\bSig_{v}^{-1}]\bv] \\
    & \qquad \qquad +\asi\log\bsi-\log\Gamma(\asi)+(\asi+\frac{m}{2}+1) (\psi(\talpha)-\log\tibeta) -\bsi \frac{\talpha}{\tibeta} \\    
    & = \bZ'\tilde{\bX}\tbmu_v -\bOne'\exp\{\tbX\tbmu_{v}+\frac{1}{2}\mbox{diag}(\tbX\tbC_{v}\tbX')\}-\bOne' \log\bZ! -\frac{(m+p)}{2}\log(2\pi)\\
         & \qquad \qquad +\frac{1}{2}\log |\bSig_\beta^{-1}|-\frac{1}{2}\Big[\tbmu_v'E[\bSig_{v}^{-1}]\tbmu_v+tr[E[\bSig_{v}^{-1}]\tbC_v]\Big] \\
         & \qquad \qquad +\asi\log\bsi-\log\Gamma(\asi)+(\asi+\frac{m}{2}+1) (\psi(\talpha)-\log\tibeta) -\bsi \frac{\talpha}{\tibeta} \\    
    & \propto \bZ'\tilde{\bX}\tbmu_v -\bOne'\exp\{\tbX\tbmu_{v}+\frac{1}{2}\mbox{diag}(\tbX\tbC_{v}\tbX')\} \\
         & \qquad \qquad -\frac{1}{2}\Big[\tbmu_v'E[\bSig_{v}^{-1}]\tbmu_v+tr[E[\bSig_{v}^{-1}]\tbC_v]\Big] \\
         & \qquad \qquad +(\asi+\frac{m}{2}+1) (\psi(\talpha)-\log\tibeta) -\bsi \frac{\talpha}{\tibeta} \\  
\end{align*}

\tb{Key Components:}
\begin{enumerate}
    \item Expectation of a log Gamma RV: $E[\text{log} \frac{1}{\ssq}] \approx \psi(\tilde{\alpha})-\log\tilde{\beta} $
    \item Expectation of Quadratic Forms: $E_{q(\bv)}\Big[\bv'\bSig_{\bv}^{-1}\bv\Big]=\tbmu_\bv'\bSig_{\bv}^{-1}\tbmu_\bv+tr[\bSig_{\bv}^{-1}\tbC_\bv]$
\end{enumerate}

\tred{Part 2:} $\tred{E_q[\log q(\bv,\ssq)]}$
\begin{align*}
    E_q[\log q(\bv,\ssq)]& \propto E_q\Big[\log[q(v)]+\log[q(\ssq)]\Big] \\
    & = E_q\Big[\log[2\pi^{\frac{-(m+p)}{2}} |\tbC_\bv|^{\frac{-1}{2}} \exp(-\frac{1}{2}(\bv-\tbmu_{\bv})'\tbC_\bv^{-1}(\bv-\tbmu_{\bv}) ) \\  
    & \qquad +\log [\frac{\tilde{\beta}^{\tilde{\alpha}}}{\Gamma(\tilde{\alpha})} (\ssq)^{-\tilde{\alpha}-1} \cdot \exp(\frac{-\tilde{\beta}}{\ssq}) ]\Big] \\
    & = 
    E_{q}\Big[-\frac{(m+p)}{2}\log2\pi-\frac{1}{2} \log|\tbC_\bv|-\frac{1}{2}(\bv-\tbmu_{\bv})'\tbC_\bv^{-1}(\bv-\tbmu_{\bv})+\tilde{\alpha}\log\tilde{\beta} \\
    & \qquad -\log\Gamma(\tilde{\alpha}) + (\tilde{\alpha}+1)\log\frac{1}{\ssq}-\tilde{\beta}\frac{1}{\ssq} \Big]\\ 
    & =
    -\frac{(m+p)}{2}\log2\pi-\frac{1}{2} \log|\tbC_\bv| -\frac{1}{2} tr[\tbC_\bv^{-1}\tbC_\bv]+\tilde{\alpha}\log\tilde{\beta}-\log\Gamma(\tilde{\alpha}) \\
    &\qquad +(\tilde{\alpha}+1)E_q[\log \frac{1}{\ssq}]-\tilde{\beta}E_q[\frac{1}{\ssq}] \\
    & =
    -\frac{(m+p)}{2}\log2\pi-\frac{1}{2} \log|\tbC_\bv| -\frac{m+p}{2} +\tilde{\alpha}\log\tilde{\beta} -\log\Gamma(\tilde{\alpha})\\
    &\qquad +(\tilde{\alpha}+1)(\psi(\tilde{\alpha})-\log\tilde{\beta}) -\tilde{\beta}\frac{\tilde{\alpha}}{\tilde{\beta}}\\
        & =
    -\frac{(m+p)}{2}(\log2\pi+1)-\frac{1}{2} \log|\tbC_\bv|  +\tilde{\alpha}\log\tilde{\beta} -\log\Gamma(\tilde{\alpha})\\
    &\qquad +(\tilde{\alpha}+1)(\psi(\tilde{\alpha})-\log\tilde{\beta}) -\tilde{\alpha}
\end{align*}

\noindent \tb{Evidence Lower Bound:}
\begin{align*}
    ELBO&=E_q\Bigg[\log \frac{p(\bZ, \bbe, \bW ,\ssq, \phi)}{q(\bbe,\bW,\ssq, \phi)}\Bigg]\\
    & = \tblue{E_q[\log p(\bZ, \bbe, \bW,\ssq, \phi)]} - \tred{E_q[log(q(\bbe,\bW,\ssq, \phi))]} \\
       & = \tblue{\bZ'\tilde{\bX}\tbmu_v -\bOne'\exp\{\tbX\tbmu_{v}+\frac{1}{2}\mbox{diag}(\tbX\tbC_{v}\tbX')\}-\bOne' \log\bZ! -\frac{(m+p)}{2}\log(2\pi)}\\
    & \qquad \tblue{+\frac{1}{2}\log |\bSig_\beta^{-1}|-\frac{1}{2}\Big[\tbmu_v'E[\bSig_{v}^{-1}]\tbmu_v+tr[E[\bSig_{v}^{-1}]\tbC_v]\Big]} \\
    & \qquad \tblue{+\asi\log\bsi-\log\Gamma(\asi)+(\asi+\frac{m}{2}+1) (\psi(\talpha)-\log\tibeta) -\bsi \frac{\talpha}{\tibeta}} \\
    & \qquad
    \tred{-\Bigg[-\frac{(m+p)}{2}(\log2\pi+1)-\frac{1}{2} \log|\tbC_\bv|+\tilde{\alpha}\log\tilde{\beta} -\log\Gamma(\tilde{\alpha})}\\
    &\qquad \tred{+(\tilde{\alpha}+1)(\psi(\tilde{\alpha})-\log\tilde{\beta}) -\tilde{\alpha} \Bigg]} \\ 
    & \propto \tblue{\bZ'\tilde{\bX}\tbmu_v -\bOne'\exp\{\tbX\tbmu_{v}+\frac{1}{2}\mbox{diag}(\tbX\tbC_{v}\tbX')\}}\\
    & \qquad \tblue{-\frac{1}{2}\Big[\tbmu_v'E[\bSig_{v}^{-1}]\tbmu_v+tr[E[\bSig_{v}^{-1}]\tbC_v]\Big]} \\
    & \qquad \tblue{+(\asi+\frac{m}{2}+1) (\psi(\talpha)-\log\tibeta) -\bsi \frac{\talpha}{\tibeta}} \\
    & \qquad
    \tred{+\frac{1}{2} \log|\tbC_\bv|-\tilde{\alpha}\log\tilde{\beta} +\log\Gamma(\tilde{\alpha})}\\
    &\qquad \tred{-(\tilde{\alpha}+1)(\psi(\tilde{\alpha})+\log\tilde{\beta}) +\tilde{\alpha}} \\ 
    & \propto \bZ'\tilde{\bX}\tbmu_v -\bOne'\exp\{\tbX\tbmu_{v}+\frac{1}{2}\mbox{diag}(\tbX\tbC_{v}\tbX')\}\\
    & \qquad
    -\frac{1}{2}\tbmu_v'E[\bSig_{v}^{-1}]\tbmu_v-\frac{1}{2}tr[E[\bSig_{v}^{-1}]\tbC_v]\\
    & \qquad -\bsi \frac{\talpha}{\tibeta}+\frac{1}{2} \log|\tbC_\bv|-\tilde{\alpha}\log\tilde{\beta}
    \end{align*}

since $-\frac{1}{2}E[\log{|\bSig_{\bv}|}]=\frac{1}{2}E[\log |\bSig_\beta^{-1}|]+ \frac{m}{2}E[\log \frac{1}{\ssq}]= \frac{1}{2}\log |\bSig_\beta^{-1}|+ \frac{m}{2}(\psi(\tilde{\alpha})-\log\tilde{\beta})$

\newpage
\section{Basis-SGLMM (Bernoulli Data Model): MFVB}
\tb{Hierarchical Model:}
% \vspace{-0.3in}
\begin{align*}\vspace{-0.1in}
    \tb{Data Model:} &\qquad  Z_i|p_i \sim \mbox{Bern}(p_i)\\
    & \qquad p_i = (1+\exp\{-\bX_{i}^\prime\bbe -\bgPhi_{i}^\prime\bdel\})^{-1}\\
    \tb{Process Model:}& \qquad \bdel|\ssq \sim \mcN(\bzero,\ssq\bI)\\
    \tb{Prior Model:}& \qquad \bbe \sim \mcN(\bzero,\bSig_{\beta}), \qquad \ssq \sim IG(\alpha_\sigma, \beta_\sigma)
\end{align*}

\tb{Probability Density Functions}
% \vspace{-0.3in}
\begin{align*}
    \mbox{Joint :}& \qquad p(\bZ, \bbe, \bdel,\ssq)= p(\bZ|\bbe, \bdel)p(\bbe)p(\bdel|\ssq)p(\ssq)\\
    \mbox{Likelihood :}& \qquad p(\bZ|\bbe, \bdel)= \prod_{i=1}^{n}p_{i}^{z_i}(1-p_{i})^{1-z_i}, \qquad \mbox{where } p_i=(1+\exp\{-\bX_{i}^\prime\bbe -\bgPhi_{i}^\prime\bdel\})^{-1}\\
    \mbox{Process :}&\qquad p(\bdel|\ssq) = (2\pi)^{-m/2}(\ssq)^{-m/2}\exp\{-\frac{1}{2\ssq}\bdel'\bdel\}\\
    \mbox{Prior :}& \qquad p(\bbe) = (2\pi)^{-p/2}|\bSig_{\beta}|^{-1/2}\exp\{-\frac{1}{2}\bbe'\bSig_{\beta}^{-1}\bbe\}\\
    & \qquad  p(\ssq)= \frac{\bsi^{\asi}}{\Gamma(\asi)}(\ssq)^{-\asi-1}\exp\{-\frac{\bsi}{\ssq}\}
\end{align*}

\noindent \tb{Log joint posterior density (Original)}:
\begin{align*}
    \log[p(\bZ, \bbe, \bdel , \ssq)]& = \log[p(\bZ|\bbe, \bdel)]+\log[p(\bbe)]+\log[p(\bdel|\ssq)]+\log[p(\ssq)]\\
    &= \bZ^{\prime}\bX\bbe+\bZ^{\prime}\bgPhi\bdel -\bOne^\prime \log (1+\exp\{\bX\bbe+\bgPhi\bdel\})\\
    & \qquad -\frac{1}{2}\Big(\bbe'\bSig_{\beta}^{-1}\bbe+m\log\ssq+(m+p)\log(2\pi)+\log|\bSig_{\beta}| +\frac{1}{\ssq}\bdel'\bdel\Big)\\
    & \qquad +\asi\log\bsi-\log\Gamma(\asi) -(\asi+1) \log\ssq -\frac{\bsi}{\ssq}
\end{align*}
\noindent \tb{Quadratic Approximation (Jaakola and Jordan, 1997):}
$$
-\log(1+e^{x})=\argmax_{\xi}\Big\{ \lambda(\xi)x^{2}-\frac{1}{2}x +\psi(\xi)\Big\}
$$
where $\lambda(\xi)=-\tanh(\xi/2)/(4\xi)$ and $\psi(\xi)=\xi/2-\log(1+e^{\xi})+\xi\tanh(\xi/2)/4$
\\
\\
\noindent \tb{Computing the optimal Auxiliary Variables} (Jaakola and Jordan, 1997)

$$
\pmb{\xi}= \sqrt{\mbox{Diagonal}\Big\{\tilde{X}(\tbC_{\bv}+\tbmu_{\bv}\tbmu_{\bv}')\tilde{X}'\Big\}}
$$

\noindent \tb{Log joint posterior density (Modified)}:\\
Let $\bv=(\bbe, \bdel)^\prime$, $\tilde{\bX}=[\bX\quad \bgPhi]$, and $\bSig_{\bv}=\begin{bmatrix}
\bSig_{\beta}  & \bzero \\
\bzero & \ssq\bI
\end{bmatrix}$, then 
\begin{align*}
    \log[p(\bZ, \bbe, \bdel , \ssq)]& = \bZ^{\prime}\tilde{\bX}\bv 
    + \bOne'\Big[\bv'\tilde{\bX}'\bD\tilde{\bX}\bv-\frac{1}{2}\tilde{\bX}\bv+\psi(\xi)\Big]\\
    & \qquad -\frac{1}{2}\Big(\bbe'\bSig_{\beta}^{-1}\bbe+m\log\ssq+(m+p)\log(2\pi)+\log|\bSig_{\beta}| +\frac{1}{\ssq}\bdel'\bdel\Big)\\
    & \qquad +\asi\log\bsi-\log\Gamma(\asi)-(\asi+1) \log\ssq -\frac{\bsi}{\ssq}\\
    &= \bv'\tilde{\bX}'\bD\tilde{\bX}\bv - \frac{1}{2}\bOne'\tilde{\bX}\bv+\bOne'\psi(\xi)+\bZ'\tilde{\bX}\bv-\frac{1}{2}\bv'\bSig_{\bv}^{-1}\bv \\
    & \qquad -\frac{1}{2}\Big(m\log\ssq+\log|\bSig_\beta|+(m+p)\log(2\pi) \Big) \\
    & \qquad +\asi\log\bsi-\log\Gamma(\asi)-(\alpha_\sigma+1)\log\ssq-\frac{\bsi}{\ssq}\\
    & = -\frac{1}{2}\Big(\bv'(-2\tilde{\bX}'\bD\tilde{\bX}+\bSig_{\bv}^{-1})\bv-2(\bZ'-\frac{1}{2}\bOne')\tilde{\bX}\bv\Big)+\bOne'\psi(\xi)\\
    & \qquad -\frac{1}{2}\Big(m\log\ssq+\log|\bSig_\beta|+(m+p)\log(2\pi)\Big) \\
    & \qquad +\asi\log\bsi-\log\Gamma(\asi)-(\alpha_\sigma+1)\log\ssq-\frac{\bsi}{\ssq}
\end{align*}
where $\bf{D}=\mbox{diag}(\lambda(\pmb{\xi}))$. 

\subsection{Variational Function for \texorpdfstring{$\bbe$ and $\bW$}{Lg} }
We represent $\bv=(\bbe,\bW)'$ to preserve dependece between $\bbe$ and $\bW$.  The distribution that minimizes the KL divergence is $q(\bv)\propto \exp\{E_{-\bv}[ \log p(\bZ, \bv , \ssq)]\}$, or:
$$q(\bv)\sim \mcN(\tbmu_{\bv}, \tbC_{\bv})$$
where $\tbC_{\bv}= (-2\tilde{\bX}'\bf{D}\tilde{\bX}+E[\bSig_{\bv}^{-1}])^{-1}$ and $\tbmu_{\bv}=\tbC_{\bv}\tilde{\bX}'(\bZ-\frac{1}{2}\bOne')$ and 

$$
E[\bSig_{\bv}^{-1}]=\begin{bmatrix}
\bSig_{\beta}^{-1}  & \bzero \\
\bzero & \frac{\tilde{\alpha}_\sigma}{ \tilde{\beta}_\sigma}\bI
\end{bmatrix}
$$

Note that the covariance matrix and mean vector are split as follows:
$$
\tbC_{\bv}=\begin{bmatrix}
\tbC_\beta  & \tbC_{\beta,\delta} \\
\tbC_{\delta,\beta} & \tbC_{\delta}
\end{bmatrix}, \qquad \tbmu_{\bv} = (\tbmu_\beta , \tbmu_\delta)'
$$

\subsection{Variational Distribution for \texorpdfstring{$\sigma^2$:}{Lg}} 
The distribution that minimizes the KL divergence is $q(\ssq)\propto \exp\{E_{-\ssq}[ \log p(\bZ, \bv , \ssq)]\}$
\begin{align*}
    \log [p(\bZ, \bv , \ssq)]& \propto -(\alpha_\sigma +\frac{m}{2}+1)\log(\ssq)- \frac{1}{\ssq}[\beta_\sigma+\frac{1}{2}E[\bdel^{\prime}\bdel]]\\
    & \qquad \propto -(\alpha_\sigma +\frac{m}{2}+1)\log(\ssq)- \frac{1}{\ssq}[\beta_\sigma+\frac{1}{2}(tr(\tbC_{\delta})+\tbmu_\delta'\tbmu_\delta)]
\end{align*}
It follows that $q(\ssq)= IG(\tilde{\alpha},\tilde{\beta})$ where $\tilde{\alpha}= \alpha_\sigma+m/2$ and $\tilde{\beta}= \beta_\sigma+\frac{1}{2}(tr(\tbC_{\delta})+\tbmu_\delta'\tbmu_\delta)$.

\subsection{Evidence Lower Bound}
\begin{align*}
    ELBO&=E_q\Bigg[\log \frac{p(\bZ, \bv ,\ssq)}{q(\bv,\ssq)}\Bigg]\\
    & = \tblue{E_q[\log p(\bZ, \bv,\ssq)]} - \tred{E_q[log(q(\bv,\ssq))]}
\end{align*}

\tblue{Part 1:} $\tblue{E_q[\log p(\bZ, \bv,\ssq)]}$
\begin{align*}
    E_q[\log p(\bZ, \bv,\ssq)]& = E_q\Big[-\frac{1}{2}\Big(\bv'(-2\tilde{\bX}'\bD\tilde{\bX}+\bSig_{\bv}^{-1})\bv-2(\bZ'-\frac{1}{2}\bOne')\tilde{\bX}\bv\Big)+\bOne'\psi(\xi)\\
    & \qquad -\frac{1}{2}\Big(\log{|\bSig_{\bv}|}+(m+p)\log(2\pi) \Big) \\
    & \qquad +\asi\log\bsi-\log\Gamma(\asi)-(\alpha_\sigma+1)\log\ssq-\frac{\bsi}{\ssq}\Big] \\
    & =
    E_{q}[\bv'\tilde{\bX}'\bD\tilde{\bX}\bv]-\frac{1}{2}E_{q}[\bv'\bSig_{\bv}^{-1}\bv]+E_{q}[(\bZ'-\frac{1}{2}\bOne')\tilde{\bX}\bv]+\bOne'\psi(\xi) \\
    & \qquad +\frac{m}{2}E[\log\frac{1}{\ssq}]-\frac{1}{2}\Big(\log|\bSig_\beta|+(m+p)\log(2\pi) \Big)] \\
    & \qquad +\asi\log\bsi-\log\Gamma(\asi)+(\alpha_\sigma+1)
    E_{q}[\log\frac{1}{\ssq}]-\bsi E_{q}[\frac{1}{\ssq}] \\
    & =\tbmu_{\bv}'\tilde{\bX}'\bD\tilde{\bX}\tbmu_{\bv}+tr[\tilde{\bX}'\bD\tilde{\bX} \tbC_{\bv}]-\frac{1}{2}\tbmu_{\bv}'\bSig_{\bv}^{-1}\tbmu_{\bv}-\frac{1}{2} tr[\bSig_{\bv}^{-1}\tbC_{\bv}]+(\bZ'-\frac{1}{2}\bOne')\tilde{\bX}\tbmu_{\bv} \\
    & \qquad +\bOne'\psi(\xi) -\frac{1}{2}\Big(\log|\bSig_\beta|+(m+p)\log(2\pi) \Big) \\
    & \qquad +\asi\log\bsi-\log\Gamma(\asi)+(\alpha_\sigma+\frac{m}{2}+1)
    E_{q}[\log\frac{1}{\ssq}]-\bsi E_{q}[\frac{1}{\ssq}]\\
    & = \tbmu_{\bv}'\tilde{\bX}'\bD\tilde{\bX}\tbmu_{\bv}+tr[\tilde{\bX}'\bD\tilde{\bX} \tbC_{\bv}]-\frac{1}{2}\tbmu_{\bv}'\bSig_{\bv}^{-1}\tbmu_{\bv}-\frac{1}{2} tr[\bSig_{\bv}^{-1}\tbC_{\bv}] +(\bZ'-\frac{1}{2}\bOne')\tilde{\bX}\tbmu_{\bv}\\
    & \qquad +\bOne'\psi(\xi) -\frac{1}{2}\Big(\log|\bSig_\beta|+(m+p)\log(2\pi) \Big) \\
    & \qquad +\asi\log\bsi-\log\Gamma(\asi)+(\alpha_\sigma+\frac{m}{2}+1)
    (\psi(\tilde{\alpha})-\log\tilde{\beta}) -\bsi \frac{\tilde{\alpha}}{\tilde{\beta}}
\end{align*}

\tb{Key Components:}
\begin{enumerate}
    \item Expectation of a log Gamma RV: $E[\text{log} \frac{1}{\ssq}] \approx \psi(\tilde{\alpha})-\log\tilde{\beta} $
    \item Expectation of Quadratic Forms: $E_{q(\bv)}\Big[\bv'\bSig_{\bv}^{-1}\bv\Big]=\tbmu_\bv'\bSig_{\bv}^{-1}\tbmu_\bv+tr[\bSig_{\bv}^{-1}\tbC_\bv]$
    \item  $-\frac{1}{2}E[\log{|\bSig_{\bv}|}]=-\frac{1}{2}E[\log |\bSig_\beta|]+ \frac{m}{2}E[\log \frac{1}{\ssq}]= -\frac{1}{2}\log |\bSig_\beta |+ \frac{m}{2}(\psi(\tilde{\alpha})-\log\tilde{\beta})$
\end{enumerate}

\tred{Part 2:} $\tred{E_q[\log q(\bv,\ssq)]}$
\begin{align*}
    E_q[\log q(\bv,\ssq)]& \propto E_q\Big[\log[q(v)]+\log[q(\ssq)]\Big] \\
    & = E_q\Big[\log[2\pi^{\frac{-(m+p)}{2}} |\tbC_\bv|^{\frac{-1}{2}} \exp(-\frac{1}{2}(\bv-\tbmu_{\bv})'\tbC_\bv^{-1}(\bv-\tbmu_{\bv}) ) \\  
    & \qquad +\log [\frac{\tilde{\beta}^{\tilde{\alpha}}}{\Gamma(\tilde{\alpha})} (\ssq)^{-\tilde{\alpha}-1} \cdot \exp(\frac{-\tilde{\beta}}{\ssq}) ]\Big] \\
    & = 
    E_{q}\Big[-\frac{(m+p)}{2}\log2\pi-\frac{1}{2} \log|\tbC_\bv|-\frac{1}{2}(\bv-\tbmu_{\bv})'\tbC_\bv^{-1}(\bv-\tbmu_{\bv})+\tilde{\alpha}\log\tilde{\beta} \\
    & \qquad -\log\Gamma(\tilde{\alpha}) + (\tilde{\alpha}+1)\log\frac{1}{\ssq}-\tilde{\beta}\frac{1}{\ssq} \Big]\\ 
    & =
    -\frac{(m+p)}{2}\log2\pi-\frac{1}{2} \log|\tbC_\bv| -\frac{1}{2} tr[\tbC_\bv^{-1}\tbC_\bv]+\tilde{\alpha}\log\tilde{\beta}-\log\Gamma(\tilde{\alpha}) \\
    &\qquad +(\tilde{\alpha}+1)E_q[\log \frac{1}{\ssq}]-\tilde{\beta}E_q[\frac{1}{\ssq}] \\
    & =
    -\frac{(m+p)}{2}\log2\pi-\frac{1}{2} \log|\tbC_\bv| -\frac{m+p}{2} +\tilde{\alpha}\log\tilde{\beta} -\log\Gamma(\tilde{\alpha})\\
    &\qquad +(\tilde{\alpha}+1)(\psi(\tilde{\alpha})-\log\tilde{\beta}) -\tilde{\beta}\frac{\tilde{\alpha}}{\tilde{\beta}}\\
        & =
    -\frac{(m+p)}{2}(\log2\pi+1)-\frac{1}{2} \log|\tbC_\bv|  +\tilde{\alpha}\log\tilde{\beta} -\log\Gamma(\tilde{\alpha})\\
    &\qquad +(\tilde{\alpha}+1)(\psi(\tilde{\alpha})-\log\tilde{\beta}) -\tilde{\alpha}
\end{align*}

\noindent \tb{Evidence Lower Bound:}
\begin{align*}
    ELBO&=E_q\Bigg[\log \frac{p(\bZ, \bbe, \bW ,\ssq, \phi)}{q(\bbe,\bW,\ssq, \phi)}\Bigg]\\
    & = \tblue{E_q[\log p(\bZ, \bbe, \bW,\ssq, \phi)]} - \tred{E_q[log(q(\bbe,\bW,\ssq, \phi))]} \\
       & = \tblue{\tbmu_{\bv}'\tilde{\bX}'\bD\tilde{\bX}\tbmu_{\bv}+tr[\tilde{\bX}'\bD\tilde{\bX} \tbC_{\bv}]-\frac{1}{2}\tbmu_{\bv}'\bSig_{\bv}^{-1}\tbmu_{\bv}-\frac{1}{2} tr[\bSig_{\bv}^{-1}\tbC_{\bv}]+(\bZ'-\frac{1}{2}\bOne')\tilde{\bX}\tbmu_{\bv} }\\
    & \qquad \tblue{+\bOne'\psi(\xi) -\frac{1}{2}\Big(\log{|\bSig_{\beta}|}+(m+p)\log(2\pi) \Big)} \\
    & \qquad \tblue{+\asi\log\bsi-\log\Gamma(\asi)+(\alpha_\sigma+\frac{m}{2}+1)
    (\psi(\tilde{\alpha})-\log\tilde{\beta}) -\bsi \frac{\tilde{\alpha}}{\tilde{\beta}}} \\
    & \qquad
    \tred{-\Bigg[-\frac{(m+p)}{2}(\log2\pi+1)-\frac{1}{2} \log|\tbC_\bv|+\tilde{\alpha}\log\tilde{\beta} -\log\Gamma(\tilde{\alpha})}\\
    &\qquad \tred{+(\tilde{\alpha}+1)(\psi(\tilde{\alpha})-\log\tilde{\beta}) -\tilde{\alpha} \Bigg]} \\ 
    & \propto \tblue{\tbmu_{\bv}'\tilde{\bX}'\bD\tilde{\bX}\tbmu_{\bv}+tr[\tilde{\bX}'\bD\tilde{\bX} \tbC_{\bv}]-\frac{1}{2}\tbmu_{\bv}'\bSig_{\bv}^{-1}\tbmu_{\bv}-\frac{1}{2} tr[\bSig_{\bv}^{-1}\tbC_{\bv}] +(\bZ'-\frac{1}{2}\bOne')\tilde{\bX}\tbmu_{\bv}}\\
    & \qquad \tblue{+\bOne'\psi(\xi) -\frac{1}{2}\log{|\bSig_{\beta}|}+(\alpha_\sigma+\frac{m}{2}+1)
    (\psi(\tilde{\alpha})-\log\tilde{\beta}) -\bsi \frac{\tilde{\alpha}}{\tilde{\beta}}} \\
    & \qquad
    \tred{-\Bigg[-\frac{1}{2} \log|\tbC_\bv|+\tilde{\alpha}\log\tilde{\beta} -\log\Gamma(\tilde{\alpha})+(\tilde{\alpha}+1)(\psi(\tilde{\alpha})-\log\tilde{\beta}) -\tilde{\alpha} \Bigg]} \\ 
    &\propto \tbmu_{\bv}'\tilde{\bX}'\bD\tilde{\bX}\tbmu_{\bv}+tr[\tilde{\bX}'\bD\tilde{\bX} \tbC_{\bv}]-\frac{1}{2}\tbmu_{\bv}'\bSig_{\bv}^{-1}\tbmu_{\bv}-\frac{1}{2} tr[\bSig_{\bv}^{-1}\tbC_{\bv}]+(\bZ'-\frac{1}{2}\bOne')\tilde{\bX}\tbmu_{\bv}\\
    & \qquad +\bOne'\psi(\xi)+(\alpha_\sigma+\frac{m}{2}-\tilde{\alpha})
    (\psi(\tilde{\alpha})-\log\tilde{\beta})-\bsi \frac{\tilde{\alpha}}{\tilde{\beta}}\\
     & \qquad
    +\frac{1}{2} \log|\tbC_\bv|-\tilde{\alpha}\log\tilde{\beta} +\tred{\log\Gamma(\tilde{\alpha})} +\tilde{\alpha} \\
    &\propto \tbmu_{\bv}'\tilde{\bX}'\bD\tilde{\bX}\tbmu_{\bv}+tr[\tilde{\bX}'\bD\tilde{\bX} \tbC_{\bv}]-\frac{1}{2}\tbmu_{\bv}'\bSig_{\bv}^{-1}\tbmu_{\bv}-\frac{1}{2} tr[\bSig_{\bv}^{-1}\tbC_{\bv}]+(\bZ'-\frac{1}{2}\bOne')\tilde{\bX}\tbmu_{\bv}\\
    & \qquad +\bOne'\psi(\xi)-\bsi \frac{\tilde{\alpha}}{\tilde{\beta}}+\frac{1}{2} \log|\tbC_\bv|-\tilde{\alpha}\log\tilde{\beta}
    \end{align*}

\newpage
\section{Basis-SGLMM (Poisson Data Model): Discretized \texorpdfstring{$\sigma^{2(j)}$}{Lg} }
\tb{Hierarchical Model:}
% \vspace{-0.3in}
\begin{align*}\vspace{-0.1in}
         \textbf{\mbox{Count Data Model:}} &\quad  Z_i|\lambda_i \sim \mbox{Pois}(\lambda_i)\\
    & \qquad \lambda_i=\exp\{\bX_{i}^\prime\bbe +\bgPhi_{i}^\prime\bdel\}\\
    \tb{Process Model:}& \qquad \bdel|\ssq \sim \mcN(\bzero,\ssq\bI)\\
    \tb{Prior Model:}& \qquad \bbe \sim \mcN(\bzero,\bSig_{\beta}), \qquad \ssq \sim IG(\alpha_\sigma, \beta_\sigma)
\end{align*}
\textbf{Hierarchical Model (Modified):}
% \vspace{-0.3in}
\begin{align*}\vspace{-0.1in}
       \textbf{\mbox{Count Data Model:}} &\quad  Z_i|\lambda_i \sim \mbox{Pois}(\lambda_i) \quad \text{where} \hspace{0.1cm} \lambda_i = \exp\{{\widetilde{\bX_{i}}'}\bv \} \\
    & \quad \textcolor{black}{\widetilde{\bX}}=[\bX \quad \bgPhi]\mbox{ and } \textcolor{black}{\bv}=\pr{(\bbe , \bdel)}\\
    \textbf{\mbox{Process Model:}} &\\
        & \quad  \textcolor{black}{\bv|\ssq  \sim \mcN \bigg(\begin{bmatrix}\mu_{\beta} \\ 0 \end{bmatrix},  \begin{bmatrix}
\Sigma_{\beta} & 0  \\
0 & \sigma^2 \textcolor{black}{I_{n}}  
\end{bmatrix} \bigg)} \\ 
    \textbf{\mbox{Prior Model:}}& \\ 
    & \quad \sigma^2 \sim \mbox{IG}(\alpha_\sigma,\beta_\sigma) \\
\end{align*}
\noindent \tb{Objective:} Obtain variational functions $q(\bv)$ and $q(\ssq)$ via Mean Field Variational Bayes(MFVB) to approximate $p(\bv|\cdot)$ and $p(\ssq|\cdot)$.  \\

\tb{Probability Density Functions}
% \vspace{-0.3in}
\begin{align*}
    \mbox{Joint :}& \qquad p(\bZ, \bbe, \bdel,\ssq)= p(\bZ|\bbe, \bdel)p(\bbe)p(\bdel|\ssq)p(\ssq)\\
    \mbox{Likelihood :}& \qquad p(\bZ|\bga)= \prod_{i=1}^{n}\frac{\lambda_i^{Z_i}e^{-\lambda_i}}{Z_i!}, \qquad \mbox{where } \lambda_i=\exp\{\bX_{i}^\prime\bbe +\bgPhi_{i}^\prime\bdel\}\\
    \mbox{Process :}&\qquad p(\bdel|\ssq) = (2\pi)^{-m/2}(\ssq)^{-m/2}\exp\{-\frac{1}{2\ssq}\bdel'\bdel\}\\
    \mbox{Prior :}& \qquad p(\bbe) = (2\pi)^{-p/2}|\bSig_{\beta}|^{-1/2}\exp\{-\frac{1}{2}\bbe'\bSig_{\beta}^{-1}\bbe\}\\
    & \qquad  p(\ssq)= \frac{\bsi^{\asi}}{\Gamma(\asi)}(\ssq)^{-\asi-1}\exp\{-\frac{\bsi}{\ssq}\} \\
    \mbox{Proposal :}& \qquad q(v)=2\pi^{\frac{-(n+p)}{2}} |\tbC_\bv|^{\frac{-1}{2}} \exp(-\frac{1}{2}(\bv-\tbmu_{\bv})'\tbC_\bv^{-1}(\bv-\tbmu_{\bv}))
\end{align*}

\noindent \tb{Log joint posterior density (Original)}:
\begin{align*}
    \log[p(\bZ, \bbe, \bdel , \ssq)]& = \log[p(\bZ|\bbe, \bdel)]+\log[p(\bbe)]+\log[p(\bdel|\ssq)]+\log[p(\ssq)]\\
    & = \bZ'(\bX\bbe+\bgPhi\bdel) -\bOne' \Big(e^{\bX\bbe+\bgPhi\bdel}\Big)- \bOne' \log\bZ!\\
    & \qquad -\frac{1}{2}\Bigg((m+p)\log(2\pi)+m\log\ssq+\log|\bSig_{\beta}|+\frac{1}{\ssq}\bdel'\bdel+\bbe'\bSig_{\beta}^{-1}\bbe
    \Bigg)\\
    & \qquad +\asi\log\bsi-\log\Gamma(\asi) -(\asi+1) \log\ssq -\frac{\bsi}{\ssq}
\end{align*}

\noindent \tb{Log joint posterior density (Modified)}:\\
Let $\bv=(\bbe, \bdel)^\prime$, $\tilde{\bX}=[\bX\quad \bgPhi]$, and $\bSig_{\bv}=\begin{bmatrix}
\bSig_{\beta}  & \bzero \\
\bzero & \ssq\bI
\end{bmatrix}$, then 
\begin{align*}
    \log[p(\bZ, \bv ,\ssq)]& =\bZ'\tilde{\bX}\bv -\bOne' \Big(e^{\tilde{\bX}\bv}\Big)-\bOne' \log\bZ!\\
    & \qquad -\frac{1}{2}\Big((m+p)\log(2\pi)+m\log \ssq +\log |\bSig_{\beta}|+\bv'\bSig_{v}^{-1}\bv\Big)\\
    & \qquad +\asi\log\bsi-\log\Gamma(\asi)-(\asi+1) \log\ssq -\frac{\bsi}{\ssq}
\end{align*}

\subsection{Variational Function for \texorpdfstring{$\bbe$ and $\bW$}{Lg} }
\noindent \tb{Variational Distribution for $\bv=(\bbe,\bdel)'$:} The distribution that minimizes the KL divergence is $q(\bv)\propto \exp\{E_{-\bv}[ \log p(\bZ, \bv , \ssq)]\}$. For the Poisson case, this distribution is not available in closed form; hence, we provide a Gaussian approximation via Laplace approximation (2nd order Taylor Expansion). The objective function is as follows:

\begin{align*}
    f(\bv)&= E_{q(-\bv)}[\log p(\bZ, \bv, \ssq)] \\
    &= E_{q(-\bv)}\Big[\bZ'\tilde{\bX}\bv -\bOne' \Big(e^{\tilde{\bX}\bv}\Big)-\bOne' \log\bZ!\\
    & \qquad \qquad -\frac{1}{2}\Big((m+p)\log(2\pi)+m\log \ssq +\log |\bSig_{\beta}|+\bv'\bSig_{v}^{-1}\bv\Big)\\
    & \qquad \qquad +\asi\log\bsi-\log\Gamma(\asi)-(\asi+1) \log\ssq -\frac{\bsi}{\ssq} \Big] \\   
    & = \bZ'\tilde{\bX}\bv -\bOne' \Big(e^{\tilde{\bX}\bv}\Big)-\bOne' \log\bZ!\\
    & \qquad \qquad -\frac{1}{2}\Big((m+p)\log(2\pi)+m\log \ssq +\log |\bSig_{\beta}|+\bv'\bSig_{v}^{-1}\bv\Big)\\
    & \qquad \qquad +\asi\log\bsi-\log\Gamma(\asi)-(\asi+1) \log\ssq -\frac{\bsi}{\ssq} 
\end{align*}

\tb{Variational Function for $\bv$:} For objective function and corresponding gradient
$$
f(\bv)\propto \bZ'\tilde{\bX}\bv -\bOne' \Big(e^{\tilde{\bX}\bv}\Big)-\frac{1}{2}\bv'\bSig_{v}^{-1}\bv
$$
$$\nabla f(\bv)= \tilde{\bX}'\bZ- \tilde{\bX}'Diag(e^{\tilde{\bX}\bv})\bOne'  - \bSig_{v}^{-1}\bv$$
we have the resulting variational function 
$$q_v(\bv)= \mcN(\tbmu_v, \tbC_{v})$$ where $\tbmu_\bv=\argmax_{\bv}f(\bv)$ and $\tbC_\bv= -(\bH)^{-1}$ where $\bH=\frac{\partial^2 f}{\partial \bv^2}\Bigr|_{\substack{\bv=\tbmu_\bv}}$.\\ 
Note that the mean vector and covariance matrix are partitioned as follows:

$$
\tbmu_{\bv} = (\tbmu_\beta , \tbmu_\delta)', \qquad \tbC_{\bv}=\begin{bmatrix}
\tbC_\beta  & \tbC_{\beta,\delta} \\
\tbC_{\delta,\beta} & \tbC_{\delta}
\end{bmatrix}, 
$$

\subsection{Evidence Lower Bound}
\begin{align*}
    ELBO&=E_q\Bigg[\log \frac{p(\bZ, \bv ,\ssq)}{q(\bv,\ssq)}\Bigg]\\
    & = \tblue{E_q[\log p(\bZ, \bv ,\ssq)]} - \tred{E_q[\log q(\bv,\ssq)]}
\end{align*}

Here, we have to decompose by partitioning the parameter space $\btheta=(\pmb{\theta}_c,{\pmb{\theta}_d})'$. We estimate $\pmb{\theta}_c$ but fix $ \pmb{\theta}_d$
$$\pmb{\theta}_c={ \{\bv, \ssq \} }, \qquad \pmb{\theta}_d={ \{\ssq \} } $$

\tblue{Part 1:} $\tblue{E_q[\log p(\bZ, \bv,\ssq )]}$
\begin{align*}
    E_q[\log p(\bZ, \bv,\ssq)] & = E_q\Big[ \bZ'\tilde{\bX}\bv -\bOne' \Big(e^{\tilde{\bX}\bv}\Big)-\bOne' \log\bZ!\\
    & \qquad \qquad -\frac{1}{2}\Big((m+p)\log(2\pi)+m\log \ssq +\log |\bSig_{\beta}|+\bv'\bSig_{v}^{-1}\bv\Big)\\
    & \qquad \qquad +\asi\log\bsi-\log\Gamma(\asi)-(\asi+1) \log\ssq -\frac{\bsi}{\ssq} \Big] \\
    & = \bZ'\tilde{\bX}\tbmu_v -\bOne' E[e^{\tilde{\bX}\bv}]-\bOne' \log\bZ! -\frac{(m+p)}{2}\log(2\pi)\\
    & \qquad \qquad -\frac{1}{2}\log |\bSig_\beta|-\frac{1}{2}E[\bv'\bSig_{v}^{-1}\bv] \\
    & \qquad \qquad +\asi\log\bsi-\log\Gamma(\asi)-(\asi+\frac{m}{2}+1)\log\ssq - \frac{\bsi}{\ssq} \\    
    & = \bZ'\tilde{\bX}\tbmu_v -\bOne'\exp\{\tbX\tbmu_{v}+\frac{1}{2}\mbox{diag}(\tbX\tbC_{v}\tbX')\}-\bOne' \log\bZ! -\frac{(m+p)}{2}\log(2\pi)\\
    & \qquad \qquad -\frac{1}{2}\log |\bSig_\beta|-\frac{1}{2}\Big[\tbmu_v'\bSig_{v}^{-1}\tbmu_v+tr[\bSig_{v}^{-1}\tbC_v]\Big] \\
    & \qquad \qquad +\asi\log\bsi-\log\Gamma(\asi)-(\asi+\frac{m}{2}+1)\log\ssq - \frac{\bsi}{\ssq} \\ 
\end{align*}

\tb{Key Components:}
\begin{enumerate}
    \item Expectation of a log Gamma RV: $E[\text{log} \frac{1}{\ssq}] \approx \psi(\tilde{\alpha})-\log\tilde{\beta} $
    \item Expectation of Quadratic Forms: $E_{q(\bv)}\Big[\bv'\bSig_{\bv}^{-1}\bv\Big]=\tbmu_\bv'\bSig_{\bv}^{-1}\tbmu_\bv+tr[\bSig_{\bv}^{-1}\tbC_\bv]$
\end{enumerate}

\tred{Part 2:} $\tred{E_q[\log q(\bv,\ssq)]}$
\begin{align*}
    E_q[\log q(\bv,\ssq)]& \propto E_q [\log q(v)] \\
    & = E_q\Big[\log[2\pi^{\frac{-(m+p)}{2}} |\tbC_\bv|^{\frac{-1}{2}} \exp(-\frac{1}{2}(\bv-\tbmu_{\bv})'\tbC_\bv^{-1}(\bv-\tbmu_{\bv}) ) \Big] \\
    & = E_{q}\Big[-\frac{(m+p)}{2}\log2\pi-\frac{1}{2} \log|\tbC_\bv|-\frac{1}{2}(\bv-\tbmu_{\bv})'\tbC_\bv^{-1}(\bv-\tbmu_{\bv}) \Big]\\
    & = -\frac{(m+p)}{2}\log2\pi-\frac{1}{2} \log|\tbC_\bv| -\frac{1}{2} tr[\tbC_\bv^{-1}\tbC_\bv] \\
    & = -\frac{(m+p)}{2}\log2\pi-\frac{1}{2} \log|\tbC_\bv| -\frac{1}{2}(m+p) \\
\end{align*}

\newpage
\noindent \tb{Evidence Lower Bound}
\begin{align*}
    ELBO&=E_q\Bigg[\log \frac{p(\bZ, \bv ,\ssq)}{q(\bv,\ssq)}\Bigg]\\
    & = \tblue{E_q[\log p(\bZ, \bv ,\ssq)]} - \tred{E_q[\log q(\bv,\ssq)]} \\
    & = \tblue{\bZ'\tilde{\bX}\tbmu_v -\bOne'\exp\{\tbX\tbmu_{v}+\frac{1}{2}\mbox{diag}(\tbX\tbC_{v}\tbX')\}-\bOne' \log\bZ! -\frac{(m+p)}{2}\log(2\pi)}\\
    & \tblue{\qquad  -\frac{1}{2}\log |\bSig_\beta|-\frac{1}{2}\Big[\tbmu_v'\bSig_{v}^{-1}\tbmu_v+tr[\bSig_{v}^{-1}\tbC_v]\Big]} \\
    & \tblue{\qquad  +\asi\log\bsi-\log\Gamma(\asi)-(\asi+\frac{m}{2}+1)\log\ssq - \frac{\bsi}{\ssq}} \\ 
    & \tred{\qquad- \Big[ -\frac{(m+p)}{2}\log2\pi-\frac{1}{2} \log|\tbC_\bv| -\frac{1}{2}(m+p) \Big]} \\
    & = \tblue{\bZ'\tilde{\bX}\tbmu_v -\bOne'\exp\{\tbX\tbmu_{v}+\frac{1}{2}\mbox{diag}(\tbX\tbC_{v}\tbX')\}-\bOne' \log\bZ!} \\
    & \tblue{\qquad  -\frac{1}{2}\log |\bSig_\beta|-\frac{1}{2}\Big[\tbmu_v'\bSig_{v}^{-1}\tbmu_v+tr[\bSig_{v}^{-1}\tbC_v]\Big]} \\
    & \tblue{\qquad  +\asi\log\bsi-\log\Gamma(\asi)-(\asi+\frac{m}{2}+1)\log\ssq - \frac{\bsi}{\ssq}} \\ 
    & \tred{\qquad +\frac{1}{2} \log|\tbC_\bv| +\frac{1}{2}(m+p) } \\
    & \propto \tblue{\bZ'\tilde{\bX}\tbmu_v -\bOne'\exp\{\tbX\tbmu_{v}+\frac{1}{2}\mbox{diag}(\tbX\tbC_{v}\tbX')\}} \\
    & \tblue{\qquad  -\frac{1}{2}\log |\bSig_\beta|-\frac{1}{2}\Big[\tbmu_v'\bSig_{v}^{-1}\tbmu_v+tr[\bSig_{v}^{-1}\tbC_v]\Big]} \\
    & \tblue{\qquad -(\asi+\frac{m}{2}+1)\log\ssq - \frac{\bsi}{\ssq}} \\ 
    & \tred{\qquad +\frac{1}{2} \log|\tbC_\bv|  } \\
\end{align*}

\newpage
\section{Basis-SGLMM (Bernoulli Data Model): Discretized \texorpdfstring{$\sigma^{2(j)}$}{Lg} }
\tb{Hierarchical Model:}
% \vspace{-0.3in}
\begin{align*}\vspace{-0.1in}
    \tb{Binary Data Model:} &\qquad  Z_i|p_i \sim \mbox{Bern}(p_i)\\
    & \qquad p_i = (1+\exp\{-\bX_{i}^\prime\bbe -\bgPhi_{i}^\prime\bdel\})^{-1}\\
    \tb{Process Model:}& \qquad \bdel|\ssq \sim \mcN(\bzero,\ssq\bI)\\
    \tb{Prior Model:}& \qquad \bbe \sim \mcN(\bzero,\bSig_{\beta}), \qquad \ssq \sim IG(\alpha_\sigma, \beta_\sigma)
\end{align*}

\textbf{Hierarchical Model (Modified):}
% \vspace{-0.3in}
\begin{align*}\vspace{-0.1in}
       \textbf{\mbox{Binary Data Model:}} &\quad  Z_i|p_i \sim \mbox{Bern}(p_i) \quad \text{where} \hspace{0.1cm} p_i = (1+\exp\{-{\widetilde{\bX_{i}}'}\bv \})^{-1} \\
    & \quad \textcolor{black}{\widetilde{\bX}}=[\bX \quad \bgPhi]\mbox{ and } \textcolor{black}{\bv}=\pr{(\bbe , \bdel)}\\
    \textbf{\mbox{Process Model:}} &\\
        & \quad  \textcolor{black}{\bv|\ssq  \sim \mcN \bigg(\begin{bmatrix}\mu_{\beta} \\ 0 \end{bmatrix},  \begin{bmatrix}
\Sigma_{\beta} & 0  \\
0 & \sigma^2 \textcolor{black}{I_{n}}  
\end{bmatrix} \bigg)} \\ 
    \textbf{\mbox{Prior Model:}}& \\ 
    & \quad \sigma^2 \sim \mbox{IG}(\alpha_\sigma,\beta_\sigma) \\
\end{align*}
\noindent \tb{Objective:} Obtain variational functions $q(\bv)$ and $q(\ssq)$ via Mean Field Variational Bayes(MFVB) to approximate $p(\bv|\cdot)$ and $p(\ssq|\cdot)$.  \\

\tb{Probability Density Functions}
% \vspace{-0.3in}
\begin{align*}
    \mbox{Joint :}& \qquad p(\bZ, \bbe, \bdel,\ssq)= p(\bZ|\bbe, \bdel)p(\bbe)p(\bdel|\ssq)p(\ssq)\\
    \mbox{Likelihood :}& \qquad p(\bZ|\bbe, \bdel)= \prod_{i=1}^{n}p_{i}^{z_i}(1-p_{i})^{1-z_i}, \qquad \mbox{where } p_i=(1+\exp\{-\bX_{i}^\prime\bbe -\bgPhi_{i}^\prime\bdel\})^{-1}\\
    \mbox{Process :}&\qquad p(\bdel|\ssq) = (2\pi)^{-m/2}(\ssq)^{-m/2}\exp\{-\frac{1}{2\ssq}\bdel'\bdel\}\\
    \mbox{Prior :}& \qquad p(\bbe) = (2\pi)^{-p/2}|\bSig_{\beta}|^{-1/2}\exp\{-\frac{1}{2}\bbe'\bSig_{\beta}^{-1}\bbe\}\\
    & \qquad  p(\ssq)= \frac{\bsi^{\asi}}{\Gamma(\asi)}(\ssq)^{-\asi-1}\exp\{-\frac{\bsi}{\ssq}\}\\
    \mbox{Proposal :}& \qquad q(v)=2\pi^{\frac{-(n+p)}{2}} |\tbC_\bv|^{\frac{-1}{2}} \exp(-\frac{1}{2}(\bv-\tbmu_{\bv})'\tbC_\bv^{-1}(\bv-\tbmu_{\bv}))
\end{align*}

\noindent \tb{Log joint posterior density Fix $\ssq$ (Original)}:
\begin{align*}
    \log[p(\bZ, \bbe, \bdel , \ssq)]& = \log[p(\bZ|\bbe, \bdel)]+\log[p(\bbe)]+\log[p(\bdel|\ssq)]+\log[p(\ssq)]\\
    &= \bZ^{\prime}\bX\bbe+\bZ^{\prime}\bgPhi\bdel -\bOne^\prime \log (1+\exp\{\bX\bbe+\bgPhi\bdel\})\\
    & \qquad -\frac{1}{2}\Big(\bbe'\bSig_{\beta}^{-1}\bbe+m\log\ssq+(m+p)\log(2\pi)+\log|\bSig_{\beta}| +\frac{1}{\ssq}\bdel'\bdel\Big)\\
    & \qquad +\asi\log\bsi-\log\Gamma(\asi) -(\asi+1) \log\ssq -\frac{\bsi}{\ssq}
\end{align*}

\noindent \tb{Quadratic Approximation (Jaakola and Jordan, 1997):}
$$
-\log(1+e^{x})=\argmax_{\xi}\Big\{ \lambda(\xi)x^{2}-\frac{1}{2}x +\psi(\xi)\Big\}
$$
where $\lambda(\xi)=-\tanh(\xi/2)/(4\xi)$ and $\psi(\xi)=\xi/2-\log(1+e^{\xi})+\xi\tanh(\xi/2)/4$
\\
\\
\noindent \tb{Computing the optimal Auxiliary Variables} (Jaakola and Jordan, 1997)

$$
\pmb{\xi}= \sqrt{\mbox{Diagonal}\Big\{\tilde{X}(\tbC_{\bv}+\tbmu_{\bv}\tbmu_{\bv}')\tilde{X}'\Big\}}
$$

\noindent \tb{Log joint posterior density Fix $\ssq$ (Modified)}:\\
Let $\bv=(\bbe, \bdel)^\prime$, $\tilde{\bX}=[\bX\quad \bgPhi]$, and $\bSig_{\bv}=\begin{bmatrix}
\bSig_{\beta}  & \bzero \\
\bzero & \ssq\bI
\end{bmatrix}$, then 
\begin{align*}
    \log[p(\bZ, \bbe, \bdel , \ssq)]& = \bZ^{\prime}\tilde{\bX}\bv 
    + \bOne'\Big[\bv'\tilde{\bX}'\bD\tilde{\bX}\bv-\frac{1}{2}\tilde{\bX}\bv+\psi(\xi)\Big]\\
    & \qquad -\frac{1}{2}\Big(\bbe'\bSig_{\beta}^{-1}\bbe+m\log\ssq+(m+p)\log(2\pi)+\log|\bSig_{\beta}| +\frac{1}{\ssq}\bdel'\bdel\Big)\\
    & \qquad +\asi\log\bsi-\log\Gamma(\asi)-(\asi+1) \log\ssq -\frac{\bsi}{\ssq}\\
    &= \bv'\tilde{\bX}'\bD\tilde{\bX}\bv - \frac{1}{2}\bOne'\tilde{\bX}\bv+\bOne'\psi(\xi)+\bZ'\tilde{\bX}\bv-\frac{1}{2}\bv'\bSig_{\bv}^{-1}\bv \\
    & \qquad -\frac{1}{2}\Big(\log|\bSig_\bv|+(m+p)\log(2\pi) \Big) \\
    & \qquad +\asi\log\bsi-\log\Gamma(\asi)-(\alpha_\sigma+1)\log\ssq-\frac{\bsi}{\ssq}\\
    & = -\frac{1}{2}\Big(\bv'(-2\tilde{\bX}'\bD\tilde{\bX}+\bSig_{\bv}^{-1})\bv-2(\bZ'-\frac{1}{2}\bOne')\tilde{\bX}\bv\Big)+\bOne'\psi(\xi)\\
    & \qquad -\frac{1}{2}\Big(\log|\bSig_\bv|+(m+p)\log(2\pi)\Big) \\
    & \qquad +\asi\log\bsi-\log\Gamma(\asi)-(\alpha_\sigma+1)\log\ssq-\frac{\bsi}{\ssq}
\end{align*}
where $\bf{D}=\mbox{diag}(\lambda(\pmb{\xi}))$. 

\subsection{Variational Function for \texorpdfstring{$\bbe$ and $\bW$}{Lg} }
We represent $\bv=(\bbe,\bW)'$ to preserve dependece between $\bbe$ and $\bW$.  The distribution that minimizes the KL divergence is $q(\bv)\propto \exp\{E_{-\bv}[ \log p(\bZ, \bv , \ssq)]\}$, or:
$$q(\bv)\sim \mcN(\tbmu_{\bv}, \tbC_{\bv})$$
where $\tbC_{\bv}= (-2\tilde{\bX}'\bf{D}\tilde{\bX}+\bSig_{\bv}^{-1})^{-1}$ and $\tbmu_{\bv}=\tbC_{\bv}\tilde{\bX}'(\bZ-\frac{1}{2}\bOne')$ and 
$$
\bSig_{\bv}^{-1}=\begin{bmatrix}
\bSig_{\beta}^{-1}  & \bzero \\
\bzero & \frac{1}{ \ssq}\bI
\end{bmatrix}
$$

Note that the covariance matrix and mean vector are split as follows:
$$
\tbC_{\bv}=\begin{bmatrix}
\tbC_\beta  & \tbC_{\beta,\delta} \\
\tbC_{\delta,\beta} & \tbC_{\delta}
\end{bmatrix}, \qquad \tbmu_{\bv} = (\tbmu_\beta , \tbmu_\delta)'
$$

\subsection{Evidence Lower Bound}
\begin{align*}
    ELBO&=E_q\Bigg[\log \frac{p(\bZ, \bv ,\ssq)}{q(\bv,\ssq)}\Bigg]\\
    & = \tblue{E_q[\log p(\bZ, \bv ,\ssq)]} - \tred{E_q[\log q(\bv,\ssq)]}
\end{align*}

Here, we have to decompose by partitioning the parameter space $\btheta=(\pmb{\theta}_c,{\pmb{\theta}_d})'$. We estimate $\pmb{\theta}_c$ but fix $ \pmb{\theta}_d$
$$\pmb{\theta}_c={ \{\bv, \ssq \} }, \qquad \pmb{\theta}_d={ \{\ssq \} } $$

\tblue{Part 1:} $\tblue{E_q[\log p(\bZ, \bv,\ssq )]}$
\begin{align*}
    E_q[\log p(\bZ, \bv,\ssq)] & = E_q\Big[-\frac{1}{2}\Big(\bv'(-2\tilde{\bX}'\bD\tilde{\bX}+\bSig_{\bv}^{-1})\bv-2(\bZ'-\frac{1}{2}\bOne')\tilde{\bX}\bv\Big)+\bOne'\psi(\xi)\\
    & \qquad -\frac{1}{2}\Big(\log|\bSig_\bv|+(m+p)\log(2\pi)\Big) \\
    & \qquad +\asi\log\bsi-\log\Gamma(\asi)-(\alpha_\sigma+1)\log\ssq-\frac{\bsi}{\ssq} \Big]\\
    & =
    E_{q}[\bv'\tilde{X}'\bD\tilde{X}\bv]-\frac{1}{2}E_{q}[\bv'\bSig_{\bv}^{-1}\bv]+E_{q}[(\bZ'-\frac{1}{2}\bOne')\tilde{X}\bv]+\bOne'\psi(\xi) \\
    & \qquad -\frac{1}{2}\Big(\log|\bSig_\bv|+(m+p)\log(2\pi)\Big) \\
    & \qquad +\asi\log\bsi-\log\Gamma(\asi)-(\alpha_\sigma+1)\log\ssq-\frac{\bsi}{\ssq} \\
    & =
    \tbmu_{\bv}'\tilde{X}'\bD\tilde{X}\tbmu_{\bv}+tr[\tilde{X}'\bD\tilde{X} \tbC_{\bv}]-\frac{1}{2}\tbmu_{\bv}'\bSig_{\bv}^{-1}\tbmu_{\bv}-\frac{1}{2} tr[\bSig_{\bv}^{-1}\tbC_{\bv}] \\
    & \qquad +(\bZ'-\frac{1}{2}\bOne')\tilde{X}\tbmu_{\bv}+\bOne'\psi(\xi) \\
    & \qquad -\frac{1}{2}\Big(\log|\bSig_\bv|+(m+p)\log(2\pi)\Big) \\
    & \qquad +\asi\log\bsi-\log\Gamma(\asi)-(\alpha_\sigma+1)\log\ssq-\frac{\bsi}{\ssq} \\
\end{align*}

\tb{Key Components:}
\begin{enumerate}
    \item Expectation of a log Gamma RV: $E[\text{log} \frac{1}{\ssq}] \approx \psi(\tilde{\alpha})-\log\tilde{\beta} $
    \item Expectation of Quadratic Forms: $E_{q(\bv)}\Big[\bv'\bSig_{\bv}^{-1}\bv\Big]=\tbmu_\bv'\bSig_{\bv}^{-1}\tbmu_\bv+tr[\bSig_{\bv}^{-1}\tbC_\bv]$
\end{enumerate}

\tred{Part 2:} $\tred{E_q[\log q(\bv,\ssq)]}$
\begin{align*}
    E_q[\log q(\bv,\ssq)]& \propto E_q [\log q(v)] \\
    & = E_q\Big[\log[2\pi^{\frac{-(m+p)}{2}} |\tbC_\bv|^{\frac{-1}{2}} \exp(-\frac{1}{2}(\bv-\tbmu_{\bv})'\tbC_\bv^{-1}(\bv-\tbmu_{\bv}) ) \Big] \\
    & = E_{q}\Big[-\frac{(m+p)}{2}\log2\pi-\frac{1}{2} \log|\tbC_\bv|-\frac{1}{2}(\bv-\tbmu_{\bv})'\tbC_\bv^{-1}(\bv-\tbmu_{\bv}) \Big]\\
    & = -\frac{(m+p)}{2}\log2\pi-\frac{1}{2} \log|\tbC_\bv| -\frac{1}{2} tr[\tbC_\bv^{-1}\tbC_\bv] \\
    & = -\frac{(m+p)}{2}\log2\pi-\frac{1}{2} \log|\tbC_\bv| -\frac{1}{2}(m+p) \\
\end{align*}
\noindent \tb{Evidence Lower Bound}
\begin{align*}
    ELBO&=E_q\Bigg[\log \frac{p(\bZ, \bv ,\ssq)}{q(\bv,\ssq)}\Bigg]\\
    & = \tblue{E_q[\log p(\bZ, \bv ,\ssq)]} - \tred{E_q[\log q(\bv,\ssq)]} \\
    &=\tblue{\tbmu_{\bv}'\tilde{X}'\bD\tilde{X}\tbmu_{\bv}+tr[\tilde{X}'\bD\tilde{X} \tbC_{\bv}]-\frac{1}{2}\tbmu_{\bv}'\bSig_{\bv}^{-1}\tbmu_{\bv}-\frac{1}{2} tr[\bSig_{\bv}^{-1}\tbC_{\bv}]} \\
    & \tblue{\qquad +(\bZ'-\frac{1}{2}\bOne')\tilde{X}\tbmu_{\bv}+\bOne'\psi(\xi)} \\
    & \tblue{\qquad -\frac{1}{2}\Big(\log|\bSig_\bv|+(m+p)\log(2\pi)\Big)} \\
    & \tblue{\qquad +\asi\log\bsi-\log\Gamma(\asi)-(\alpha_\sigma+1)\log\ssq-\frac{\bsi}{\ssq}} \\
    & \tred{\qquad- \Big[ -\frac{(m+p)}{2}\log2\pi-\frac{1}{2} \log|\tbC_\bv| -\frac{1}{2}(m+p) \Big]} \\
    &=\tblue{\tbmu_{\bv}'\tilde{X}'\bD\tilde{X}\tbmu_{\bv}+tr[\tilde{X}'\bD\tilde{X} \tbC_{\bv}]-\frac{1}{2}\tbmu_{\bv}'\bSig_{\bv}^{-1}\tbmu_{\bv}-\frac{1}{2} tr[\bSig_{\bv}^{-1}\tbC_{\bv}]} \\
    &\tblue{\qquad +(\bZ'-\frac{1}{2}\bOne')\tilde{X}\tbmu_{\bv}+\bOne'\psi(\xi)-\frac{1}{2}\log|\bSig_\bv|} \\
    & \tblue{\qquad +\asi\log\bsi-\log\Gamma(\asi)-(\alpha_\sigma+1)\log\ssq-\frac{\bsi}{\ssq}} \\
    & \tred{\qquad +\frac{1}{2} \log|\tbC_\bv| +\frac{1}{2}(m+p) } \\
    & \propto \tblue{\tbmu_{\bv}'\tilde{X}'\bD\tilde{X}\tbmu_{\bv}+tr[\tilde{X}'\bD\tilde{X} \tbC_{\bv}]-\frac{1}{2}\tbmu_{\bv}'\bSig_{\bv}^{-1}\tbmu_{\bv}-\frac{1}{2} tr[\bSig_{\bv}^{-1}\tbC_{\bv}]} \\
    &\tblue{\qquad +(\bZ'-\frac{1}{2}\bOne')\tilde{X}\tbmu_{\bv}+\bOne'\psi(\xi)-\frac{1}{2}\log|\bSig_\bv|} \\
    & \tblue{\qquad-(\alpha_\sigma+1)\log\ssq-\frac{\bsi}{\ssq}} \\
    & \tred{\qquad +\frac{1}{2} \log|\tbC_\bv|} 
\end{align*}

\end{document}